\documentclass[conference]{IEEEtran}
\usepackage{cite}
\usepackage{amsmath}
\usepackage{newtxtext}
\usepackage{newtxmath}
\makeatletter\@ifundefined{updelta}{\let\updelta\deltaup}{}\makeatother
\usepackage{algorithmic}
\usepackage{graphicx}
\usepackage{url}
\usepackage{textcomp}
\usepackage{xcolor}
\usepackage[hidelinks,breaklinks=true]{hyperref}
\newcommand{\figwidth}{\columnwidth}
\newcommand{\figwidewidth}{\textwidth}
\newcommand{\indic}{\varmathbb{1}}

\newcommand{\papertitle}{%
Learned Bow Control on a Measured Bowed-String Model:
Two Minimum Bow Forces and the Imitation of a Steady-State Tabulation}

\newcommand{\paperabstract}{%
A finite-difference model of the violin string on rigid supports, with its friction
characteristic, wave impedance and losses taken from published measurement and its
bow contact resolved implicitly, is used to map where Helmholtz motion exists. The
upper edge of the playable band recovers Schelleng's maximum bow force. Its lower
edge splits in two. Once established, Helmholtz motion is kept down to Schelleng's
minimum: the force at which it is lost goes as $\beta^{-2}$ and as the inverse of
the quality factor, at one to three times his value computed with the model's own
losses. From rest, it appears only above $\kappa Z v_b/\beta$, with $\kappa \approx
1.1$, thirty to a hundred and ten times higher at the measured losses. The
maintenance limit holds under grid refinement, while $\kappa$ changes by up to about
eleven percent. Guettler's attack regions are reproduced, and the cepstral peak
prominence of the bridge force separates Helmholtz motion from the other regimes. On
this plant, neural controllers command the bow speed and a frog force and imitate a
greedy rule on a playability map built from rest. Its labels depend only on stroke
position and previous commands, so the objective rewards neither the oscillation
inputs nor memory. Of six architectures of matched parameter count, on four strings
at twenty seeds, a gated recurrent unit has the highest mean but is not
distinguishable from a feedforward network or a state-space model on any string.
Where the map is right, the rule is a strong baseline: no learned controller
improves on it in steady-state regulation. Under a mid-stroke friction reduction,
the controller's score regressed on the rule's has slope $0.32$, crossing at a rule
score of $88$. Under a rigid finger stop the plant is invariant, and referring the
amplitude target to the training pitch removes the transfer loss over three octaves.
}
\newcommand{\paperkeywords}{%
bowed-string instruments, Helmholtz motion, stick-slip friction, finite-difference time-domain, imitation learning, recurrent neural networks, neural network control, learning-based control%
}

\def\BibTeX{{\rm B\kern-.05em{\sc i\kern-.025em b}\kern-.08em
    T\kern-.1667em\lower.7ex\hbox{E}\kern-.125emX}}

\begin{document}

\title{\papertitle}
\hypersetup{pdftitle={\papertitle}, pdfauthor={Homayoon Beigi and Grace Conneely}}

\author{
\IEEEauthorblockA{\footnotesize Columbia University Nonlinear Control Lab\\
Technical Report No.~20260913-01\vspace{0.9\baselineskip}}

\IEEEauthorblockN{
Homayoon Beigi\textsuperscript{1,2,3}
and
Grace Conneely\textsuperscript{1}
}
\IEEEauthorblockA{
\textsuperscript{1}Department of Mechanical Engineering, Columbia University, New York, NY, USA\\
\textsuperscript{2}Department of Electrical Engineering, Columbia University, New York, NY, USA\\
\textsuperscript{3}Recognition Technologies, Inc., South Salem, NY, USA\\
Email: hb87@columbia.edu and gec2147@columbia.edu
}
}

\maketitle

\begin{abstract}
\paperabstract
\end{abstract}


\section{Introduction}

\subsection{Motivation}
\label{sec:motivation}

The sound of a violin arises from a nonlinear {\it stick-slip} interaction between
the bow and the string, which the player governs through three quantities,
namely, the normal force that the bow applies to the string, the speed of the
bow, and the distance from the bridge at which the bow crosses the string. A
small change in any one of these quantities may move the string from a regime
which sounds like a tone into one which does not. In fact, a player spends years
learning to hold the first regime while listening for its edges.

The regime which sounds like a tone is known as {\it Helmholtz motion}, a cycle in
which the bow carries the string along for most of each period and then releases
it, once per period. It has been studied since Helmholtz described it in $1863$,
and the conditions under which it forms and survives have since been stated
quantitatively. Holding those conditions across a complete bow stroke is a control
problem. A player learns it by adjusting the commands while listening to the
result. Here, we ask how far a small learned controller gets by imitating a rule
which reads a tabulation of the plant.

{\it Schelleng}~\cite{r-m:schelleng-1973} bounded the bow forces at which Helmholtz motion exists between a minimum and a maximum. His minimum is
a condition for keeping a motion which is already there. How the motion is reached
in the first place is a separate question, which {\it
Guettler}~\cite{r-m:guettler-2002-2} took up for the attack.

This paper separates the two in a model. A finite-difference string, with its
friction, impedance and losses taken from published measurement, is mapped over the
bow force, the bow speed and the bow-bridge distance. The upper edge of its playable
band is Schelleng's maximum. Its lower edge splits in two. One force is needed to
establish Helmholtz motion, and a much smaller one is enough to keep it once it is
established.

The work reported here is one layer of a longer program, and it is the layer on
which everything above it rests. The ultimate objective is a complete playing
simulation of a physical violin, parameterized from measurement throughout. In
such a simulation, the string would couple through the bridge to a body and a
soundpost instead of ending rigidly, and the left hand would be modeled as flesh
of finite impedance rather than as an ideal stop. The bow would be held at the
angle and tilt that a player actually uses, instead of normal to the string at a
fixed point of contact. Such a model would also have to reverse the bow, so that
a note is no longer a single stroke, and it would have to follow the dynamics of
real playing, beyond those of a single sustained tone.

Beyond the violin, the program would take in the rest of the family, namely, the
smaller violins, the viola, the cellos in their several sizes, the double bass,
and the contrabass. We expect these instruments to differ from the violin in the
values of the quantities which parameterize this model, rather than in kind. Of
course, this expectation is untested here, and it is offered as a direction and
not as a result. None of it is attempted in this paper, and no instrument other
than the violin is modeled, simulated, or measured anywhere in it. What we
establish here is the string and its contact with the bow, the diagnostics by
which a regime is identified, and a controller which holds the regime across a
stroke. This comes first, since a simulation which cannot hold a tone on one open
string will not hold one on a violin either.

\subsection{Literature Review}

\paragraph*{Helmholtz Motion}

{\it Helmholtz}~\cite{r-m:helmholtz-1875} was the first to describe the motion
which carries his name, and it remains the reference against which a bowed
string is judged. In Helmholtz motion, the bow carries the string along ({\it stick})
until it releases it ({\it slip}). Both the release and the recapture are triggered
by the arrival at the bow of a corner which circulates on the string.

\paragraph*{Physical Modeling of Bowed Strings}
Such classical works, by Helmholtz and since then by {\it Raman}~\cite{r-m:raman-1918},
{\it Guettler}~\cite{r-m:guettler-2002-1}, and {\it Woodhouse}~\cite{r-m:woodhouse-2004},
have contributed to the thesis that a stable violin tone arises from a repeating
and stable stick-slip cycle, and they have further characterized that
relationship through experiments and specific
equations~\cite{r-m:schelleng-1973,r-m:cremer-1984,r-m:mcintyre-1979,r-m:schoonderwaldt-2008}.
Their work underlies the modern physical models of bowed strings. In these
models, a nonlinear friction law is coupled to a one-dimensional wave equation,
which reproduces the interaction whether the string is advanced as a partial
differential equation or as a {\it digital
waveguide}~\cite{r-m:smith-2010,r-m:mcintyre-1979}.

\paragraph*{Playability}
{\it Schelleng}~\cite{r-m:schelleng-1973} put the steady state on a quantitative
footing, bounding the Helmholtz regime between a minimum and a maximum bow
force, each of which is a function of the bow speed and of the bow-bridge
distance. {\it Woodhouse}~\cite{r-m:woodhouse-1993-1,r-m:woodhouse-1993-2} recast
those bounds computationally, deriving {\it reflection functions} for the string
terminations and using them to predict the minimum bow force and the transient
which precedes steady motion. {\it Mansour,
Woodhouse, and Scavone}~\cite{r-m:mansour-2017} found that the maximum depends on
the string and on the rosin and is almost independent of the body, while the
minimum depends on the small motion at the bridge. {\it Guettler}~\cite{r-m:guettler-2002-2} then moved
the question from the steady state to the onset, mapping the bow acceleration
against the bow force and grading an attack by the number of slips which precede
regular triggering. Measurement supplies the other half of the picture.
{\it Schoonderwaldt, Guettler, and
Askenfelt}~\cite{r-m:schoonderwaldt-2008} measured both bow-force limits on a
monochord and on a violin.
{\it Askenfelt}~\cite{r-m:askenfelt-1986,r-m:askenfelt-1989} registered the bow
position, bow velocity, bow-bridge distance, and bow force in normal playing.
{\it Demoucron, Askenfelt, and Causs\'e}~\cite{r-m:demoucron-2009} built a
detachable sensor which recovers the bow force from the transverse force at the
hair termination, and {\it Ablitzer, Dalmont, and Dauchez}~\cite{r-m:ablitzer-2012}
modeled the statics of the tightened bow itself. In
Section~\ref{sec:schelleng}, we separate the force needed to establish
Helmholtz motion from rest from the much lower force at which established motion is
lost, and in Section~\ref{sec:guettler}, we apply Guettler's onset test.

\paragraph*{Friction Model}
The oscillation of a bowed string depends on the friction between the bow and the
string. The {\it LuGre} dynamic friction model~\cite{r-m:canudasdewit-1995}
represents stick-slip behavior through an internal {\it bristle} state. It
captures the {\it Stribeck effect}, in which the average friction force falls as
the relative velocity rises, as well as the {\it hysteresis} which results from
the dependence of the contact on its own history. The model has since been
re-examined, and its identification has been
clarified~\cite{r-m:astrom-2008}. It belongs to a broader control literature on
the modeling and compensation of
friction~\cite{r-m:olsson-1998,r-m:armstronghelouvry-1994}. 

In practice, many bowed-string models use a simpler law, which keeps the Stribeck
dependence of friction on the sliding velocity and drops the internal state. This
is enough to reproduce the stick-slip behavior which Helmholtz motion requires.
Such laws are common in musical acoustics, since they are simple and inexpensive
to compute, and they retain the nonlinear dependence of friction on the relative
velocity of the bow and the
string~\cite{r-m:mcintyre-1979,r-m:galluzzo-2017}. One alternative replaces the
velocity-only curve with one in which the friction force follows the temperature
of the rosin layer. {\it Woodhouse}~\cite{r-m:woodhouse-2003} reports that this
{\it thermal model} establishes Helmholtz motion more readily than a
steady-sliding curve does, and that it predicts a different flattening behavior.
Therefore, the choice of a friction law is a modeling commitment as well as a
simplification.

A second alternative keeps the dependence on the sliding velocity and resolves
the transition between sticking and slipping over a finite time. The
{\it elasto-plastic} model of {\it Dupont, et al.}~\cite{r-m:dupont-2000,r-m:dupont-2002}
adds an adhesion map to the LuGre bristle state, which removes the drift that
LuGre shows at low sliding speeds, and {\it Serafin, Avanzini, and
Rocchesso}~\cite{r-m:serafin-2003} were the first to apply it to a bowed string.
It has since been implemented on a finite-difference stiff
string~\cite{r-m:willemsen-2019}, as well as on a string driven by a finite-width
bow with hair compliance and torsional motion, where the simulated Guettler
diagram was compared with one measured on a robot-bowed
monochord~\cite{r-m:matusiak-2024,r-m:lampis-2024}. {\it Matusiak, Chatziioannou,
and van Walstijn}~\cite{r-m:matusiak-2025} refine that model so that passivity
holds for any choice of parameters, and they derive a finite-difference scheme
which inherits the energy balance of the continuous system. {\it Van Walstijn,
et al.}~\cite{r-m:vanwalstijn-2026} then combine the elasto-plastic pre-sliding
with the contact temperature of the thermal model. They report that the friction
parameters may be set so that both Schelleng bow-force limits are placed against
measurement over a range of bow-bridge distances.

The friction law which we use here is the static characteristic with no internal
state. Of the three families, it is the one whose parameters may be taken from
published measurement without being fitted, which is the discipline we apply
throughout this paper. It is also the one which cannot reproduce hysteresis in
the friction-velocity plane.

\paragraph*{Control of Bowed Strings}

Most prior approaches to controlling the dynamics of a bowed string rely on
open-loop parameter maps, such as bow force-speed diagrams, to maintain Helmholtz
motion~\cite{r-m:schelleng-1973,r-m:schoonderwaldt-2008}. These maps work within
limited operating regions, and they often cannot follow the strong
nonlinearities, discontinuities, and regime transitions which frictional contact
introduces~\cite{r-m:armstronghelouvry-1994}.

Learning-based work on the bowed string has taken two complementary forms.
{\it Luan and Scavone}~\cite{r-m:luan-2025} apply physics-informed networks to the
friction nonlinearity itself, learning the contact behavior of a lumped
bow-string system instead of closing a loop around it. {\it Percival, Bailey, and
Tzanetakis}~\cite{r-m:percival-2011,r-m:percival-2013-1,r-m:percival-2013-2}, on
the other hand, do close the loop. Their support vector machines, trained on the
audio which a bowed-string physical model produces, generate the finger position,
bow-bridge distance, bow velocity, and bow force, so that the system performs
written music with the competence of a beginning student. That work, however,
treats the model as a black box and judges its own output by its timbre.

\paragraph*{Imitation Learning and Its Ceiling}

The controller in this paper is trained on labels generated by a rule, which
makes it an {\it imitation learner}, and the bound which it runs into is the one
that literature is about. Fitting a policy to the actions of a demonstrator was
the approach of {\it Pomerleau}'s ALVINN~\cite{r-m:pomerleau-1988}, and it remains
the simplest form of {\it learning from
demonstration}~\cite{r-m:argall-2009,r-m:osa-2018}. The central difficulty of
{\it behavior cloning} is {\it distribution shift}. Namely, a policy trained on the
states which the demonstrator visits is evaluated on the states which its own
errors produce. {\it Ross and Bagnell}~\cite{r-m:ross-2010} show that the resulting
error may compound quadratically in the horizon, which motivates the interactive
aggregation of {\it DAgger}~\cite{r-m:ross-2011}. Compounding error explains why a cloned policy
may fall below its demonstrator. That it does not generally exceed the demonstrator
follows from the objective, whose minimizer is the demonstrator's own action.
Imitating an algorithmic teacher, instead of a human, is also established.
{\it Guided policy search} trains a network on the output of a trajectory
optimizer~\cite{r-m:levine-2016}, which is adapted to the policy as training
proceeds, whereas the teacher here is fixed, a greedy rule on a playability map. Imitation
learning has also been applied to violin bowing. {\it Jin, et
al.}~\cite{r-m:jin-2024} learn bowing motion from recorded human performance,
which places their teacher on the demonstrator side of that distinction and ours
on the other.

What this plant adds is a measurable bound, rather than a new one. The teacher reads
an exhaustive tabulation of the steady state which the plant reaches from rest, so
its competence may be scored at every operating point, and the controller's score
may be regressed on it. That turns the domain on which the bound holds from an
assumption into a measurement.

The controller developed here commands the bow velocity and a force at the frog,
which an assumed {\it transmission factor} maps to the contact force. It is scored
on the oscillation regime, quantified by the {\it stick fraction}, instead of a
judgment of timbre, although its training objective is the imitation of the rule.
Neural networks approximate the piecewise-continuous nonlinearities of frictional
contact~\cite{r-m:selmic-2002}, and the rule imitated here is discontinuous in the
state, so a small network is a reasonable starting point.

\subsection{Our Contributions}

\begin{itemize}

\item{{\it A quantitative demonstration that the contact must be resolved
implicitly} -- Implicit and energy-balanced treatments of the bow-string contact
are well established~\cite{r-m:desvages-2016}. What we show here is the size of
the consequence of not using one. A point force on a finite-difference grid acts
on a lumped mass on the order of $10^{-6}$~kg, so a friction force which lags by
one time step changes the bow-point velocity by several meters per second. In
Section~\ref{sec:friction}, we show that no bow force whatsoever produces a stick
phase under such a treatment.}

\item{{\it The stick fraction as a diagnostic, and the failure it exposes} -- In
Section~\ref{sec:diagnostics}, we define the {\it stick fraction}, the proportion
of a window during which the bow carries the string. We show that a regime
classifier which lacks an explicit stick test will report Helmholtz motion for
small-amplitude continuous slipping, which has the right period and the right
spectral peak, while having none of the underlying physics. The check is
inexpensive, and it should precede any use of a playability map.}

\item{{\it A model from measured parameters, and its verification} -- The friction
characteristic, the wave impedance, and the quality factor are taken from published
measurement, rather than fitted, and the contact is resolved implicitly. The
characteristic is the static one, chosen since, of the three families reviewed
above, it is the one whose parameters may be taken from measurement without being
fitted, and it is also the one which cannot reproduce hysteresis. The four
open strings are one non-dimensional problem, scaled in force by their impedance,
and the model is verified on them in Section~\ref{sec:params}. Driven at the
midpoint of its own band, each of the four open strings returns Helmholtz motion
with a stick fraction of $89.1\%$, against an ideal $90.0\%$. In
Section~\ref{sec:params}, we also show why a single operating point cannot
be used for this purpose, since the force chosen on A sits at $99\%$ of the
maximum on E and at $36\%$ on G.}

\item{{\it Two minimum bow forces} -- In Section~\ref{sec:schelleng}, we show that
the lower edge of the playable band is two limits. From rest, Helmholtz motion
appears only above $\kappa Z v_b/\beta$, with $\kappa = 1.112 \pm 0.017$ at the
discretization used, and raising the force slowly lowers that little. Once
established, it is kept down to a force which goes as $\beta^{-2}$ and as the
inverse of the quality factor, at $1.0$ to $2.7$ times Schelleng's minimum computed
with the model's own losses. At the measured losses, the first lies $31$ to $108$
times above the second. The upper edge recovers Schelleng's maximum. The constant
reproduces on sixty-four held-out operating points which share neither reference
axis, at the same discretization, to $0.41$ standard errors.}

\item{{\it Guettler's attack regions} -- In Section~\ref{sec:guettler}, the model
reproduces Guettler's three regions. At the grid used, the band of clean attacks
closes as the bow moves from a tenth to a twentieth of the string length from the
bridge. That test is run against Guettler's description of the plane, rather than
against a measured one, which is the weaker of the two forms that it now
takes~\cite{r-m:lampis-2024,r-m:matusiak-2024}.}

\item{{\it A closed-loop formulation, and what its supervision trains} -- The
controller commands the bow velocity and a frog force, mapped to the contact by an
assumed transmission factor, and it is scored on the oscillation regime, measured
by the stick fraction. It imitates a greedy rule on a playability map built from
rest, whose labels depend only on the stroke position and the previous commands, so
the objective rewards neither the oscillation inputs nor memory
(Section~\ref{sec:training}). In Section~\ref{sec:alpha}, we quantify the cost of
getting that transmission wrong, which is strongly asymmetric.}

\item{{\it A controller comparison at matched capacity, and a failure mode with a
mechanism} -- In Section~\ref{sec:architecture}, we train six architectures at
$1362$ to $1410$ parameters on all four strings, with twenty seeds each. A gated
recurrent unit leads on the mean of all four, by margins which survive Holm correction on two strings against the minimal variant and on none against the feedforward network or the state-space model. From the in-band start, it leads the feedforward network on all four strings, at $p = 0.002$ on their mean, before any correction for the choice of start, and at thirty-seven of the forty positions, at $p = 0.0015$ averaged over
them. The other three positions lie within one standard error of zero. Its
margin over the mean of the other five architectures is largest on the
disturbance start, at $19.2$ points against $6.4$, $6.4$, and $2.2$ on the
in-band, cold, and hot starts. Over the feedforward network alone, it leads there by $11.9$, $15.5$, and $8.2$
points on G, D, and E, and on A it trails by $1.3 \pm 2.3$ points, at $p = 0.59$.
None of these margins survives Holm correction over the five comparisons on its
string.
The feedforward network completes more strokes on three of the four starts, but
by $20.0$ points on the cold one against $6.2$ and $1.2$ on the others. Of
course, the cold start is one which no player would use, and in
Section~\ref{sec:coldplayable}, we find its Helmholtz target out of reach only on G. The minimal gated variant, whose gates are computed from the input alone, can latch after an abrupt velocity override, since its gates cannot make the reset depend on the state itself. Namely,
eight of its twenty seeds on the E string hold below $9\%$ Helmholtz motion for
the remainder of the stroke, and forcing a state reset at every control step
recovers all eight. Depth compounds this on two strings and changes nothing on
the other two.}

\item{{\it Three ways of selecting a controller which do not work, namely, by
parameter count, by training-context length, and by training loss} -- Over a
four-rung ladder of parameter budgets on all four strings and four
architectures, which amounts to one hundred and ninety-two fits, the training
loss falls in every one of the sixteen cells. Meanwhile, the closed-loop quality does not rise. The mean falls in fourteen of the sixteen cells, by more than four
points in nine, and it rises in two. At three seeds, however, the standard error of
a change runs from $2.2$ to $19.3$ points, and only one change, a fall of $11.7$
points for the state-space model on D, is significant by Welch's $t$-test, at
$p = 0.0062$. None survives Holm correction over the sixteen. The seed spread widens, instead of narrowing, in thirteen. The training loss also falls with the length of the
training context, while the closed-loop quality moves in every direction.
Therefore, a search driven by the quality of fit picks the largest model every time, although the largest model is not, in general, the better controller (Sections~\ref{sec:context} and~\ref{sec:capacity}).}

\item{{\it A ceiling located in the supervision, rather than in the model} -- The
rule which generates the training labels of the controller reads an exhaustive
tabulation of the steady state which the plant reaches from rest, at $1089$
operating points. Where the map is right, the learned controller does not exceed
the rule in steady-state regulation, and over forty positions at twenty seeds, the
lookup leads in all $160$ cells, significantly in $156$ after Holm correction. Since
the lookup reads the map of the very position at which it plays, that is an
oracle bound, rather than a baseline. What the controller does better is finish
(Sections~\ref{sec:baselines}, \ref{sec:fingering}, and~\ref{sec:capacity}).}

\item{{\it The domain of that ceiling, which is where the tabulation fails} -- The controller overtakes the tabulation wherever the tabulation itself fails. On G, it trails the tabulation at all four friction levels where the tabulation holds, and two of those four margins differ from zero at $p < 0.05$, while the other two, of $1.5$ and $2.4$ points, do not, at $p = 0.052$. The controller overtakes the failing tabulation with its string-state inputs frozen as well, so the advantage lies in its learned map from stroke position and commands, rather than in feedback from the oscillation. The boundary is located, rather than argued.
Namely, $57$ of the map's own $1089$ cells return Helmholtz motion on a properly
settled plant while its labels call them unplayable. The lookup commands two of
them in $593$ control steps, and both are forced by its rate limit, whereas the
controller commands them on one step in five and reaches Helmholtz motion on four
of five. Regressing the score of the controller on that of the tabulation, over a
friction sweep, gives a slope of $0.319 \pm 0.082$, more than eight standard errors below the unit slope which inheriting the teacher's competence would produce,
with the two crossing at a tabulation score of $88$. Where the tabulation scores
below $60$, it averages $41.9$ and the controller $74.4$, and above $90$, the
order reverses, at $98.2$ against $90.9$. The operative variable is the
tabulation's own health, and not how far the contact has moved, since the
tabulation is not monotone in the friction removed. In fact, the reduction at
which it collapses on three strings is a local minimum, and it recovers before
falling again. Measured across two independent families of fit, none of twenty-four
cells separates the two at $p < 0.05$ by Welch's $t$-test over the seeds
(Section~\ref{sec:contactdist}).}

\item{{\it Two attempts to lift that ceiling, which fail the same way} -- In
Section~\ref{sec:rl}, we optimize against the simulator, instead of the teacher,
and we find no net gain over sixteen architecture and string combinations. We do find a significant separation in mechanism, however. The conditions which the search
is shown improve in fifteen of sixteen, and those which it is not shown degrade
in eleven, a separation of $9.2$ points at $p = 0.0008$. A residual controller,
which begins at the tabulation and learns a correction to it, shows the same trade, from a different architecture and a different search, improving most
on the string which it was shown and losing most on the conditions which it was
not. The bound which a residual architecture appears to offer holds at
initialization, and it is destroyed by training on a distribution which does not
cover the evaluation conditions.}

\item{{\it How a network uses inputs on which its labels do not depend} --
Retraining without each input group in turn, over four strings and four
architectures, spreads the effect of removing the four string-state inputs by
$19.3$ points across strings, against $6.8$ across architectures. Since the labels
do not depend on the string-state inputs, this measures input representation.
Whether that belongs to the strings, or to the fixed window and command box, is not
separated here (Section~\ref{sec:ablation}).}

\item{{\it An acoustic check on the mechanical measure} -- Each condition is
rendered from the bridge force and scored with descriptors from the voice
literature. The {\it cepstral peak prominence} separates Helmholtz motion from
every other regime by a factor of eight to twenty-four, and it tracks the
mechanical regime measure at a rank correlation of $+0.803$ over seventy-two
renderings, which span four strings and four controllers. On the other hand, a
plain harmonicity measure reaches $+0.324$, and it rates a string which the bow
never grips above Helmholtz motion (Section~\ref{sec:acoustic}).}

\item{{\it Fingered notes across three octaves, on a plant which is provably
invariant} -- In Section~\ref{sec:fingering}, we evaluate forty positions from G3
to G6 at twenty seeds each, four of them open and thirty-six fingered. Under a
rigid stop at a constant bow-bridge fraction, stopping the string leaves every
coefficient of the discrete scheme unchanged and rescales only time, so that two
positions on one string are the same simulation relabeled, as confirmed to
$6.8 \times 10^{-13}$~m/s over sixteen operating points. Therefore, one
playability map serves every stopped position on a string. Since the plant is
identical, every point of loss when an open-string controller is deployed at a
stopped position belongs to the controller. In Section~\ref{sec:fingering}, we
trace most of it to a single feature, namely, an amplitude target which is
specified as a displacement, where the displacement scales with the reciprocal
of the fundamental and the velocity does not.}

\item{{\it Three measurement choices which decide results, identified by making
more than one} -- A scoring window can decide whether an effect exists at all.
Scored over the whole stroke, a mid-stroke change in the contact condition
separates the full controller from its variant with frozen string-state inputs on
both recurrent architectures, at twenty seeds, by $14.7$ and $18.5$ points, at $p$ below
$0.001$. Scored over the steps after the onset, which is the more specific
measure, only the minimal variant reaches significance, and the gated
recurrent unit does not, at $p = 0.64$. Reporting one window would
have allowed either a finding or a null to be presented as the result.
Likewise, a start condition can decide whose failure a failure is. The cold
start begins at a commanded speed whose entire force window lies below the
model's own force floor on one string, so a class of cold-start stalls belongs to
the condition, rather than to the controller, and the completion figures are
reported both with and without it (Section~\ref{sec:coldplayable}). In addition,
a choice of protocol can decide which way an effect points, rather than whether it exists. The feature ablation and a freeze at inference agree about which
strings the string-state inputs help, on the disturbance start, while a third
measurement, under a different disturbance, reverses the sign on G
(Sections~\ref{sec:ablation}, \ref{sec:baselines}, and~\ref{sec:contactdist}).}

\end{itemize}

Each half of this paper also appears in shorter form, the bowed-string model and
its two minimum forces~\cite{r-m:beigi-2026a}, and the learned
controller~\cite{r-m:beigi-2026b}.

\section{Problem Formulation}

\subsection{Nomenclature}
\label{sec:nomenclature}

Table~\ref{tab:nomenclature} lists every symbol which we use in this paper,
together with its meaning and its unit. Each symbol carries a single meaning
throughout, so that a reader who meets an unfamiliar symbol anywhere in the text
may resolve it there.

\setlength{\tabcolsep}{0.35em}
\renewcommand{\arraystretch}{1.05}
\begin{table}[htbp]
\caption{Symbols used throughout this paper. Each symbol carries a single meaning, and
bold lowercase denotes a vector.}
\label{tab:nomenclature}
\centering
\footnotesize
\begin{tabular}{|l|l|l|}
\hline
\textbf{Symbol} & \textbf{Meaning} & \textbf{Unit} \\
\hline
\multicolumn{3}{|l|}{{\it String and grid}} \\
\hline
$u(x,t)$, $u^n_l$ & transverse displacement & m \\
$u_B$ & displacement at the bow point & m \\
$x$, $L$ & position along the string, length & m \\
$x_B$, $\beta$ & bow position, $x_B/L$ & m, -- \\
$c$ & transverse wave speed & m/s \\
$f_0$, $f_0^\star$ & nominal fundamental, its target & Hz \\
$f_s$ & sample rate, $1/k$ & Hz \\
$\rho A$ & linear mass density & kg/m \\
$T$ & string tension & N \\
$Z=\rho A c$ & wave impedance & N$\cdot$s/m \\
$\sigma_0$, $\sigma_1$ & damping coefficients & s$^{-1}$, m$^2$/s \\
$Q_1$, $Q_2$ & quality factors of modes 1, 2 & -- \\
$Q$ & common value of $Q_1$ and $Q_2$ & -- \\
$h$, $k$ & space step, time step & m, s \\
$n$, $l$, $l_B$ & time index, space index, bow node & -- \\
$N_x$ & number of grid points & -- \\
$\lambda=ck/h$ & Courant number & -- \\
$\updelta_{xx}$ & undivided second spatial difference & -- \\
$\delta(\cdot)$ & Dirac delta & 1/m \\
$\indic_{l=l_B}$ & discrete indicator of the bow node & -- \\
$\Gamma=(1+\sigma_0 k)^{-1}$ & damping denominator of the update & -- \\
\hline
\multicolumn{3}{|l|}{{\it Contact}} \\
\hline
$v_b$, $\dot v_b$ & bow speed, its rate at the attack & m/s, m/s$^2$ \\
$v_{\mathrm{rel}}$, $v^\star$ & relative velocity, its force-free value & m/s \\
$F_N$ & normal force at the contact point & N \\
$F_{\mathrm{frog}}$ & normal force at the frog & N \\
$f_{\mathrm{bow}}$ & friction force on the string & N \\
$\varphi$ & friction characteristic & -- \\
$a_1$, $a_2$, $\mu_c$ & friction curve coefficients & -- \\
$\mu_s$, $\Delta\mu$ & effective static limit, $\mu_s-\mu_c$ & -- \\
$v_1$, $v_2$ & Stribeck velocity scales & m/s \\
$v_\varepsilon$ & sign-regularization velocity & m/s \\
$a$ & implicit-solve coefficient, Equation~\ref{eq:implicit} & m/s \\
$\alpha(s)$, $\alpha_{\mathrm{tip}}$ & transmission factor, its value at the tip & -- \\
$T_h$ & bow-hair tension & N \\
$K_s$ & stick stiffness at the tip & N/m \\
$F_W$, $d$ & bow weight, its center of gravity & N, m \\
$\tau$, $\tau_{\max}$ & moment applied at the frog, its bound & N$\cdot$m \\
$\theta$ & bow inclination from the horizontal & rad \\
\hline
\multicolumn{3}{|l|}{{\it Diagnostics}} \\
\hline
$\eta$ & stick fraction & -- \\
$W$, $\epsilon$ & analysis window, stick tolerance & s, m/s \\
$f_{\mathrm{slip}}$ & slip-interval frequency & Hz \\
\hline
$C(q)$, $\widehat{C}(q)$ & cepstrum, its fitted line & -- \\
\hline
$q$, $q^\star$ & quefrency, its value at the peak & s \\
$A$, $A^\star$ & bow-point amplitude, its target & m \\
$r$ & oscillation regime label, 1--6 & -- \\
$F_{\min}$, $F_{\max}$ & Schelleng bow-force limits & N \\
$F_{\mathrm{est}}$ & establishment threshold & N \\
$\kappa$ & constant of the establishment threshold & -- \\
$G$ & dimensionless threshold function & -- \\
$R$ & termination resistance & N$\cdot$s/m \\
\hline
\multicolumn{3}{|l|}{{\it Control}} \\
\hline
$m$, $M$ & control-step index, step count & -- \\
$s$, $L_{\mathrm{bow}}$ & stroke position, bow length & --, m \\
$\Delta t$ & control-step duration & s \\
$\mathbf{z}_m$ & feature vector, Equation~\ref{eq:features} & -- \\
$\mathbf{y}_m$ & commanded increments & N, m/s \\
$\Delta f$, $\Delta A$ & frequency and amplitude error & Hz, -- \\
$J$ & reinforcement reward, Equation~\ref{eq:reward} & -- \\
\hline
\multicolumn{3}{|l|}{{\it Statistics}} \\
\hline
$p$, $R^2$ & significance level, fit coefficient & -- \\
\hline
\end{tabular}
\end{table}

\subsection{String Model}

Following {\it Desvages and Bilbao}~\cite{r-m:desvages-2016}, we represent the
string by a damped wave equation with frequency-dependent losses, as follows,
\begin{equation}
\begin{aligned}
u_{tt}(x,t) =\;& c^2 u_{xx}(x,t) - 2\sigma_0 u_t(x,t) \\
&+ 2\sigma_1 u_{xxt}(x,t) + \frac{1}{\rho A}\, f_{\mathrm{bow}}(t)\,\delta(x-x_B)
\end{aligned}
\label{eq:string}
\end{equation}
where $u(x,t)$ is the transverse displacement, $c$ is the wave speed, $\rho A$ is
the linear density, and $\sigma_0$ and $\sigma_1$ are the frequency-independent
and the frequency-dependent damping coefficients, respectively. The bow acts at
$x_B = \beta L$.

Discretizing in space and time, on a grid of spacing $h$ with a time step $k$, we
get the following update,
\begin{equation}
\begin{aligned}
(1+\sigma_0 k)\,u^{n+1}_l =\;& 2u^n_l - (1-\sigma_0 k)\,u^{n-1}_l \\
&+ \lambda^2 \updelta_{xx} u^n_l
+ \frac{2\sigma_1 k}{h^2}\,\updelta_{xx}\!\left(u^n_l - u^{n-1}_l\right) \\
&+ \frac{k^2}{\rho A\,h}\, f^n_{\mathrm{bow}}\,\indic_{l=l_B}
\end{aligned}
\end{equation}
where $\lambda = ck/h$ is the Courant number, and $\updelta_{xx}$ is the undivided
second difference, $\updelta_{xx} u^n_l = u^n_{l+1} - 2u^n_l + u^n_{l-1}$. The ends
are fixed, $u^n_0 = u^n_{N_x-1} = 0$. The index $l_B$ is the grid point nearest to the bow
position $x_B$, and $\indic_{l=l_B}$ is its discrete indicator, which is equal to
one at $l = l_B$ and zero at every other grid point. Taken with the factor $1/h$,
it is the discrete counterpart of $\delta(x-x_B)$ in Equation~\ref{eq:string}, and
it converts the lumped bow force, in newtons, into the force density which
Equation~\ref{eq:string} requires. Both damping terms are retained. The coefficients $\sigma_0$ and $\sigma_1$ are
obtained by matching the quality factors of the first two modes, $Q_1$ and $Q_2$,
which are equal here, with the common value $Q$. The velocity of the bow point at
step $n$ is the centered difference $(u^{n+1}_{l_B} - u^{n-1}_{l_B})/(2k)$. So the
relative velocity at that step depends on $u^{n+1}$, and the contact force must be
solved together with the update.

\subsection{Friction and the Stick Phase}
\label{sec:friction}

The bow-string contact is described by a friction characteristic which is odd in
the relative velocity, $v_{\mathrm{rel}} = u_t(x_B,t) - v_b$, as follows,
\begin{equation}
\varphi(v) = \Bigl[a_1 e^{-|v|/v_1} + a_2 e^{-|v|/v_2} + \mu_c\Bigr]
\frac{2}{\pi}\arctan\!\left(\frac{v}{v_\varepsilon}\right)
\end{equation}
The expression is that of {\it Smith and Woodhouse}~\cite[Eq.~(31)]{r-m:smith-2000}, in form
as well as in value. Namely, the bracketed term is the steady-sliding curve which
they measured for rosin-coated surfaces, and the arctan factor is the device
which they use to give the sticking phase a finite slope. The constants are also
theirs, with $a_1 = 0.40$, $v_1 = 0.01$ m/s, $a_2 = 0.45$, $v_2 = 0.10$ m/s, and
$\mu_c = 0.35$. The bracketed term rises to $1.20$ as the sliding speed approaches zero, and it falls to the Coulomb asymptote of $0.35$ at high speeds. The arctan factor caps the product itself at $1.048$, which it reaches at
$2$~mm/s. That cap, rather than the nominal $1.20$, is the static limit which the
model has, $\mu_s = 1.048$, and the friction difference which enters Schelleng's
limits is $\Delta\mu = \mu_s - \mu_c = 0.698$.
The width of the arctan transition is set by $v_\varepsilon = 2\times10^{-4}$
m/s, which is also theirs, and the factor makes $\varphi$ odd, with
$\varphi(0)=0$. The force applied to the string is then given by,
\begin{equation}
f_{\mathrm{bow}} = -F_N\,\varphi(v_{\mathrm{rel}})
\label{eq:fbow}
\end{equation}
where the sign expresses the fact that friction opposes the relative sliding.

It is not viable to resolve Equation~\ref{eq:fbow} explicitly. A point force on
the grid acts on a lumped mass $\rho A h$, on the order of $10^{-6}\,$kg for the
discretization which we use here, so a friction force which lags by one step
changes the bow-point velocity by several meters per second in a single sample.
The bow cannot then capture the string, and no stick phase forms, whatever the
bow force. Therefore, the friction is resolved implicitly. Writing
$\Gamma=(1+\sigma_0 k)^{-1}$, and letting $v^\star$ be the relative velocity which
the bow point would acquire with no bow force, we see that the new relative
velocity satisfies the scalar equation,
\begin{equation}
v_{\mathrm{rel}} + a\,\varphi(v_{\mathrm{rel}}) = v^\star,
\qquad a = \frac{\Gamma\,k\,F_N}{2\rho A h} = \frac{\Gamma\,\lambda\,F_N}{2Z}
\label{eq:implicit}
\end{equation}
which is solved at each step by Newton iteration, with a bracketed bisection as a
fallback. At a fixed Courant number, $a$ does not depend
on the grid spacing, so refining the grid does not cure a lagged force. Implicit
treatments of this contact are well
established~\cite{r-m:mcintyre-1983,r-m:bilbao-2009}, and energy-balanced schemes
adopt them for the same reason~\cite{r-m:desvages-2016}. Equation~\ref{eq:implicit} may admit more than
one root, which is known as {\it Friedlander's
ambiguity}~\cite[pp.~518--519]{r-m:friedlander-1953}. Its left hand side is monotone in
$v_{\mathrm{rel}}$, and therefore the root is unique, whenever
$a\,\varphi'(v) > -1$ holds for every $v$ in the range searched. Since $a$ is
proportional to the bow force, that condition fails at high forces on the falling
part of the characteristic, which the contact crosses at each release and capture. {\it Friedlander} resolved it by giving the bow a finite width, which
attaches a small mass at the contact and turns the boundary condition into a
differential equation with a unique solution. Following {\it McIntyre,
Schumacher, and Woodhouse}~\cite{r-m:mcintyre-1983} instead, we retain the root
which continues the current phase of the motion, sticking or slipping, and we
never select the intermediate root. This is what makes sticking possible. In
other words, for a small $|v^\star|$, the solution is driven to
$|v_{\mathrm{rel}}| \ll v_\varepsilon$, and the bow carries the string.

Selecting a root is what a static characteristic requires. A dynamic friction
law in the LuGre family, on the other hand, may be arranged so that no selection
is needed. {\it Matusiak, Chatziioannou, and van Walstijn}~\cite{r-m:matusiak-2025}
refine the elasto-plastic model so that its update equation has a single root,
and they report that the unrefined form admits more than one root at every
sampling rate which they tested, so oversampling does not remove the ambiguity.
Their scheme is also solved by Newton iteration, and it may meet a singular
Jacobian, which is the failure mode that the bracketed bisection above exists to
catch. Therefore, the two treatments differ in where the ambiguity is resolved.
Here, it is resolved by a rule which is applied to the solution, and there, by a
change to the model which leaves a single solution to find.

\subsection{Diagnostics}
\label{sec:diagnostics}

Since Helmholtz motion is defined by its stick phase, the diagnostic which we use
throughout this paper is the {\it stick fraction},
\begin{equation}
\eta = \frac{1}{W}\,\bigl|\{\, t \in W : |v_{\mathrm{rel}}(t)| < \epsilon \,\}\bigr|
\label{eq:stick}
\end{equation}
which is the proportion of the analysis window $W$ during which the bow carries the
string, with $\epsilon = 1$~mm/s. On the grid, the measure $|\cdot|$ is the number
of samples which pass the test, times $k$. Ideal Helmholtz motion gives
$\eta \approx 1-\beta$. The {\it slip onsets}
are the instants at which $|v_{\mathrm{rel}}|$ crosses $\epsilon$ upward. Their
mean spacing gives the slip frequency, $f_{\mathrm{slip}}$, and their relative
standard deviation gives the {\it jitter}. Since $f_{\mathrm{slip}}$ is obtained
from the timing of events, rather than from a spectral peak, it is continuous,
and it is not quantized to the width of an FFT bin.

Each map cell starts from rest. The string is at rest, and the bow force and the
bow speed are applied at their fixed values from the first step. Except where a
window fixed in seconds is named, the motion settles for $140$ nominal periods, and
$W$ is the $120$ periods which follow. Where windows
settled for three decay times are used instead, the same $120$ periods follow three
decay times of settling. The window is set in string periods, rather than in
seconds, so that the four strings settle for the same number of periods. Under a
window fixed in seconds, they receive $15.7$ to $52.8$ periods to settle, and the
bow-bridge dependence of the G string cannot be measured at all.

The regime is the one named by the first of these tests which the window passes.
\begin{enumerate}
\item $\eta > 0.98$: constant sticking.
\item $\eta < 0.05$: constant slipping.
\item Too few slip onsets to measure a spacing: constant sticking.
\item More than $1.5$ slips per nominal period: multiple slipping.
\item Fewer than $0.6$ slips per nominal period: sub-harmonic or ALF motion.
\item $\eta < 0.35$: raucous or irregular motion.
\item A jitter above $0.25$: raucous or irregular motion.
\item Otherwise: Helmholtz motion.
\end{enumerate}
The floor on $\eta$ is what keeps small-amplitude continuous slipping, which has the
right period but none of the right physics, out of the Helmholtz class. In
Section~\ref{sec:schelleng}, the establishment threshold is shown not to change
when $\epsilon$, the floor on $\eta$, or the ceiling on the slip count is varied.

\subsection{Parameters and Verification}
\label{sec:params}

The values which we use throughout this paper are collected in
Table~\ref{tab:params}, and the symbols in Table~\ref{tab:nomenclature}.

\setlength{\tabcolsep}{0.30em}
\renewcommand{\arraystretch}{1.15}
\begin{table}[htbp]
\caption{Parameter values used throughout this paper. The upper part is per string,
with the sample rate at each of the two grids. On A, $Z$ is the value quoted by
{\it Pickering}~\cite{r-m:pickering-1992}. The lower part is common to the four
strings, except for $\sigma_0$ and $\sigma_1$, which are given for A and scale with
the wave speed.}
\label{tab:params}
\centering
\footnotesize
\begin{tabular}{|c|c|c|c|c|c|c|}
\hline
 & $f_0$ & $\rho A$ & $T$ & $Z$ & $f_s$ (Hz) & $f_s$ (Hz) \\
 & (Hz) & ($10^{-3}$ kg/m) & (N) & (N$\cdot$s/m) & $N_x = 51$ & $N_x = 101$ \\
\hline
G & $196.00$ & $2.457$ & $41.12$ & $0.3178$ & $21382$ & $42764$ \\
D & $294.00$ & $1.283$ & $48.31$ & $0.2490$ & $32073$ & $64145$ \\
A & $440.00$ & $0.689$ & $58.10$ & $0.2001$ & $48000$ & $96000$ \\
E & $660.00$ & $0.412$ & $78.18$ & $0.1795$ & $72000$ & $144000$ \\
\hline
\end{tabular}

\vspace{0.6em}
\setlength{\tabcolsep}{0.35em}
\begin{tabular}{|l|c|l|}
\hline
\textbf{Parameter} & \textbf{Value} & \textbf{Note} \\
\hline
String length $L$ & $0.33$ m & violin \cite{r-m:cremer-1984} \\
\hline
Grid points $N_x$ & $51$ ($h=6.60$ mm), $101$ & \\
\hline
Courant number $\lambda$ & $0.9167$ & $\lambda^2 + 4\sigma_1 k/h^2 \le 1$ \\
\hline
Damping $\sigma_0$ (A) & $1.024$ s$^{-1}$ & $Q_1\!=\!900$ \cite{r-m:bavu-2005} \\
\hline
Damping $\sigma_1$ (A) & $5.649\times10^{-3}$ m$^2$/s & $Q_2\!=\!900$ \cite{r-m:bavu-2005} \\
\hline
Friction $a_1$, $v_1$ & $0.40$, $0.01$ m/s & \cite{r-m:smith-2000} \\
\hline
Friction $a_2$, $v_2$ & $0.45$, $0.10$ m/s & \cite{r-m:smith-2000} \\
\hline
Coulomb $\mu_c$ & $0.35$ & \cite{r-m:smith-2000} \\
\hline
Static limit, nominal, $\mu_s$ & $1.20$, $1.048$ & \cite{r-m:smith-2000} \\
\hline
Regularization $v_\varepsilon$ & $2\times10^{-4}$ m/s & \cite{r-m:smith-2000} \\
\hline
Bow position $\beta$ & $1/10$ & maps \\
\hline
\end{tabular}
\end{table}

Every physical parameter is taken from published measurement, rather than
fitted. The friction characteristic is the steady-sliding curve which {\it Smith
and Woodhouse}~\cite{r-m:smith-2000} measured on rosin-coated surfaces. On A,
the linear density is set so that the wave impedance, $Z = \rho A c$, equals the
$0.2$~N$\cdot$s/m quoted for a typical violin string~\cite{r-m:pickering-1992},
which gives a tension of $58$ N. The sample rate scales with the fundamental, so the
four strings share $\lambda$, $N_x$ and $Q$. Two grids are used. The playability
maps and the verification below run at $N_x = 51$. The band-edge sweeps of
Section~\ref{sec:schelleng} and the attack sweeps of Section~\ref{sec:guettler} run
at $N_x = 101$, so that $\beta$ is not quantized coarsely by the grid spacing. The quality factors follow the
measured decay of a string on a monochord, $7.5$ s at that instrument's own
fundamental, which corresponds to $Q \approx 900$, since $Q$ is $\pi$ times the frequency times the decay time~\cite{r-m:bavu-2005}. That correspondence fixes the fundamental
of the monochord near $38$ Hz if the $7.5$ s is an amplitude e-folding time, and
near $260$ Hz if it is a $T_{60}$, and the source does not say which. That figure
is the appropriate one here, since the model uses rigid terminations, whereas a
string mounted on an instrument loses more energy through the bridge and has a
lower $Q$. The same $Q$ is used on all four strings, which is an assumption and
not a measurement. Namely, since $Q$ is proportional to the frequency times the decay time, holding $Q$ fixed makes the
decay time string-dependent, from $1.46$ s on G to $0.43$ s on E. The assumption
is bounded, rather than merely noted. Refitting the establishment threshold of
Section~\ref{sec:schelleng} at half and at twice this $Q$, and at per-string
values taken from published bandwidths, moves its constant by at most $4.1$
percent. Therefore, a factor of four in $Q$ costs about half of what the law's
own residual already allows. The command box spans $0.05$ to $4$ N, and within
it, each string carries its own force floor, $0.40$ N on G and $0.05$ N on the
other three. The box lies inside the $0.1$--$10$ N reported for ordinary
playing~\cite{r-m:askenfelt-1989}, although its upper part corresponds to
forceful playing, rather than to a typical sustained note.

Two notational conventions hold throughout this paper. Differences quoted in the
text are taken from the stored values and then rounded, and not from the rounded
figures printed in the tables, so the difference of two table cells may disagree
with the text in the last digit.

 A figure written with a plus-minus term gives the standard error of the
estimate, in every place but three. Tables~\ref{tab:board} and~\ref{tab:cpp}, and
the gap figures of Section~\ref{sec:contactdist}, give a standard deviation
instead, namely, across seeds in the first and the third, and across the
analysis frames of a single recording in the second.

\begin{figure}[htbp]
\centerline{\includegraphics[width=\figwidth]{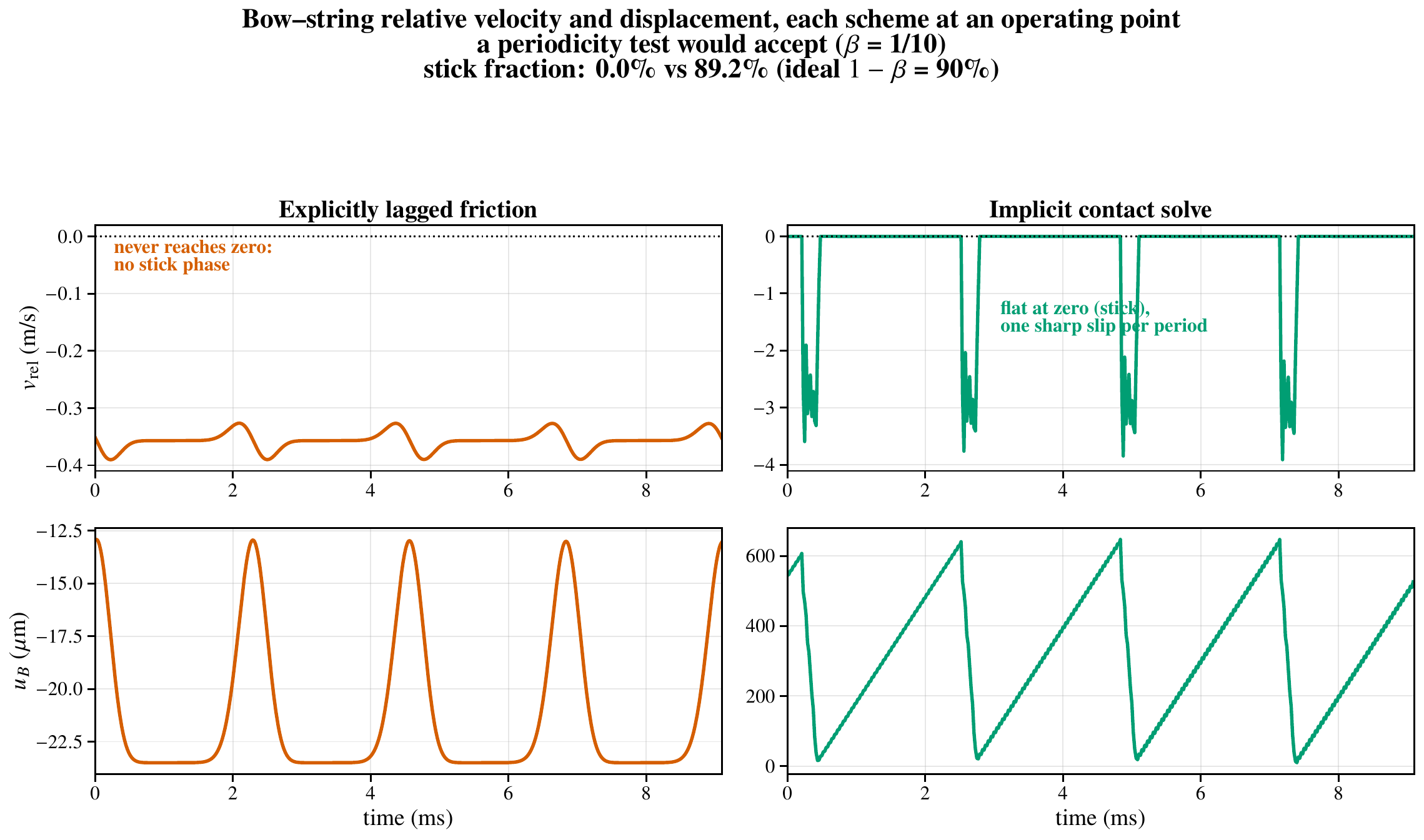}}
\caption{Bow-string relative velocity (top) and bow-point displacement (bottom), with each
scheme shown at an operating point which a periodicity test would accept. The
implicit scheme shows the Helmholtz signature, namely, $v_{\mathrm{rel}}$ held at
zero through the stick phase, with one sharp slip per period, and a sawtooth
displacement. An explicitly lagged friction force (left) produces neither, which is
the trap of Section~\ref{sec:diagnostics}, since the motion is periodic at the
nominal period and never grips the string.}
\label{fig:waveform}
\end{figure}

Three checks verify the implementation. The free vibration of the undriven A
string is at $440.00$ Hz, and the measured quality factor of its first mode is
$900$, matching its target. At $v_b = 0.30$ m/s and $F_N = 1.5$ N, the A string
settles into one slip per nominal period with a stick fraction of $89.2\%$, against
an ideal $1-\beta = 90\%$. The slip frequency is $432.4$ Hz, rather than $440$ Hz,
which is in the direction of the {\it flattening effect} expected under bow force.
That effect was identified by {\it McIntyre, Schumacher, and
Woodhouse}~\cite{r-m:mcintyre-1983} and analyzed by
{\it Boutillon}~\cite{r-m:boutillon-1991}. Its magnitude here grows with the bow
force, from $-14$ cents at $F_N = 0.8$ N to $-30$ cents at $F_N = 1.5$ N, which is the direction those studies report. The same difference is familiar to players as the discrepancy between an open string tuned pizzicato and the pitch which it sounds under the bow, although, on a real instrument, the bending stiffness of the string contributes to that discrepancy as well, and it is absent here. The magnitude is not converged, however. At $0.8$ N it falls to $-6.5$ and $-4.0$ cents on grids of twice and four times the resolution. At the grid used, the slip period locks
to a whole number of samples, so the slip jitter and the size of the flattening of
the slip frequency reflect that lock, and neither is used here as a check.
Figure~\ref{fig:waveform} shows the resulting waveforms.

That operating point is a fixed reference condition, rather than a verification
claim, and it is specific to A. Measured against each string's own
band, $F_N = 1.5$ N sits at $36\%$ of the band on G, $59\%$ on D, $83\%$ on A, and
$99\%$ on E, which places it at the maximum force for E. Held at the same window,
which is fixed in seconds while the sample rate scales with the fundamental, only
A returns Helmholtz motion at all. G, D, and E are classified as multiple
slipping, and they return slip frequencies which carry no meaning. Neither
failure is a property of those strings. Of course, a single force and a single
duration are not comparable across a set whose impedance ranges over a factor of
$1.77$ and whose periods range over a factor of $3.4$.

Therefore, verifying the simulator on four strings requires two changes
together, and both of them are changes of measurement, rather than of the model.
Each string is driven at the midpoint of its own band, and the analysis
window is set in string periods, instead of in seconds. Under those conditions,
all four strings give the same answer, namely, Helmholtz motion, with a stick
fraction of $89.1\%$ against an ideal $90.0\%$. These are checks of consistency,
since a stick fraction near $1-\beta$ with one slip per period is what any model
in Helmholtz motion returns. The slip frequency is $0.83\%$ flat of nominal, and
the slip jitter is at machine precision, between $1.2$ and $1.5 \times 10^{-14}$,
which reflects a period locked to a whole number of samples. The bow-point displacement
scales as the reciprocal of the fundamental, with the product of the two constant
across the four strings to within a factor of $1.043$, which is the same scaling
that governs the stopped-string results of Section~\ref{sec:fingering}.

Run under those conditions, A returns a stick fraction of $89.2\%$ and a slip
frequency of $432.4$ Hz, which reproduces the single-string figures above
exactly. Therefore, the four-string measurement is an extension of the
verification, rather than a revision of it, and nothing which we have reported for
A changes.

\subsection{Playability}
\label{sec:playability}

\begin{figure*}[htbp]
\centerline{\includegraphics[width=\figwidewidth]{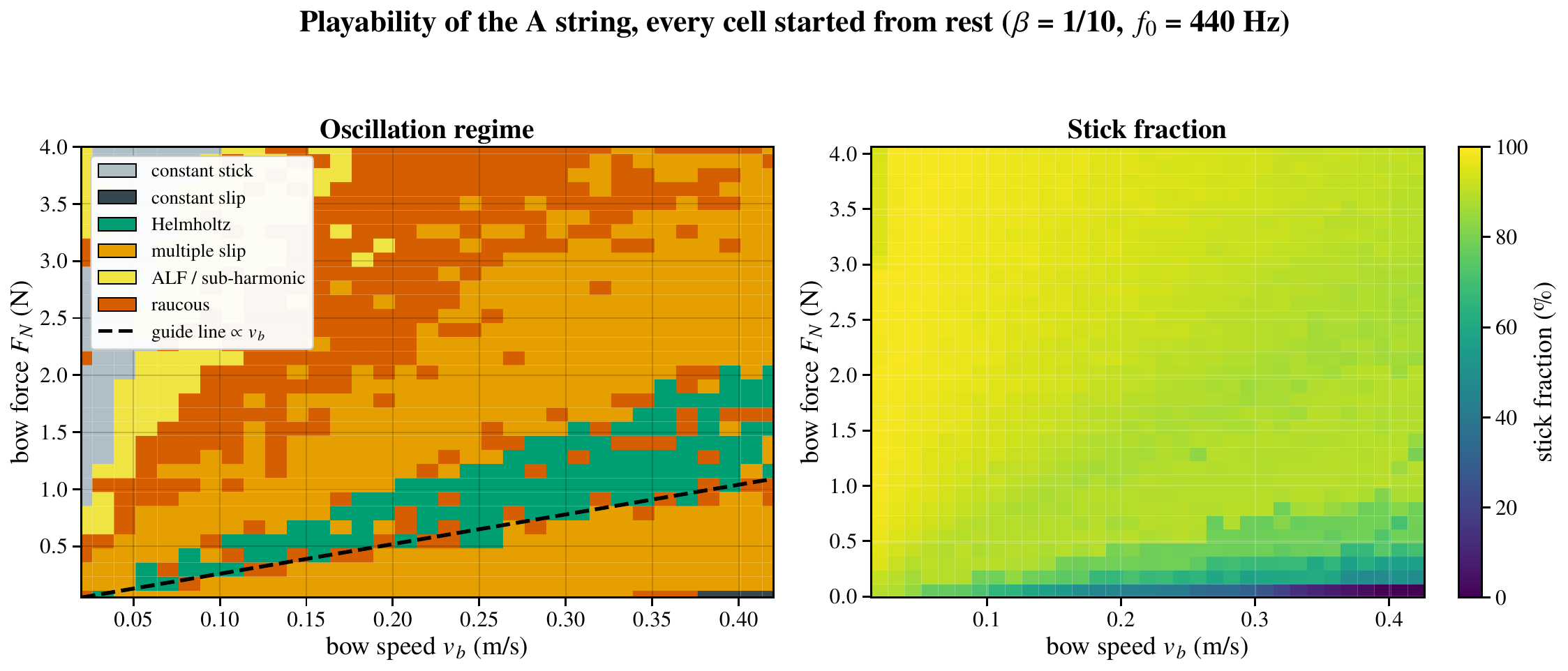}}
\caption{Playability of the model on the A string, with every cell started from rest
and measured over a window fixed in seconds. Left: the oscillation regime, with a line proportional to the bow speed drawn as a guide to its lower edge. Since every cell starts from rest, that edge is the establishment threshold of Section~\ref{sec:schelleng}, and not Schelleng's minimum. Right: the stick fraction, which is the quantity that
distinguishes Helmholtz motion from periodic slipping.}
\label{fig:playmap}
\end{figure*}

Sweeping $(F_N, v_b)$ on A and classifying each cell gives the map of
Figure~\ref{fig:playmap}. Every one of the $1089$ cells of a $33\times33$ grid
starts from rest, as Section~\ref{sec:diagnostics} describes, so the map records
which regime a step attack at a fixed bow force and bow speed settles into. It is an
exhaustive tabulation of the steady state which the plant reaches from rest, rather
than a model of it. The rule which reads this map is the strongest
baseline in the paper, and it is also what generates the labels on which the
controller is trained. Its structure is that of a Schelleng diagram. Namely,
Helmholtz motion occupies a band whose lower edge rises roughly linearly with the
bow speed, multiple slipping lies below that edge, and raucous motion lies above
the band. Of the $1089$ points, $118$ are Helmholtz, $459$
are multiple slipping, $324$ are raucous, and $106$ are sub-harmonic, with the
remaining $82$ being either constant sticking or constant slipping.

\begin{figure}[htbp]
\centerline{\includegraphics[width=\figwidth]{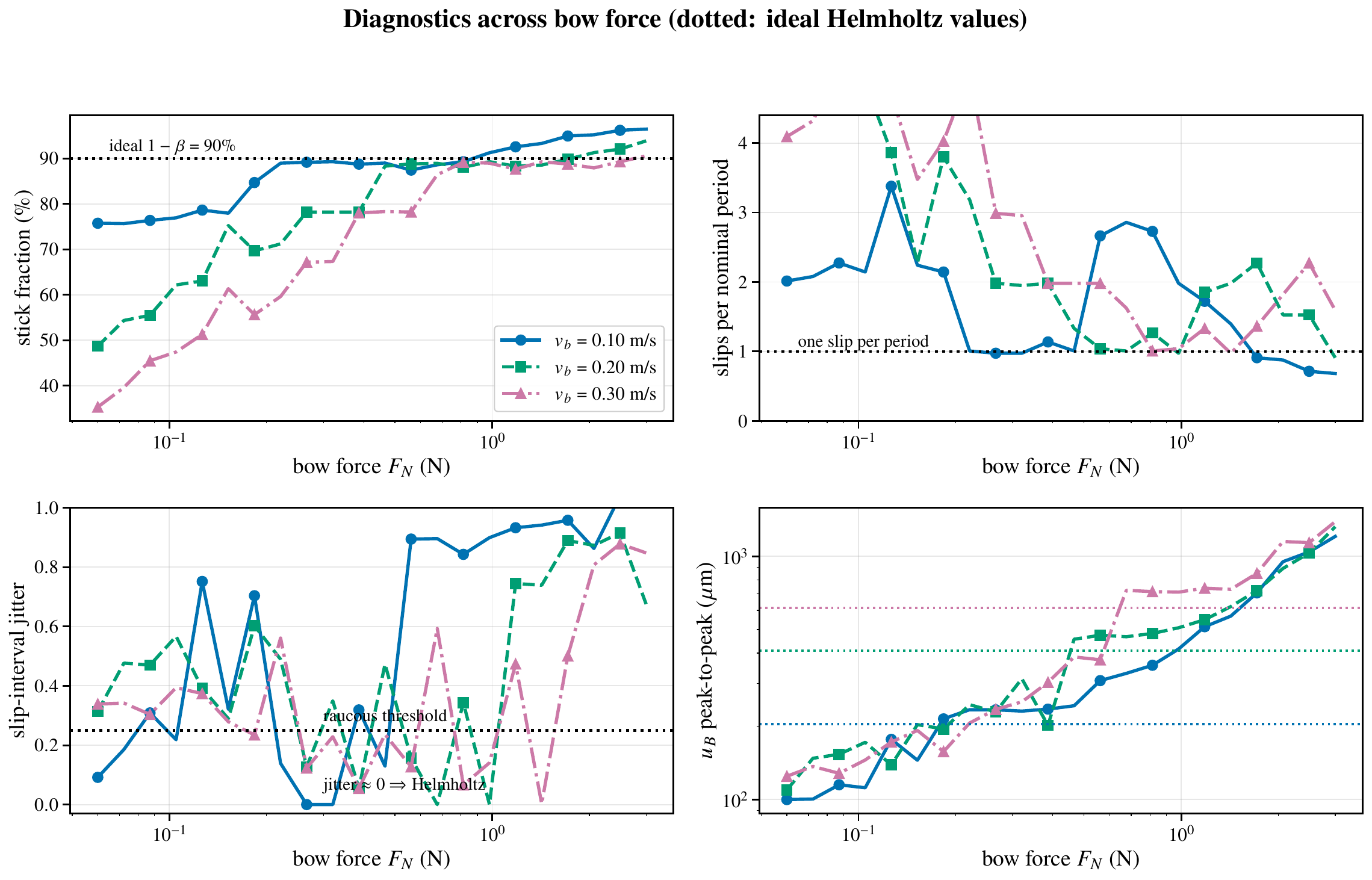}}
\caption{Diagnostics across the bow force at three bow speeds. The Helmholtz window is
where the jitter collapses to zero, the slip count reaches one per period, and the
stick fraction approaches $1-\beta$. It moves to higher forces as the bow speed
increases.}
\label{fig:sweep}
\end{figure}

\begin{figure*}[htbp]
\centerline{\includegraphics[width=\figwidewidth]{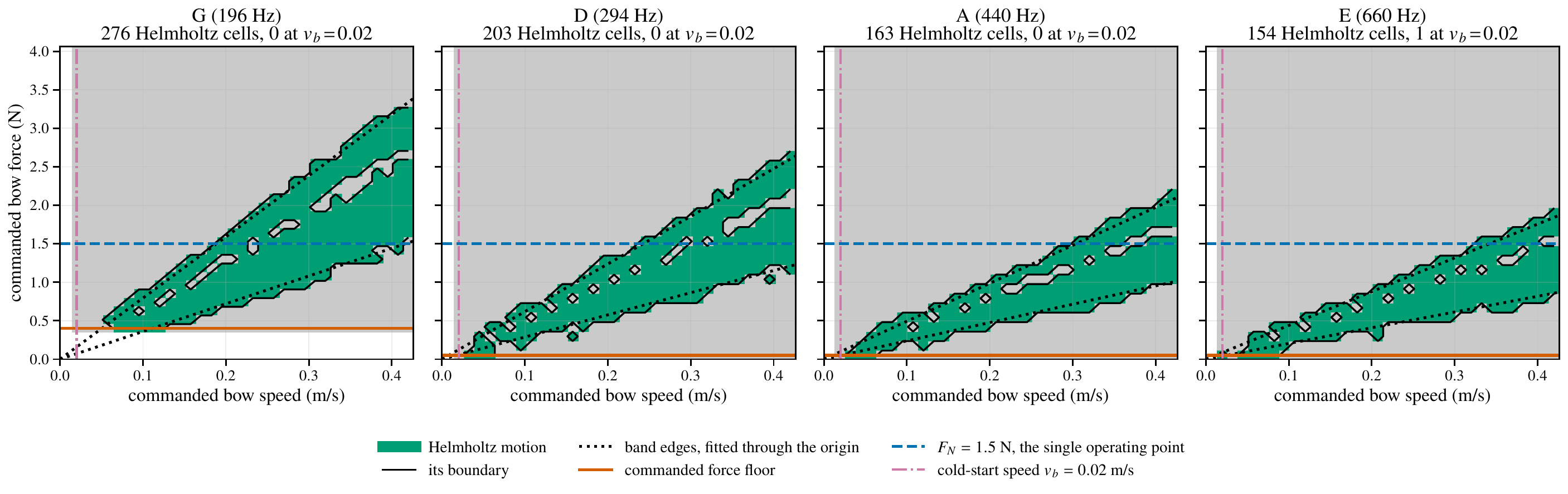}}
\caption{Playability of all four open strings, at the discretization of Table~\ref{tab:params}, measured over a window set in string periods, rather than in seconds, at $140$ settling periods. Helmholtz motion is
filled and outlined, and every other regime is recessive, so that the region is
identifiable without color. The dotted lines are the band edges, fitted
through the origin, which is the form that the proportionality to the bow speed
implies, and which may therefore be evaluated below the resolution of the grid.
The solid line is each string's commanded force floor, and the dashed line is
$F_N = 1.5$ N, the single operating point of Section~\ref{sec:params}, which sits at
$36$, $59$, $83$, and $99$ percent of the four bands. On G, the fitted window at the
cold-start speed is $[0.072, 0.159]$ N, entirely below the $0.40$ N floor, so
Helmholtz motion is unreachable there by command, rather than by resolution. The
floor and the $0.02$~m/s line are the bounds of the controller's command set, and
they are drawn here since they decide what part of each map is reachable. These
maps are reported and not adopted, since no controller in this paper was trained
on them.}
\label{fig:fourplaymap}
\end{figure*}

Those counts are a measurement at one analysis window and one discretization, and not a property of the string. This map is settled and measured over a window fixed in seconds, which is
the convention that we argue against in Section~\ref{sec:params}, and it is kept
since it is the tabulation from which the training labels of the controller were
drawn. Rebuilt at $140$ settling periods, the same string returns $163$ Helmholtz
points, rather than $118$, a difference of $38$ percent of the count. The
exchange is not one-way. In fact, $27$ cells arrive from multiple slipping and
$30$ from raucous motion at the edges of the band, while $12$ leave for raucous
motion, and none arrives from the sub-harmonic regime. The ordering of the four
strings by playable area is unchanged under every window from $35$ to $560$
periods, so comparisons across strings survive the choice, while absolute areas
do not. Figure~\ref{fig:sweep} cuts the same surface along the bow force at three
bow speeds. It shows the Helmholtz window as the interval in which the slip
jitter collapses, the slip count reaches one per period, and the stick fraction
approaches $1-\beta$, and it shows that window moving to higher forces as the bow
speed rises. Figure~\ref{fig:fourplaymap} gives the four open strings at $140$
settling periods.

\subsection{What the Map Records}
\label{sec:reach}

Two properties of the map bear on every result below. The first is its initial
condition. Since each cell starts from rest, a cell which the map calls unplayable
may still hold Helmholtz motion which arrives already established.
Section~\ref{sec:schelleng} shows that Helmholtz motion, once established,
persists far below the force needed to establish it from rest. A controller carries
the string state from one control step to the next, so it may arrive in a cell with
the motion established and hold it there.

The second is its settling window, which is set in seconds, rather than in string
periods. Its cost is measurable, rather than notional. Rebuilt at $140$ settling
periods, $57$ cells move into Helmholtz motion, $27$ from multiple slipping and $30$
from raucous motion, and $12$ move out, all of them to raucous motion. The net of
$45$ is the figure quoted above, and the $57$ are what matter here, namely, cells
which the shipped map calls unplayable, on a plant which is periodic there and not
raucous. Where this paper calls a cell mislabeled, it means one of these two
senses, a window too short to settle, or a start from rest which a closed loop does
not repeat.

Those cells are not a curiosity. Over eighty fits on the A string, namely, the
four architectures of Section~\ref{sec:architecture} at twenty seeds and four
starts each, the controller spends $20.0\%$ of its $38{,}506$ control steps
commanding one of the $57$, and it reaches Helmholtz motion on $82.2\%$ of those
steps. It visits $36$ of the $57$ distinct cells. The map lookup, on the same
four starts and the same plant, commands two of them in $593$ control steps. Both
are transit steps which are forced by the rate limit, rather than chosen. Namely,
the increment clips at $0.05$ N and $0.012$ m/s, so a target further than one
step away forces the intermediate points. Two of $593$, against the controller's
seven thousand six hundred, is the measured version of what the design of the
rule already implies.

This is what bounds the ceiling of Section~\ref{sec:baselines}. The tabulation
is correct over most of its domain, but not over all of it, and where
it is wrong, the lookup cannot follow, since the error is in the very thing which
it reads. The learned controller is not beating the tabulation at the
tabulation's own game. It goes where the tabulation declines to go. This has the
same shape as the overtaking reported in Section~\ref{sec:contactdist}, although
it is a different mechanism. On the A string, under the reduced-friction
protocol of that section, the controller does command these cells more often as
the friction falls, from $4$ to $19$ percent of its steps. However, they stop
playing once it does. Helmholtz motion in them falls from $91$ percent at full
friction to $48$ at the lowest level, which tracks the rest of the map, rather
than exceeding it. The overtaking margin does not follow the count, with a rank
correlation of $-0.54$ across the six levels, and the largest margin, $+34.3$ at
a reduction to $0.40$, is the tabulation collapsing from $100$ to $41.4$ percent,
rather than the controller reaching mislabeled cells. In other words, the cells
here are a property of the undisturbed plant, whereas the overtaking is a
property of a disturbed one.

Two things bound the reading. The regime figure is the control loop's own
classification over the last $60$ percent of each $35$ ms step, which is neither
map's window, so the $82.2$ percent answers whether the controller achieved
Helmholtz motion there, and it does not restate either map. Those periods observe
a running oscillation, rather than a settling one, since the string state is
carried from one control step to the next, which is why a short window may call
a regime which the map needed a long one to settle. Repeating the measurement at
a window two thirds longer moves the figure to $82.6$ percent. Furthermore,
$5367$ of the $7692$ arriving-cell steps are from the cold start, in the
low-force corner where that start begins. The error of the map is dense where
the cold start begins. Namely, four of the fifteen cells below $0.30$ N and
$0.08$ m/s are mislabeled, five times the average rate of the map, and two thirds
of the cold start's steps into these cells fall inside that corner. It is not
where the map is most wrong, which is at $0.4$ to $1.0$ N and mid-range bow
speeds. Of course, that coincidence cuts both ways. Seven tenths of the effect
sits on the cold start, a $32.5$~s frog-to-tip stroke at $0.02$~m/s which no
player would use, and it is there that the controller goes furthest beyond the
tabulation. From that start, the gated recurrent unit,
pooled over twenty fits each on A and D, holds Helmholtz motion on $67.0$ and
$72.8$ percent of its steps. Of those, $21.7$ and $13.6$ percent are neither in
nor next to any cell which the properly settled map calls Helmholtz, whereas
every Helmholtz step which the lookup takes from the same start lies in a cell
which that map calls Helmholtz. Not all of this motion is visible even to a
finer map. In Section~\ref{sec:coldplayable}, we hold each command at the
cold-start speed across forces from $0.03$ to $0.13$~N, in steps of $5$~mN. Over
that force range, at bow speeds up to $0.025$~m/s, the controller's own step on A
has $7.6$ times the odds of being Helmholtz on a force where held motion is
Helmholtz as on one where it is not, and on D, $8.0$ times. However, on A, $779$
of its $949$ Helmholtz steps in that range sit where held motion is not
Helmholtz, so part of what it reaches is sustained only along its own path, as the result of
Section~\ref{sec:schelleng} allows.
The cold start is the strongest instance of the result, rather than a discount on
it, though no player would use that start. The composition is also
mixed. The low-force account covers the $27$ cells arriving from multiple
slipping, and it is not shown to cover the $30$ arriving from raucous motion,
which sit above the map's upper edge, rather than below its lower one.
Section~\ref{sec:schelleng} finds the upper edge unchanged when the motion is
established first.

The rigid stop of Section~\ref{sec:fingering} does not carry these cells
unchanged to every stopped position. It makes the map position-invariant only
when its settling window is set in string periods, while the window of the
shipped map is set in seconds. The properly settled map is exactly invariant,
returning $163$ Helmholtz cells at every position, with none differing. The
shipped one is not, so the mislabeled set shrinks as the string is stopped,
namely, $57$ cells at A4, $40$ at C5, and $28$ at E5, each nested inside those of
the open string. On a fixed set of cells, the controller shows no detectable
pitch dependence. Paired over eighty fits at A4, C5, and E5, its share of steps
in the open string's $57$ is $20.4$, $20.8$, and $21.4$ percent, and its
Helmholtz fraction there is $79.5$, $76.9$, and $81.5$, with every difference
within $1.2$ standard errors. Therefore, the pitch dependence belongs to the
map.

\subsection{The Two Minimum Forces}
\label{sec:schelleng}

\begin{figure*}[htbp]
\centerline{\includegraphics[width=\figwidewidth]{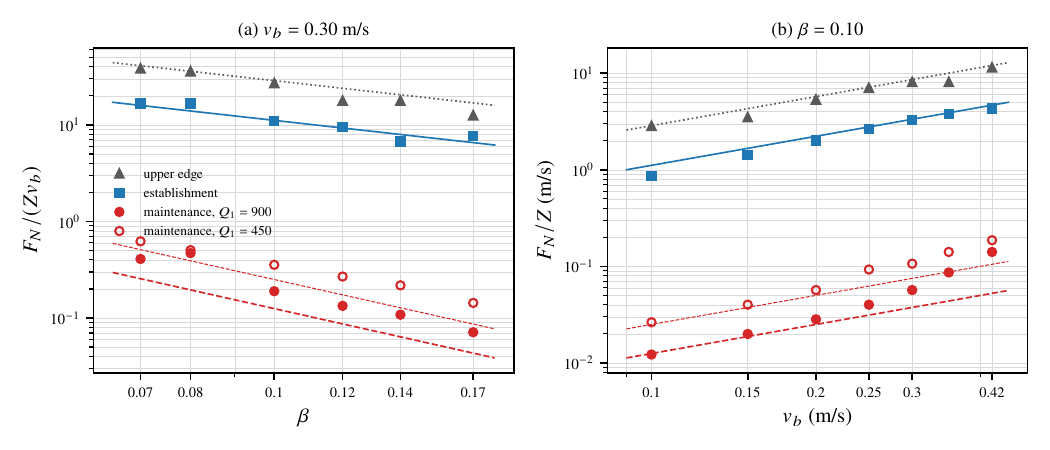}}
\caption{The two lower limits and the upper edge of the playable band on the A string,
at the discretization of Table~\ref{tab:params}, on windows settled for three decay
times. (a) Against the bow-bridge distance at $v_b = 0.30$ m/s, divided by
$Z v_b$. (b) Against the bow speed at $\beta = 0.10$, divided by $Z$. Squares: the
establishment threshold, from rest, with the law $\kappa Z v_b/\beta$ (solid).
Triangles: the upper edge, with Schelleng's maximum (dotted). Circles: the force at
which established Helmholtz motion is lost, at $Q_1 = 900$ (filled) and $450$
(open), with Schelleng's minimum computed with the model's losses (dashed).}
\label{fig:twolimits}
\end{figure*}

{\it Schelleng}~\cite{r-m:schelleng-1973} bounds the Helmholtz band above and below
by forces which are linear in the bow speed, as follows,
\begin{equation}
F_{\max} \propto \frac{Z v_b}{\beta}, \qquad
F_{\min} \propto \frac{Z^2 v_b}{R\,\beta^2}
\label{eq:schelleng}
\end{equation}
where $R$ is the resistance of the termination. The maximum is governed by the
capture and release of the string at the contact point. The minimum is a condition
for keeping an established Helmholtz motion, and it carries the loss at the
termination through $R$. This section measures both on the model, and it finds that
the lower edge of the band is two limits, rather than one
(Figure~\ref{fig:twolimits}).

\paragraph*{The Non-Dimensional Form}
At a fixed quality factor, the model fixes the form of every threshold. In
Equation~\ref{eq:string}, the bow force and the linear density enter only through
their ratio, and with time scaled by $1/f_0$ and length by $L$, both damping terms
depend on $Q$ alone. Every threshold on the bow force therefore lies at
\begin{equation}
F_N = Z v_b\, G\!\left(\beta, \frac{v_b}{v_1}, \frac{v_b}{v_2}, \frac{v_b}{v_\varepsilon}\mathpunct{\mbox{$;$}} Q\right)
\label{eq:G}
\end{equation}
with $G$ dimensionless. The sample rate scales with the fundamental, so $N_x$ and
$\lambda$ are shared, and the four open strings are one solution with the force
scaled by $Z$. The impedance exponent of every threshold is one by construction,
and the scheme preserves it. Varying the linear density alone, over a factor of ten
in impedance, gives an exponent of $+1.009 \pm 0.014$. So the four strings are
replicate readings of one experiment, and what the measurements below carry is the
dependence of $G$ on $\beta$ and $v_b$, and its scale.

\paragraph*{The Establishment Threshold}
Table~\ref{tab:schelleng} gives the exponents of both band edges. They were located
by a dense geometric scan over eighty forces from $0.04$ to $10.0$ N, at $7.2\%$ per
step, at the $N_x = 101$ grid of Table~\ref{tab:params}, with every force a separate
start from rest. On each string, six bow-bridge distances, with $\beta$ from $0.07$
to $0.17$, are crossed with $v_b = 0.30$ m/s, and
seven bow speeds, from $0.10$ to $0.42$ m/s, are crossed with $\beta = 0.10$, giving forty-eight cells over the four
strings. On a $330$ mm string, the first range is
$23.1$ mm to $56.1$ mm from the bridge, which spans mid-range ordinary bowing at
one end and reaches just past the {\it sul tasto} edge at the other. The second
range spans the commanded speeds, from a $6.5$ s frog-to-tip stroke at the slow
end to the fastest which the command box permits, and it is kept well clear of
the cold-start floor discussed in Section~\ref{sec:coldplayable}.

An earlier bisection of the same range
assumed that the Helmholtz test is monotone in force, so that one crossing
separates a failing region below from a passing region above. However, it is
not. When the whole range is scanned and every outcome is kept, instead of
bisecting to a single crossing, every one of the forty-eight cells changes its
verdict more than once, namely, twice at least, four times at the median, and
sixteen times at the worst. The raggedness rises as the bow slows, from a median
of two crossings at $v_b = 0.42$ m/s to eight at $0.10$, and it reaches seven at
the extreme bow-bridge distance, $\beta = 0.17$. A bisection run on such a cell
returns a number and discards the evidence that the number is not an edge. A band
edge is therefore taken from the longest contiguous run of Helmholtz verdicts, and
a cell counts as well resolved when it has four crossings or fewer.

\begin{table}[htbp]
\caption{Power-law exponents of the two band edges on each open string, from rest,
with the standard error of the least-squares slope. The $\beta$ fits use six points
each, and the $v_b$ fits use seven. The last row gives Schelleng's exponents for his
maximum and his minimum.}
\label{tab:schelleng}
\centering
\scriptsize
\setlength{\tabcolsep}{2pt}
\begin{tabular}{|l|c|c|c|c|}
\hline
 & \multicolumn{2}{c|}{\textbf{against $\beta$}} & \multicolumn{2}{c|}{\textbf{against $v_b$}} \\
\hline
\textbf{String} & \textbf{upper} & \textbf{lower} & \textbf{upper} & \textbf{lower} \\
\hline
G & $-1.02 \pm 0.17$ & $-0.83 \pm 0.16$ & $+0.89 \pm 0.05$ & $+0.82 \pm 0.10$ \\
D & $-1.17 \pm 0.08$ & $-1.00 \pm 0.18$ & $+0.95 \pm 0.08$ & $+1.14 \pm 0.09$ \\
A & $-1.12 \pm 0.12$ & $-1.08 \pm 0.18$ & $+0.99 \pm 0.08$ & $+1.13 \pm 0.03$ \\
E & $-0.98 \pm 0.15$ & $-0.60 \pm 0.28$ & $+1.26 \pm 0.15$ & $+1.06 \pm 0.03$ \\
\hline
Schelleng & $\mathbf{-1}$ & $\mathbf{-2}$ & $\mathbf{+1}$ & $\mathbf{+1}$ \\
\hline
\end{tabular}
\end{table}

The upper edge recovers Schelleng's exponents on both axes and on every string,
though not tightly. Against the bow-bridge distance, the four exponents span $0.19$,
and each sits within two and a half standard errors of $-1$, the worst being D at
$2.2$. Against the bow speed, they span $0.37$ about $+1$. Those
intervals are wider than the fitted edges alone would suggest, and the reason is
the same one which bounds the law below. Namely, the band edge is resolved to
about eight percent, so a per-string slope through six or seven of them carries
that scatter. The lower edge does not
follow Schelleng's minimum. Its exponent on $\beta$ falls short of $-2$ on every
string, by between $4.9$ and $7.5$ standard errors. Since the four strings are one
problem, the spread of their exponents measures the error of locating the edge.
Against the bow speed, that spread exceeds the per-string standard errors. The lower
edge gives $+1.13 \pm 0.03$ on A and $+0.82 \pm 0.10$ on G.

Fitted jointly over the thirty-three well-resolved cells, the lower edge follows the
relation,
\begin{equation}
F_{\mathrm{est}} \propto Z^{1.064 \pm 0.073}\, v_b^{1.035 \pm 0.068}\,
\beta^{-1.009 \pm 0.082}
\label{eq:fminfit}
\end{equation}
Every one of the three exponents sits within one standard error of an integer, at
$0.9$, $0.5$, and $0.1$. The impedance exponent is the identity of
Equation~\ref{eq:G}. The other two make the threshold a law,
\begin{equation}
F_{\mathrm{est}} = \kappa\, \frac{Z v_b}{\beta}, \qquad \kappa = 1.112 \pm 0.017
\label{eq:fminlaw}
\end{equation}
with $\kappa$ dimensionless, at the discretization used. It does not drift with the
bow speed or with the bow-bridge distance, at $p = 0.81$ and $0.66$. Its residual
standard deviation is $8.4\%$, and its median error is $3.8\%$. The constant absorbs
the friction difference. Scaling $a_1$ and $a_2$ together, at five levels over a
fourfold range, gives an exponent on their sum, the nominal friction difference, of
$-1.007 \pm 0.054$. So the establishment threshold has the same form as
Schelleng's maximum, proportional to $Z v_b/\beta$ and inversely proportional to the
friction difference, and the two edges run parallel in
Figure~\ref{fig:twolimits}.

The agreement is not exact. Multiplied by the friction scale, the constant should
be flat, and instead it dips by about nine percent at intermediate values before
returning, with two levels sitting more than three combined standard errors below
the control. That is a real second-order residual against an effect of a factor
of four, and it does not move the exponent.

Scaling the whole characteristic instead, friction and Coulomb term together,
measures nothing, since the plant depends on $F_N$ and $\varphi$ only through
their product, so scaling one by any factor and the other by its reciprocal leaves the equations identical. That was run over $288$ cells, and it returned the control value to
four decimals at every level, with bit-identical regime vectors, which is a check
on the simulator, rather than on the law.

Against the three-exponent fit of Equation~\ref{eq:fminfit}, which has four free
parameters where Equation~\ref{eq:fminlaw} has one, the law gives up $0.0017$ of
$R^2$, $0.948$ against $0.950$. A nested $F$ test on the restriction to integer
exponents gives $F = 0.32$ on three and twenty-nine degrees of freedom, at
$p = 0.81$, so the three free exponents are not buying anything which the data
asks for. This shows that the restriction is not worse, and not yet that it is
better. A criterion which charges for parameters does show it. The Bayesian
information criterion favors the one-parameter law by $9.4$, which is strong
evidence on the usual scale, and the Akaike criterion favors it by $4.9$.
Therefore, the one-parameter form is not a tolerable simplification of the fit.
It is the better description of the same data.

These exponents and the joint fit are measured over a window scaled to a fixed
number of string periods, rather than to a fixed duration, following
Section~\ref{sec:playability}. This choice is not cosmetic. Under a window fixed
in seconds, the four strings receive $15.7$, $23.5$, $35.2$, and $52.8$ periods to
settle. The number of bow positions at which the band cannot be located at all
follows that ordering exactly, at five of six on G, two of six on D, and none on
A or E. At the fixed-duration setting, the bow-bridge exponent of the G string
cannot be measured, since only one bow position returns a band. Its bow-speed
exponents can, however, at $+0.95 \pm 0.11$ for the lower edge and $+0.85 \pm 0.09$
for the upper edge, from four of seven speeds. The two conventions agree closely on
the bow-speed exponents, which move by $+0.04$ and $-0.09$ on A, and they disagree
on the bow-bridge exponent, which moves from $-1.14$ to $-1.58$ on the same
string.

The impedance dependence may also be read from the magnitude of the threshold,
rather than from a fitted slope. At $\beta = 0.10$ and $v_b = 0.30$ m/s, the
establishment threshold is $0.996$, $0.808$, $0.702$, and $0.570$ N on G, D, A, and
E. Regressing its logarithm on the logarithm of the impedance gives
$+0.91 \pm 0.13$, at a correlation of $+0.981$, which is a check of the first-power
similarity of Equation~\ref{eq:G}. Dividing each threshold by $Z$ leaves it varying
by a factor of $1.12$ across the four strings, and dividing it by $Z^2$ leaves it
varying by $1.79$.

Four strings give four impedances spanning a factor of $1.77$, which is little
leverage for an exponent, and no amount of stopping adds more, since a rigid stop
leaves $Z$ exactly where it was. Only a different string does. Therefore, the
exponent was measured again on strings which do not exist. They hold
$f_0 = 440$ Hz and $L = 0.33$ m, so that the wave speed is fixed, and they vary the
linear density alone over ten values, which makes $Z$ proportional to it by
construction, while no other geometry moves. Over a factor of ten in impedance,
at four combinations of bow speed and bow-bridge distance, the exponent is $+1.009 \pm 0.014$, $0.64$ standard errors from exactly one. It holds separately
at each of the four conditions, at $+0.969$, $+1.027$, $+1.022$, and $+1.018$, so
it is not borrowing leverage from the other two variables. The constant fitted
there is $1.094 \pm 0.011$ against $1.112 \pm 0.017$ from the four real strings,
$0.87$ standard errors apart.

That sweep tests the model, and not the instrument. The linear densities, $0.28$
to $2.75$ g/m, bracket the $0.412$ to $2.457$ of the four strings modeled here.
The tensions which follow from holding the wave speed fixed do not stay near
theirs, since the heaviest reaches $232$ N, where those strings carry between $41$
and $78$. What it establishes is that the linearity is a property of the model
across a decade of impedance, rather than an artifact of four points, and the
figure made entirely of violin strings remains the $+1.06 \pm 0.07$ above.

One boundary of the law is worth naming, since it is not where a reader would
look for it. Every cell behind it was measured on an open string, so the law was
fitted at four pitches and never at a stopped one. It covers the rest by
implication, rather than by omission. A rigid stop at constant $\beta$ leaves the
wave speed and the linear density alone, so it leaves $Z$ exactly where it was,
and the remaining two variables are commanded. The law therefore predicts the
same establishment threshold at every stopped position on a string. The invariance of
Section~\ref{sec:fingering} confirms it. Two positions on one string run the same
simulation, their relative velocities agreeing to $6.8 \times 10^{-13}$~m/s, and
the properly settled maps of all thirty-six stopped positions agree with their
open strings in every one of the $1089$ cells. Their band edges are therefore the
same at every stopped position, to the map's resolution. The three octaves which
the controller plays are covered by that identity, and not by a separate sweep.

The forty-eight cells form a cross, rather than a grid. Namely, every one of them
holds either $v_b = 0.30$ m/s or $\beta = 0.10$, so a constant fitted on them has
never been asked to predict a combination of the two. Sixty-four further cells
were measured, sixteen on each of the four strings, at four bow-bridge distances
crossed with four bow speeds, none of them on either reference axis and none used to estimate $\kappa$, with the prediction fixed in advance. Refitted on those cells
alone, the constant is $1.099 \pm 0.031$, which is $0.41$ standard errors from the
value carried in from the cross. The held-out cells lie inside the range spanned by
the cross, so they test whether the law separates in $\beta$ and $v_b$, and not
whether it extrapolates.

One selection effect is worth stating, since it runs the other way from what it
looks like. Restricted to the well-resolved held-out cells, the constant is $1.159 \pm 0.014$, which is $2.1$ standard errors above the fitted value, and it
reads as a four percent under-prediction off the reference axes. That gap is a
property of the crossing cut, rather than of the law. The four-crossing threshold
was not fixed in advance. It is the median crossing count in both populations,
and $2.1$ standard errors is the largest disagreement which any threshold in the
range produces. If it is relaxed, the gap closes monotonically. At eight crossings, the two agree to $0.3$ standard errors, and over the full forty-eight and sixty-four cells, they agree to $0.4$.

One qualification bounds this. The establishment threshold is the lower band edge,
where the regime classifier is least certain, and the number of crossings per cell
is an objective measure of that uncertainty, rather than a judgment. It predicts the
residual. Regressed on the absolute error of Equation~\ref{eq:fminlaw}, the slope
is $+0.016$ per crossing, at $t = +3.61$ and $p = 0.0008$. The cells whose edge is
poorly resolved are the cells which miss the law, which is what licenses quoting
the well-resolved subset, rather than the whole. Over all forty-eight, the
constant is $1.115 \pm 0.022$ and the residual standard deviation is $12.8\%$, so
the law survives the ragged cells and is simply looser through them. What the
crossings are is a separate question, and it is open. Refining the raggedest
cells changes $29$ percent of individual verdicts between the published
resolution and twice it. The crossing count falls and then rises, rather than
settling toward one, so the classification there is not resolution-converged,
and nothing here shows the raggedness to be physical.

The threshold does not depend on the classifier. The A string's cells were
reclassified with $\epsilon$ from $0.5$ to $2$~mm/s, a floor on $\eta$ from $0.35$
to $0.70$, and a ceiling of $1.2$ to $1.5$ slips per period, eighteen combinations
in all. In every one of them, $\kappa$ and both exponents are unchanged, at $-1.08$
on $\beta$ and $+1.13$ on $v_b$ (Table~\ref{tab:sensitivity}).

\begin{table}[htbp]
\caption{Sensitivity of the two lower limits on the A string. Upper part: the
constant $\kappa$ of Equation~\ref{eq:fminlaw}, fitted on the A string's seven
well-resolved cells, under refinement, longer settling, and changes to the
classifier's thresholds. The four-string value is $1.112 \pm 0.017$. Lower part:
the median exponent on $Q$ and the exponent on $\beta$ of the maintenance limit, at
the published grid and at a refined one.}
\label{tab:sensitivity}
\centering
\footnotesize
\setlength{\tabcolsep}{3pt}
\begin{tabular}{|l|l|c|}
\hline
\textbf{Grid} & \textbf{Window or change} & $\kappa$ \\
\hline
$\lambda = 0.9167$, $N_x = 101$ & $140$ periods & $1.1143$ \\
$\lambda = 0.9167$, $N_x = 201$ & $140$ periods & $1.1032$ \\
$\lambda = 0.9167$, $N_x = 401$ & $140$ periods & $1.0495$ \\
$\lambda = 0.95$, $N_x = 101$ & $140$ periods & $1.0391$ \\
$\lambda = 0.95$, $N_x = 201$ & $140$ periods & $1.0391$ \\
$\lambda = 0.95$, $N_x = 401$ & $140$ periods & $1.0185$ \\
$\lambda = 0.9167$, $N_x = 101$ & settled & $1.0923$ \\
$\lambda = 0.98$, $N_x = 101$ & settled & $0.9689$ \\
$\lambda = 0.9167$, $N_x = 101$ & classifier, $18$ settings & $1.1143$ in all \\
\hline
\textbf{Grid} & \textbf{Maintenance limit} & $Q$, $\beta$ \\
\hline
$\lambda = 0.9167$, $N_x = 101$ & settled, $12$ cells & $-1.01$, $-2.17$ \\
$\lambda = 0.95$, $N_x = 201$ & settled, $6$ cells & $-1.01$, $-1.71$ \\
\hline
\end{tabular}
\end{table}

The threshold does depend on the discretization. Refining the grid, with the
Courant number held exactly invariant, moves individual band edges by $7.2$ percent
at the median, which is the same size as the residual, while moving the mean
constant by $3.0$ percent. The scatter will not shrink by fitting the law better.
It shrinks by resolving the edge in force. Doubling the grid from $N_x = 101$ to
$201$ at a fixed Courant number moves the A string's constant by $1.0$ percent, a
further doubling moves it by $4.9$ percent, and at a Courant number of $0.98$ it is
six percent lower on the window used here, and eleven percent lower on windows
settled for three decay times. So the constant is not converged. Its exponent on
$\beta$ is steadier. At $\lambda = 0.95$, it is about $-1.0$ at $N_x = 201$ and at
$401$, while at $\lambda = 0.9167$ ragged cells pull the fit at $N_x = 401$ away
from it.

Raising the force slowly lowers the threshold little. In the twelve cells of the A
string's cross, the bow force was raised from well below the from-rest edge one step
of $7.24$ percent at a time, with the state of the string carried from step to step
and each step settled for three decay times. Helmholtz motion first appears within
three steps of the from-rest edge in $10$ of the $12$ cells, and five and nine steps
below it, a factor of $1.4$ and $1.9$, in the other two. So a slow rise lowers the
establishment threshold by at most about a factor of two, far less than the gap,
measured below, to the force at which established Helmholtz motion is lost.

\paragraph*{The Maintenance Limit}
Schelleng's minimum needs a termination resistance, and on rigid supports the only
loss is the string's own. A near-rigid end of resistance $R$ loses a fraction $4Z/R$
of the energy at each reflection, so an equivalent resistance which reproduces a
round-trip energy loss of $2\pi/Q$ scales as $ZQ$. Substituted into
Equation~\ref{eq:schelleng}, it gives $F_{\min} \propto Z v_b/(Q \beta^2)$, and with
the constant of Schelleng's derivation,
\begin{equation}
F_{\min} = \frac{\pi Z v_b}{4 Q \beta^2 \Delta\mu}
\label{eq:schellengmin}
\end{equation}
with $\Delta\mu = 0.698$. Its impedance exponent is one, as it is for any threshold
in this model. This is not the establishment threshold. That threshold goes as
$\beta^{-1}$, twelve standard errors from $\beta^{-2}$, and it does not move with $Q$
at the measured losses. Measured over a sixteen-fold range of the quality factor,
from $Q = 225$ to $3600$, at eight fixed operating points spanning all four strings,
its exponent on $Q$ is $+0.050 \pm 0.018$, against the $-1$ of
Equation~\ref{eq:schellengmin}, which is $60$ standard errors away. On windows
settled for three decay times, it is $+0.01 \pm 0.05$, consistent with zero.

Equation~\ref{eq:schellengmin} does describe the force at which an established
Helmholtz motion is lost. That force is measured by continuation. In each of the
twelve cells, Helmholtz motion is established from rest in the middle of the band,
and the bow force is then lowered one step of $7.24$ percent at a time, with the
state of the string carried from step to step, and with each step settled for three
decay times before it is classified. The force grid extends down to $0.00049$ N. In
several cells, Helmholtz motion returns below its first break, so the limit is taken
as the bottom of the lowest run of three consecutive Helmholtz verdicts. The sweep
is run at $Q_1 = Q_2 = 900$, and again at $450$.

Established Helmholtz motion is kept far below the establishment threshold. At
$Q = 900$, it is lost between $0.00244$ N, at $\beta = 0.10$ and $v_b = 0.10$ m/s,
and $0.02820$ N. That is $1.0$ to $2.7$ times the value of
Equation~\ref{eq:schellengmin} (Figure~\ref{fig:twolimits}). Its exponent on $\beta$
is $-2.17$, against Schelleng's $-2$. Its exponent on $Q$, from $450$ against $900$,
is $-1.01$ at the median over the twelve cells, against Schelleng's $-1$. Its
exponent on $v_b$ is $+1.66$, steeper than Schelleng's $+1$. The weaker loss
dependence sits in the four cells with the largest $v_b/\beta^2$, from $35$ to $61$.
Their exponents on $Q$ are $-0.10$ to $-0.71$, so they are the least loss-limited,
they lie at $1.6$ to $2.7$ times Schelleng's value, and they steepen the exponent on
$v_b$. The other cells lie at $1.0$ to $1.7$ times Schelleng's value, with exponents
on $Q$ between $-0.91$ and $-1.21$.

The maintenance limit holds under refinement of the grid. At $\lambda = 0.95$ and
$N_x = 201$, on six of the cells, its median exponent on $Q$ is still $-1.01$, and
its exponent on $\beta$ is $-1.71$, over four cells. At $Q = 900$, the limit is
unchanged in four of the six cells. In the two with the largest $v_b/\beta^2$, it
falls by $19$ and $24$ percent, and it becomes more loss-limited, at $-0.91$.

{\it Woodhouse}~\cite{r-m:woodhouse-1993-2} replaced Schelleng's single termination
resistance with a measured bridge admittance and obtained note-by-note predictions,
and {\it Mansour, Woodhouse, and Scavone}~\cite{r-m:mansour-2017} found that the
bow-bridge dependence then stops being a fixed power at all, and in extreme
cases, it splits the playable range in two. Neither mechanism operates in this
model, since its terminations are rigid.

\paragraph*{Between the Two}
At $Q = 900$, the establishment threshold lies $31$ to $108$ times above the
maintenance limit. In that range, Helmholtz motion is kept if it is established, and
it is not reached otherwise. A map built from rest, such as those of
Section~\ref{sec:playability}, records the establishment threshold as its lower
edge.

\paragraph*{The Upper Edge}
The upper edge is the same either way. Continuation moves it by $0$ to $+4$ force
steps, and by $-2$ in one cell. It behaves as the existence limit which Schelleng's
maximum describes, and its exponents follow his (Table~\ref{tab:schelleng}). That
suits a model on rigid supports, since the maximum depends little on the
body~\cite{r-m:mansour-2017}. Under refinement of the grid, the upper edge falls by
up to a quarter on the A string, more than the lower edge, and at $\lambda = 0.95$
its exponent on $\beta$ drifts from $-1.08$ to $-0.62$ as $N_x$ rises from $101$ to
$401$. So the maximum, too, is recovered at the discretization used.

\paragraph*{The Domain in Loss}
The establishment law holds at the measured losses and as they are raised. On the
twelve cells of the A string's cross, the lower edge lies within twenty-three
percent of the law at ten or eleven of them, at the measured losses and at two and
four times them. When the losses are halved, it does so at seven of twelve, and
where $v_b/\beta^2$ is smallest the edge falls nearly sevenfold below the law. As the losses fall further, Helmholtz motion also appears below that edge, interleaved with multiple slipping, first and furthest where $v_b/\beta^2$ is smallest, and the lower edge leaves the law. This persists at a Courant number of $0.98$ and with a settling window half as long, so it is neither numerical dispersion nor a settling artifact. Nor
does the law extend beyond the bow-bridge distances measured. At $\beta = 0.20$ and
$0.25$, three of four points have their lower edge a factor of one and a half to
fourteen below it, even at the measured losses.

The exponent of the establishment threshold on $\beta$ coincides with an
experimental result of {\it Schoonderwaldt, Guettler, and
Askenfelt}~\cite{r-m:schoonderwaldt-2008}, who reach $\beta^{-1}$ for the minimum
force as a low-speed limit, once the shape of the friction curve is taken into
account. Their route is an analytic limit from a modified model with a
hyperbolic friction curve. The present result, on the other hand, is an empirical
joint fit over the full measured range, on four strings, with a two-exponential
Stribeck curve. At the bow speeds measured here, their mechanism predicts
Schelleng's $\beta^{-2}$, so the agreement is in the exponent only.

The bow-speed dependence, however, does not agree, and the disagreement is large.
The same experiments found the minimum force to be independent of the bow
velocity over the range measured, five to twenty centimeters per second, which
they describe as being in clear contradiction to Schelleng's prediction. The fit
here gives $v_b^{1.035 \pm 0.068}$, fifteen standard errors from independence and
half a standard error from Schelleng's linear prediction. The two are not
measuring the same object, since theirs is an experiment on steel D and E strings
mounted on a monochord and on a violin, and this is a simulation of four strings.
The two ranges of bow speed also barely overlap, theirs running from $0.05$ to
$0.20$ m/s and this one from $0.10$ to $0.42$, so the two mostly probe different
speeds. Neither limit here is independent of the bow speed where the two ranges
overlap. Nothing here resolves it.

One recent result bears on the lower edge from a direction other than the
termination. {\it Van Walstijn, et al.}~\cite{r-m:vanwalstijn-2026} combine
elasto-plastic pre-sliding with the contact temperature of the thermal model, on
a modal string with torsional motion and a finite-width bow. They report that the
friction parameters can be set so that both the minimum and the maximum bow force
are placed against measurement over a range of bow-bridge distances. They fit to
a measured instrument, rather than to Schelleng's exponents, so the two results
are not the same quantity, and nothing measured here is confirmed or refuted by
them. What their result establishes is that a model of this kind can place the
lower edge. Which of the ingredients absent here does that work is identified by
neither paper, and the candidates are several, namely, the contact temperature,
the pre-sliding state, torsional motion, the finite bow width, and the compliance
of the hair.

\begin{figure*}[htbp]
\centerline{\includegraphics[width=\figwidewidth]{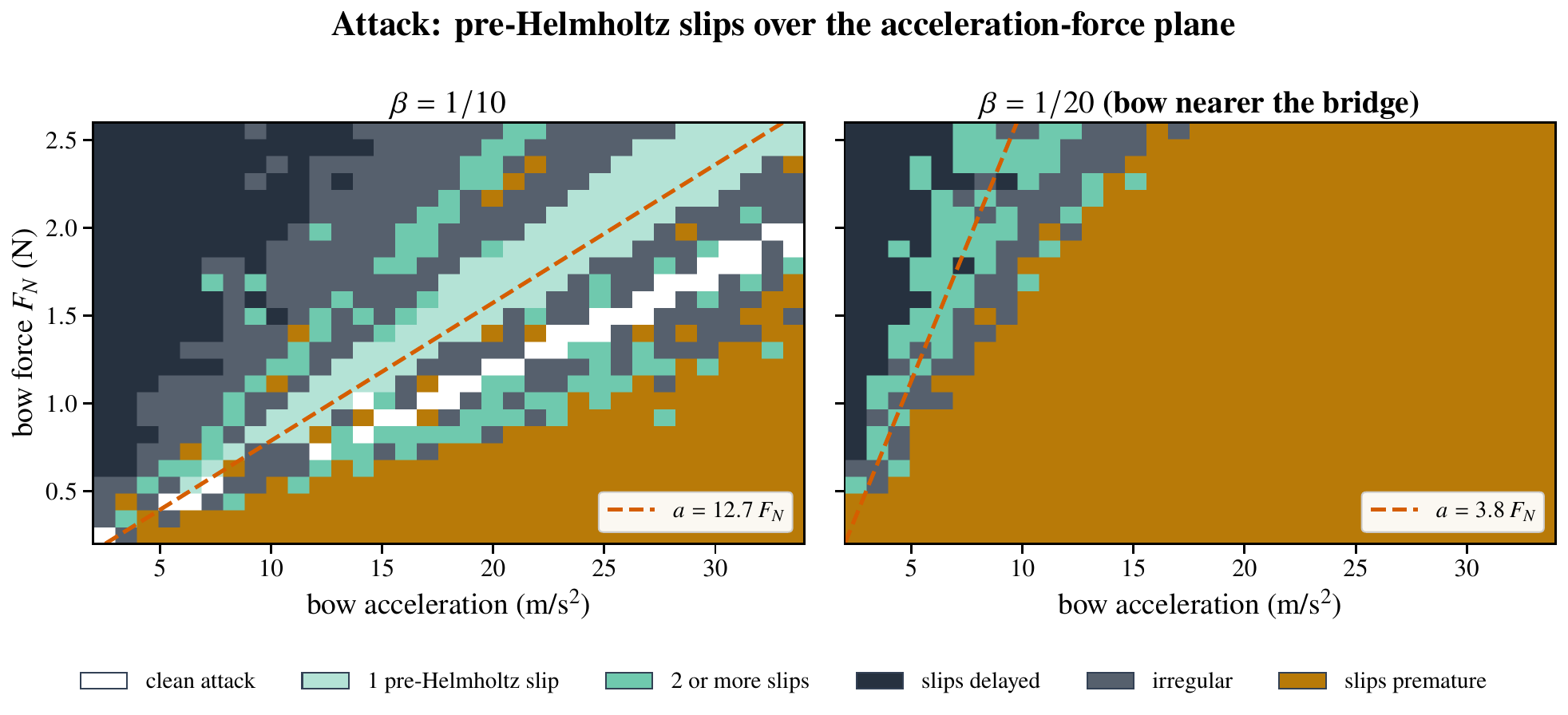}}
\caption{The {\it Guettler diagram}~\cite{r-m:guettler-2002-2} of the model, namely, attacks
from rest over the acceleration-force plane, at two bow-bridge distances. Color
gives the number of slips preceding regular triggering at the nominal period, and
luminance falls monotonically from the clean attack to the worst outcome, so that
the figure keeps its meaning in grayscale. Below the band, the contact force is
large for the acceleration, and the first slips are delayed. Above it, the force is
small, and the string is released early and repeatedly. The dashed line is the
median ratio of acceleration to force among the points which establish, in
m\,s$^{-2}$\,N$^{-1}$, and the legend writes the bow acceleration as $a$.}
\label{fig:guettler}
\end{figure*}
\subsection{Attack Transients and the Guettler Diagram}
\label{sec:guettler}

The preceding tests are of the steady state, whereas
{\it Guettler}'s test~\cite{r-m:guettler-2002-2} is of the onset. The bow starts at
rest, on a string at rest, and it accelerates uniformly,
$v_b(t) = \dot v_b\,t$, at a constant contact force. The attack is {\it clean}
when periodic slipping at the nominal period is established from the first slip,
and the number of slips preceding that point, mapped over the acceleration-force
plane, is the standard playability test for a model of this kind.

Figure~\ref{fig:guettler} gives that plane, on a $33 \times 25$ grid over
$\dot v_b \in [2, 34]$~m/s$^2$ and $F_N \in [0.2, 2.6]$~N, measured over $22$
nominal periods. The sweep runs at the grid of Section~\ref{sec:schelleng}, since
the bow sits on a grid point there, and both $\beta = 1/10$ and $\beta = 1/20$ are
exact multiples of $1/(N_x - 1)$.

The three regions which Guettler describes all appear. Where the force is large
for the acceleration, the first slips are delayed. Where it is small, the string
is released early and repeatedly. Between them, the attack establishes itself in
a narrow diagonal band. At $\beta = 1/10$, $224$ of the $825$ points establish at all,
at a median ratio $\dot v_b/F_N$ of $12.7$~m\,s$^{-2}$\,N$^{-1}$, which is the dashed
line of the figure. Of those, $36$ are clean, and a further $102$ are preceded by a
single slip. The clean ridge itself lies at $17.0$~m\,s$^{-2}$\,N$^{-1}$. At
$\beta = 1/10$, the clean region extends beyond the plotted window, to $4.1$~N and
$68$~m/s$^2$.

Run on the other three open strings at $\beta = 1/20$, the same sweep finds no
clean attack anywhere, which is the only comparison across strings which the grid
supports. At $\beta = 1/10$, the clean fractions read $2.1\%$ on G, $4.0\%$ on D,
$4.4\%$ on A, and $2.4\%$ on E, and those four numbers are not comparable with each
other. The grid is centered on A, so it clips each string differently. Namely, G
and D reach the upper force edge at $2.60$ N, while A and E reach the lower edge
and the maximum acceleration. Therefore, every count is a lower bound, and ranking
the strings on them would be ranking the grid.

Moving the bow to $\beta = 1/20$ closes the clean band, and it stays closed on a plane extended to $5.2$~N and $68$~m/s$^2$. No point in the plane
reaches Helmholtz motion with fewer than two pre-Helmholtz slips, and a probe which
samples the plane four times more finely along the ridge returns the same floor of
two. {\it Guettler}~\cite{r-m:guettler-2006}
reports that bowing closer to the bridge narrows the range of accelerations which
give a clean onset. At half the bow-bridge distance, in this model, the range
closes altogether. That is a result at the grid used, which resolves the short
bow-bridge segment at $\beta = 1/20$ with few cells, and it has not been tested
under refinement of the string's grid.

This plane has also been used as a measured validation target, rather than as a
qualitative check. {\it Lampis, Mayer, and Chatziioannou}~\cite{r-m:lampis-2024}
recorded a Guettler diagram on a monochord bowed by a robot arm, and {\it Matusiak
and Chatziioannou}~\cite{r-m:matusiak-2024} compared a simulated diagram against
it, under an elasto-plastic friction law. What we report here is the three
regions and the closing of the clean band, against Guettler's description, rather
than against a measured plane, which is the weaker of the two forms of this
test.

Two measurement details decide the outcome, and both of them were found by
checking, rather than by assumption. The steady-state detector of
Section~\ref{sec:diagnostics} thresholds $|v_{\mathrm{rel}}|$ at a fixed $1$~mm/s,
which suits a bow running at $0.3$~m/s. During an attack, however, the bow starts
from rest, the stick-phase relative velocity is a fraction of a millimeter per
second, and a fixed threshold chatters across it and reports slips which are a
hundredth of a period apart. Slip excursions run at about ten times the
instantaneous bow speed, so a threshold proportional to it separates them, and
the counts are identical for constants of $0.3$, $0.5$, and $0.7$. Periodicity
alone also admits free ringing. Where the force is small, the string is barely
gripped and rings at $f_0$, which is periodic at exactly the nominal period and
passes any test of regularity. A scan of the low-force corner at four times the
production resolution finds $934$ such points, all of them below $0.11$~N,
carrying stick fractions from $0.02$ to $0.21$, with a median of $0.11$, against
$1 - \beta = 0.90$ for real Helmholtz motion. Every one of them is below the
classifier's floor of $0.35$, so pairing the periodicity test with a stick test
removes the whole class. In other words, it is the trap of
Section~\ref{sec:diagnostics} in a second place.

\subsection{Bow Forces}
\label{sec:bowforces}

The friction model responds to the normal force at the contact point, but a
robot, like a player, commands force at the frog. The two are related by a
{\it transmission factor},
\begin{equation}
F_N = \alpha(s)\,F_{\mathrm{frog}}, \qquad
\alpha(s) = 1 - \tfrac{1}{2}s
\label{eq:alpha}
\end{equation}
where $s\in[0,1]$ is the normalized position of the contact point along the bow,
from the frog to the tip.

The origin of this factor is not a transmission ratio in the mechanical sense,
since the force at the frog is not a single quantity. It may mean the vertical
reaction on a rigid bow, the transverse force at the hair termination, the force
implied by a controlled height, or the force which a bounded hand moment can
reach, and these behave differently.

\begin{figure}[htbp]
\centerline{\includegraphics[width=\figwidth]{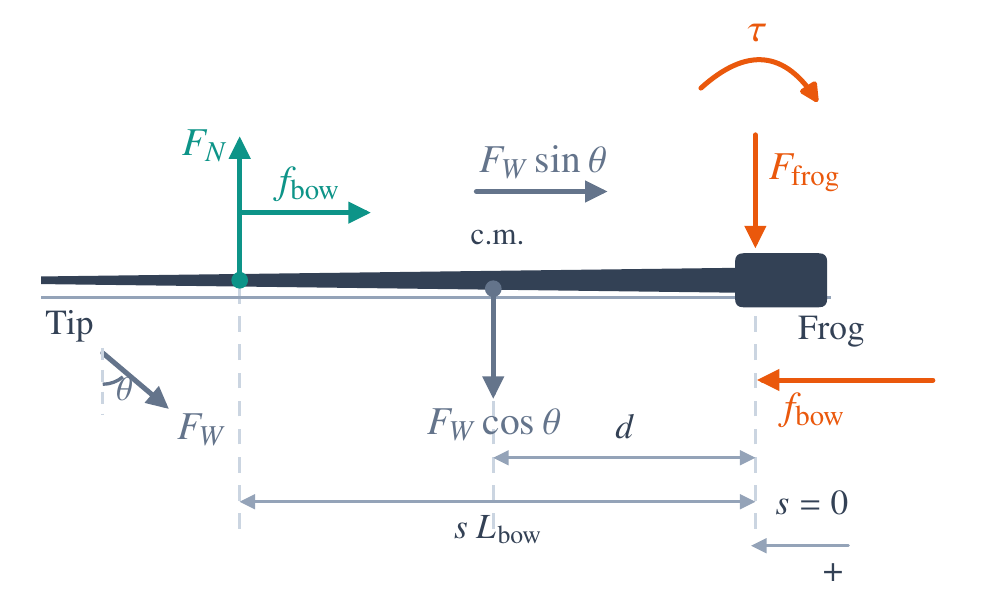}}
\caption{Free-body diagram of the forces and moments on a violin bow. Forces exerted by the
string are shown in teal, those exerted by the hand at the frog in orange, and the
bow weight in gray. Drawn after the analysis of
{\it Testa}~\cite{r-m:testa-2000}.}
\label{fig:fbd}
\end{figure}

On a rigid, massless bow, the free-body balance of Figure~\ref{fig:fbd} reduces to
$F_{\mathrm{frog}} = F_N$, with a horizontal reaction equal to $f_{\mathrm{bow}}$,
and $\tau = s\,L_{\mathrm{bow}}F_N$. Therefore, vertical equilibrium alone fixes
the contact force at the applied force, and it yields no dependence on $s$
whatsoever.

Measured at the hair termination, the dependence runs the other way.
{\it Demoucron, Askenfelt, and Causs\'e}~\cite{r-m:demoucron-2009} hung known
loads at successive points along the hair and recorded the transverse force at
the frog. A rigid stick predicts a fall as $(1-s)$, to zero at the tip. The
measurement rises instead, and it is larger with the load at the tip than at the
frog, since the bending of the stick lowers the hair termination at the tip faster
than the hair angle at the frog decreases. The gradient changes its sign once the
hair tension exceeds the product of the tip stiffness and the hair length, which
for the bow they measured is $91$ N/m times $0.53$ m, or $48$ N, against playing
tensions near $60$ N. On that reading, $\alpha$ falls from $1.00$ to $0.83$,
monotonically but more shallowly than in Equation~\ref{eq:alpha}, while its
rigid-stick counterpart rises without bound.

Compliance gives a third answer. If the hair alone is treated as a taut ribbon of
tension $T_h$, loaded at the fraction $s$, its transverse stiffness is
$T_h/(L_{\mathrm{bow}}\,s(1-s))$, which is symmetric about mid-bow. The hair is one
of two terms. Adding the stick as a spring of stiffness $K_s$ at the tip gives a
deflection per unit force of $s(1-s)L_{\mathrm{bow}}/T_h + s^{2}/K_s$, whose second
term dominates toward the tip~\cite{r-m:ablitzer-2012}. Therefore, the compliance
of an assembled bow rises monotonically from the frog to the tip, rather than
peaking at mid-bow, and for $L_{\mathrm{bow}} = 0.65$ m, $T_h = 45$ N, and
$K_s = 72$ N/m, the stiffness falls by a factor of $4.8$ over $s\in[0.2,1]$.

The fourth is the largest force which a player can apply. A grip is a
finger-and-thumb couple acting over a few centimeters, so the moment which it
supplies is bounded. Taking moments about the frog, we have
$s L_{\mathrm{bow}} F_N = \tau + F_W d$ for a bow of weight $F_W$ with its center of gravity at $d$, with the bow held horizontal so that its inclination $\theta$ in Figure~\ref{fig:fbd} is zero, and the greatest contact force which may be reached is,
\begin{equation}
F_N^{\max}(s) = \frac{\tau_{\max} + F_W d}{s\,L_{\mathrm{bow}}}
\label{eq:leverage}
\end{equation}
This falls as $1/s$, by a factor of $5$ over $s\in[0.2,1]$, against $1.8$ for
Equation~\ref{eq:alpha}. {\it Guettler}~\cite{r-m:guettler-2006} gives the same
balance for a $60$ g bow with its center of gravity $19$ cm ahead of the thumb,
where holding $0.5$ N from frog to tip takes a torque running from $-0.11$ to
$+0.21$ N$\cdot$m, and Equation~\ref{eq:leverage} reproduces both endpoints to
within $2\%$.

Which of the four the stroke follows is not settled by measurement.
{\it Askenfelt}~\cite{r-m:askenfelt-1986,r-m:askenfelt-1989} recorded the bow force
and the bow position simultaneously under normal playing conditions, but the frog
and tip gauges were wired into one bridge whose branch sensitivities were set to
give a constant signal for a given bow force at every position along the
hair~\cite{r-m:demoucron-2009}. That record is position-independent by
construction, and it cannot supply $\alpha(s)$.

\begin{figure}[htbp]
\centerline{\includegraphics[width=\figwidth]{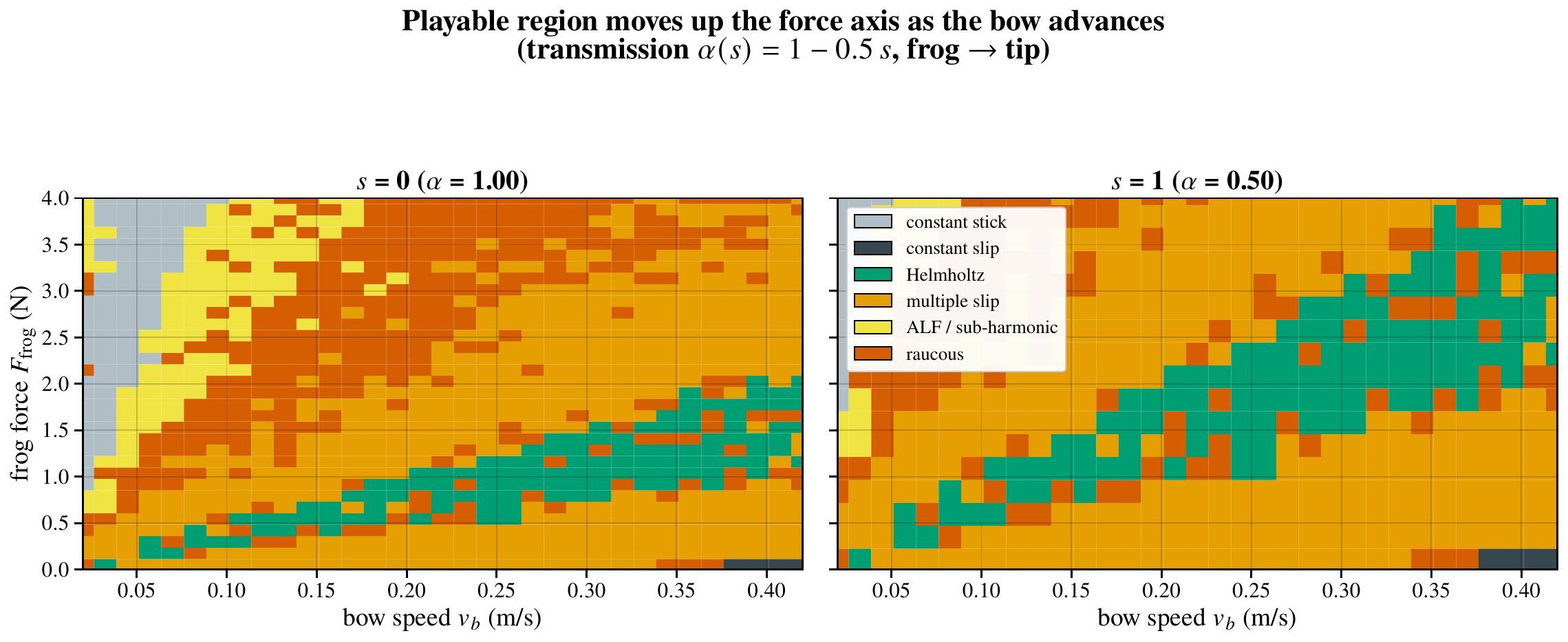}}
\caption{The playable region in frog-force coordinates at the start and at the end of the
stroke. Since the physics depends only on the contact force, the map at any $s$
follows from the contact-force map by rescaling the force axis by $\alpha(s)$.}
\label{fig:stroke}
\end{figure}

Equation~\ref{eq:alpha} is an assumed command map, a known gain schedule on the
action, and not a property of the bow read from its statics. The four readings
disagree among themselves in shape, magnitude, and sign by more than any of them
disagrees with the linear form. In Section~\ref{sec:alpha}, we measure the cost of
getting the profile wrong, which is the appropriate test for an assumption of this
kind. Since $\alpha$ is known and $s$ is an input, the schedule reaches the
controller through its labels, which read the map with its force axis rescaled by
$\alpha(s)$, and a controller which commands $F_{\mathrm{frog}}$ must raise its
command as the stroke proceeds to hold the contact force constant. The playable
region moves up the frog-force axis accordingly (Figure~\ref{fig:stroke}).

\section{Controller}

\subsection{Formulation}

The control objective is to hold Helmholtz motion across a complete bow stroke,
while commanding only the quantities which a robot can command, namely, the
normal force at the frog and the bow velocity. The contact point, $x_B$, is held
fixed.

At control step $m$, the string is advanced for $\Delta t = 35$ ms under the
current commands, the last $60\%$ of that window is analyzed, and the controller
proposes increments. The string state is carried between steps, so that the
oscillation continues across the stroke, instead of restarting from rest, and the simulated interval matches the interval over which the bow advances. The analysis window is fixed in seconds, and the last $60$ percent of a step spans about $4$ periods on G and $14$ on E.

\begin{figure}[htbp]
\centerline{\includegraphics[width=\figwidth]{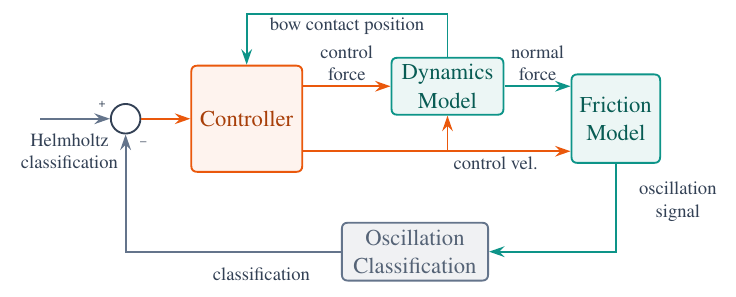}}
\caption{Closed-loop control architecture. The classifier and the stick-fraction
diagnostic are computed from the relative-velocity signal over each control window, and they are returned to the controller. The reference at the left names the goal of the loop. The controller is trained by imitation (Section~\ref{sec:training}), and not on the error which the junction forms.}
\label{fig:block}
\end{figure}

Figure~\ref{fig:block} shows the loop.

The feature vector is given by,
\begin{equation}
\mathbf{z}_m =
\begin{bmatrix}
s_m & r_m & \eta_m & \Delta f_m & \Delta A_m & F^{m-1}_{\mathrm{frog}} & v^{m-1}_b
\end{bmatrix}^{\!\top}
\label{eq:features}
\end{equation}
where $s_m$ is the stroke position, $r_m$ is the regime label, $\eta_m$ is the
stick fraction, $\Delta f_m = f_0^\star - f_{\mathrm{slip},m}$ is the frequency
error, and $\Delta A_m = \log_{10}(A^\star/A_m)$ is the amplitude error. The output
is $\mathbf{y}_m = [\Delta F_{\mathrm{frog}}, \Delta v_b]^\top$, which is applied as
increments, rate-limited to $0.05$ N and $0.012$ m/s per step.

The stick fraction is a continuous measure of how nearly Helmholtz the current
motion is, and it carries information which a categorical regime label does not.
The frequency error is derived from the timing of slips, so it varies
continuously. On the other hand, a spectral estimate over a short window is
quantized to the bin width, and it may be identically zero for an entire stroke,
in which case that input is inert.

\subsection{Training}
\label{sec:training}

The labels come from the playability map. At each state, the target increment
points toward the nearest $(F_{\mathrm{frog}}, v_b)$ pair classified as Helmholtz
at the current stroke position, clipped to the rate limits. Three kinds of
trajectory are generated, namely, {\it tracking} trajectories, which start inside
the playable region, {\it boundary-crossing} trajectories, in which a deliberate
excursion is forced at mid-stroke, and {\it recovery} trajectories, which start
from a random point anywhere in the command space. Five hundred trajectories of thirty steps give $15\,000$ labeled samples.

The label is therefore a function of the stroke position and the previous commands
alone, $(s_m, F^{m-1}_{\mathrm{frog}}, v^{m-1}_b)$, and not of the oscillation. All
three are inputs, and $s$ itself follows from the command history. A network which
reproduced its labels exactly would ignore the four string-state inputs, $r$,
$\eta$, $\Delta f$, and $\Delta A$, and it would need no memory, so the objective
requires neither. Any use which a trained network makes of them comes from finite
capacity, finite data, and their correlation with the commands along the teacher's
trajectories. It is not a feedback law which the networks were trained to learn. It
follows that the feature ablation of Section~\ref{sec:ablation} and the freezing of
Sections~\ref{sec:baselines} and~\ref{sec:contactdist} measure how a network uses
inputs on which its labels do not depend. The architecture comparison of
Section~\ref{sec:architecture} compares the inductive biases of networks fitted to
the same open-loop labels.

Two more expressive families were considered before we settled on a plain
feedforward network. A {\it Kolmogorov-Arnold network}~\cite{r-m:liu-2024} replaces
linear weights with learnable univariate splines, and it is attractive on
low-dimensional problems, both for its sample efficiency and since the learned
functions may be inspected. A recurrent policy, such as an
{\it LSTM}~\cite{r-m:hochreiter-1997}, would let the controller carry state between
control steps. Neither was adopted. Both are ways of fitting the training labels
more closely, and in Section~\ref{sec:capacity}, we show that the closeness of fit
is not what limits performance here, since raising the parameter budget lowers the
training loss in every cell, while the closed-loop quality does not follow. That reasoning holds for the spline network. Recurrence is tested directly in Section~\ref{sec:architecture}, which retrains six architectures at a matched parameter count. There, the gated recurrent unit has the highest mean on every string, but at twenty seeds it cannot be distinguished from the feedforward network. The feedforward network of this section is retained as the baseline against which that comparison is made, and the spline network is revisited in Section~\ref{sec:futurearch}.

\begin{figure}[htbp]
\centerline{\includegraphics[width=\figwidth]{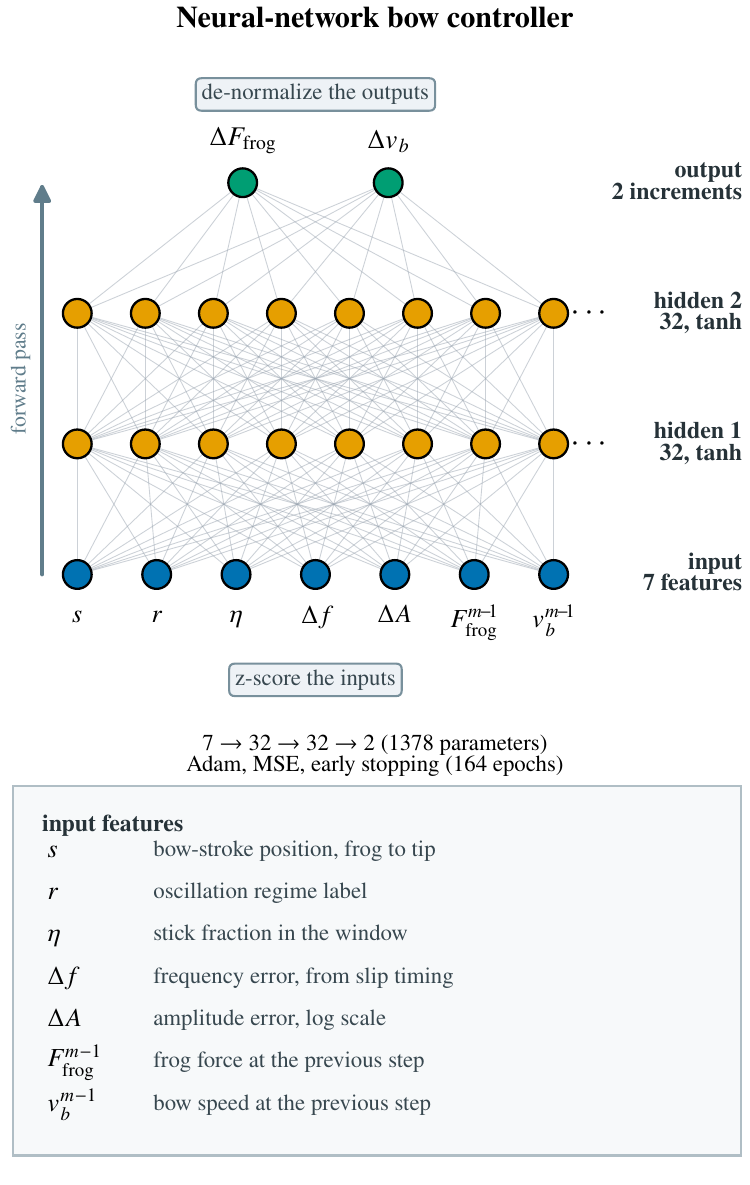}}
\caption{Architecture of the feedforward bow controller. The input
features are those of Equation~\ref{eq:features}.}
\label{fig:nn}
\end{figure}

The controller is therefore a feedforward network with two hidden layers of $32$
units and $\tanh$ activation, linear outputs, and $1378$ parameters
(Figure~\ref{fig:nn}). The inputs and outputs are z-scored. Training uses Adam~\cite{r-m:kingma-2017} on a mean-squared-error objective, with early stopping on a held-out $15\%$ split,
which halted after $164$ epochs.

\section{Results}

\subsection{Sustained Helmholtz Motion}

\begin{figure}[htbp]
\centerline{\includegraphics[width=\figwidth]{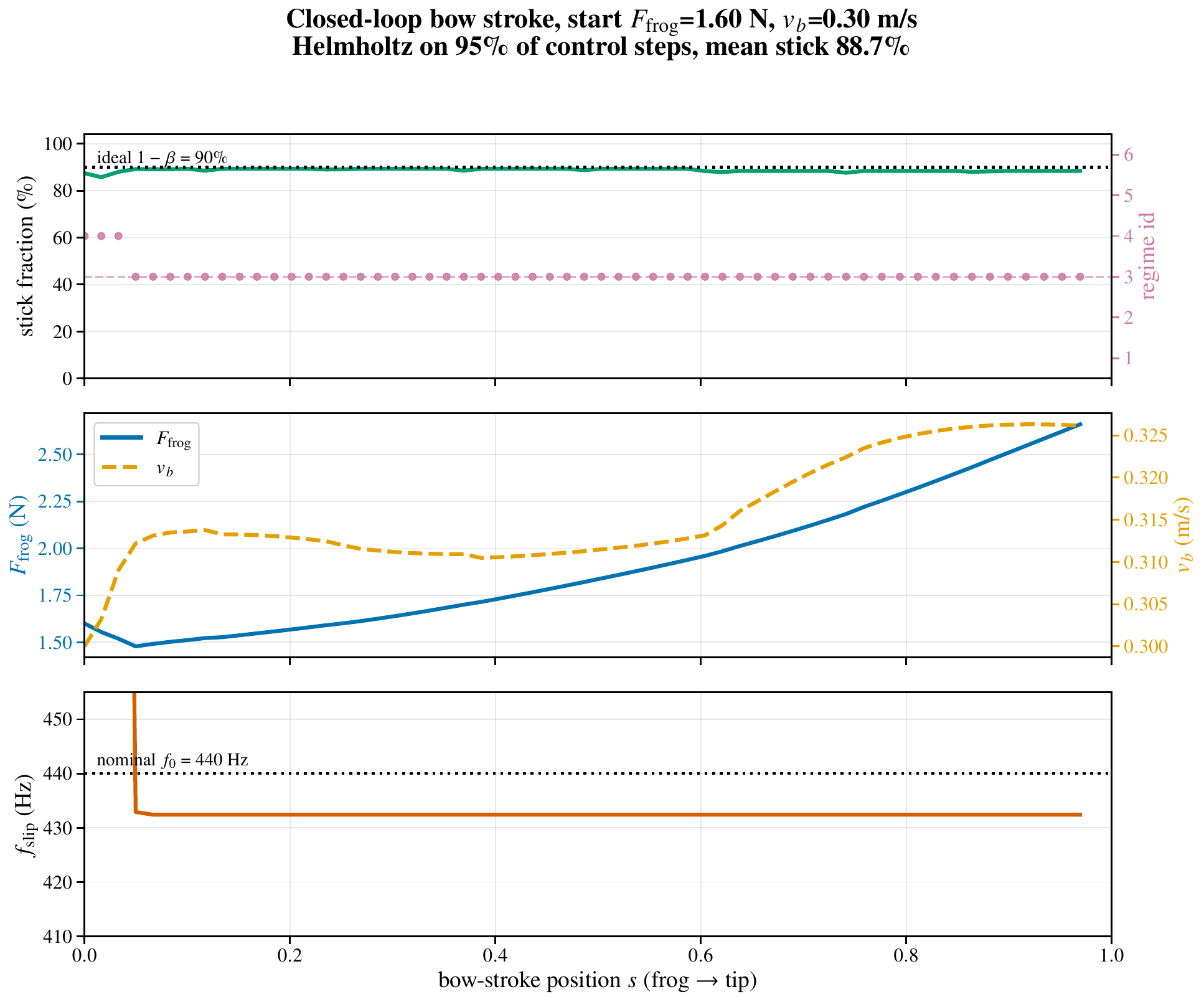}}
\caption{Closed-loop bow stroke of the feedforward baseline from the in-band start on the A string. Top: stick fraction and regime label.
Middle: commanded frog force and bow speed. Bottom: slip frequency.}
\label{fig:closed}
\end{figure}

Started inside the playable region, at $F_{\mathrm{frog}} = 1.60$ N and
$v_b = 0.30$ m/s, the feedforward baseline of Section~\ref{sec:training} holds
Helmholtz motion on $94.8\%$ of the $58$ control steps of a full stroke, with a
mean stick fraction of $88.7\%$ against the ideal $90\%$ (Figure~\ref{fig:closed}).
The commanded frog force rises from $1.6$ N to $2.7$ N over the stroke, which is
most of what holding the contact force constant would require as $\alpha(s)$ falls
from $1$ to $0.5$. Holding the contact force at $1.6$ N to the end of the stroke would take $3.1$ N at the frog, so the contact force falls by about a seventh over the stroke, instead of staying level. That rise is the schedule which the labels encode through $\alpha(s)$. The slip frequency stays between $432.4$ and $432.9$ Hz over the fifty-five Helmholtz steps, about seven hertz flat of $440$ Hz. At the grid used, the slip period locks to a whole number of samples, so the size of that offset reflects the lock (Section~\ref{sec:params}). It leaves that band only on the three
opening steps, which are exactly the three which the regime measure counts
against it.

\begin{figure}[htbp]
\centerline{\includegraphics[width=\figwidth]{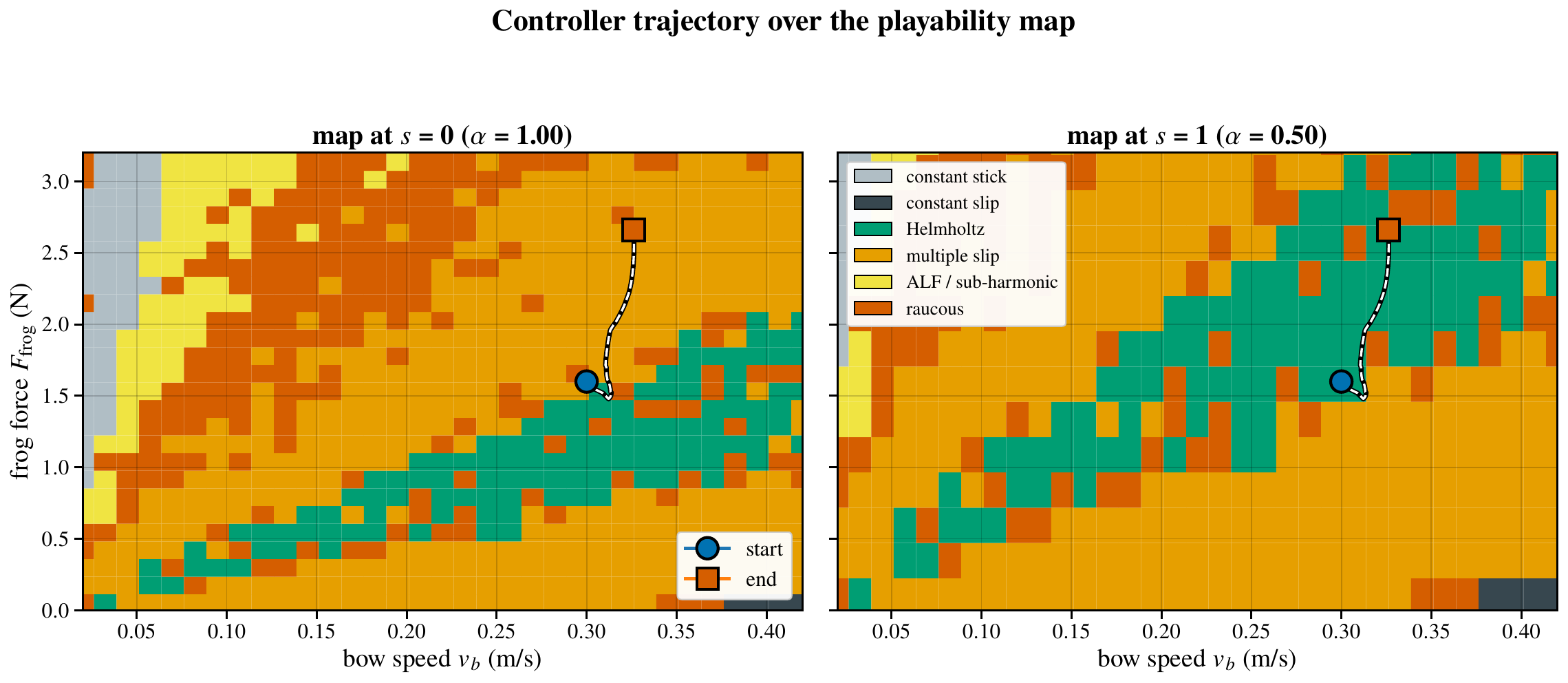}}
\caption{Command trajectory over the playability map at the start and at the end of the
stroke.}
\label{fig:traj}
\end{figure}

Figure~\ref{fig:traj} overlays the command trajectory on the playability map at
the two ends of the stroke. The path begins inside the playable band and ends
inside it, and the band itself has moved upward between the two panels.

\subsection{Disturbance Rejection}

\begin{figure}[htbp]
\centerline{\includegraphics[width=\figwidth]{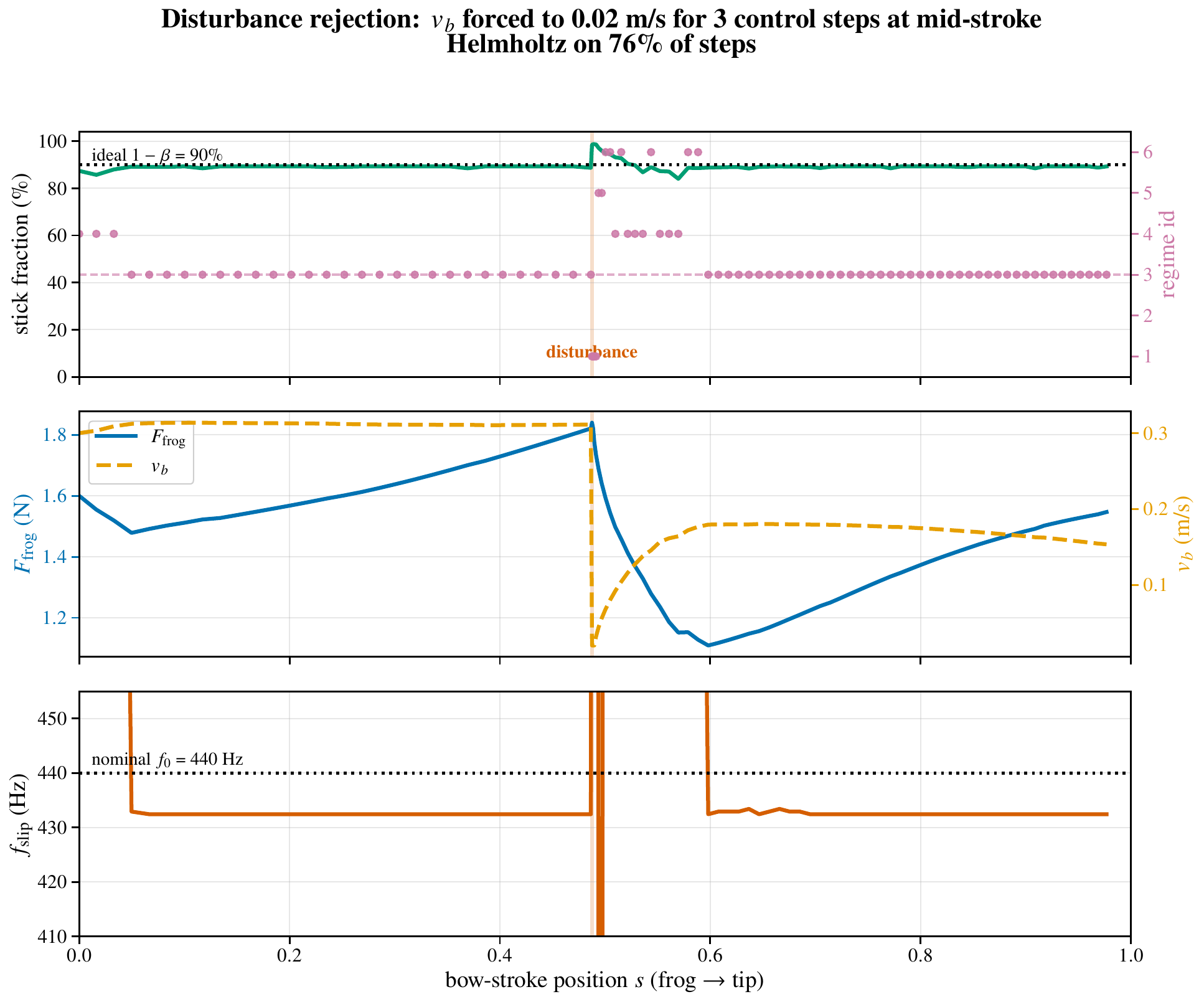}}
\caption{Disturbance rejection. The shaded band marks the three control
steps during which $v_b$ was overridden.}
\label{fig:dist}
\end{figure}

The bow velocity was overridden to $0.02$ m/s for three control steps at
mid-stroke. The regime leaves Helmholtz motion for the duration of the override,
and it is recovered eighteen control steps after the release, after which the
stroke completes normally. Helmholtz motion is held on $75.8\%$ of the steps
overall (Figure~\ref{fig:dist}). Recovery is dominated by the rate limit on
$\Delta v_b$, and not by the string dynamics.

\subsection{Dependence on Initial Conditions}

\begin{figure}[htbp]
\centerline{\includegraphics[width=\figwidth]{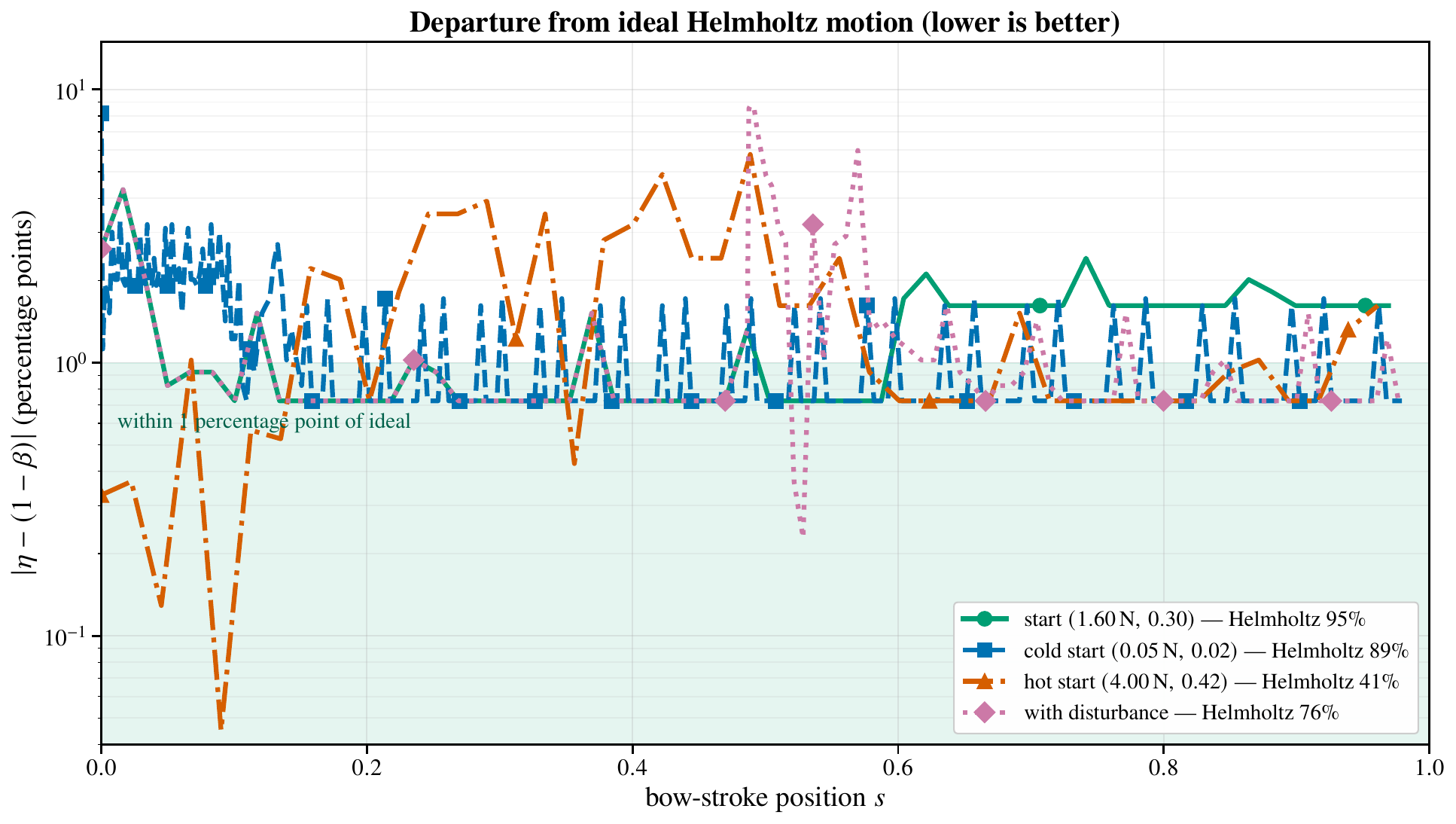}}
\caption{Departure from ideal Helmholtz motion across the stroke for four
initial conditions. The stick fraction itself spans only $81$--$100\%$, so
the runs are indistinguishable on a linear axis. Its deviation from the
ideal $1-\beta$ spans about $2.2$ decades, which a logarithmic axis
resolves. Lower is better, and the shaded band marks agreement to within one
percentage point.}
\label{fig:ic}
\end{figure}

Figure~\ref{fig:ic} compares four starts. It plots the deviation of the stick
fraction from the ideal $1-\beta$ on a logarithmic axis, rather than the stick
fraction itself. The runs all sit between $81\%$ and $100\%$, and they are not
separable on a linear scale, whereas their departure from the ideal covers rather
more than two decades. The run started inside the playable region holds a
deviation of about $0.7$ percentage points for almost the whole stroke, the cold
start settles near $2$, and the hot start remains erratic between $0.07$ and $7$.
From a cold start at the lower command limits ($0.05$ N, $0.02$ m/s), the
controller reaches the playable region and holds Helmholtz motion on $89.5\%$ of
the steps, although the stroke takes $238$ steps, since the bow advances slowly
while $v_b$ is small. From a hot start at the upper limits ($4.00$ N,
$0.42$ m/s), it recovers only partially, and the stroke completes in $44$ steps.
With $|\Delta F_{\mathrm{frog}}| \le 0.05$ N per step, the controller cannot bring
the force down fast enough to spend the early part of the stroke inside the
playable band. Helmholtz motion is held on $40.9\%$ of the steps. Both halves of
that explanation were tested, and they do not behave the same way
(Table~\ref{tab:ratelimit}).

The stroke itself cannot be lengthened, since a violin bow is a fixed object.
What may be changed is the control interval, and shortening it buys more
decisions over the same stroke. Halving $\Delta t$ takes the network from $40.9\%$
to $59.8\%$, and quartering it takes it to $78.9\%$, while it takes the map lookup
from $47.7\%$ to $74.7\%$ to $82.2\%$. Both gain, by $38.0$ points and $34.5$, and
the lookup leads at every interval. Therefore, shortening the interval does not
let the network overtake it. That gain is not an artifact of the shorter analysis
window. On the in-band start, which is already near ideal and has no room to
improve, the same change moves the network from $94.8\%$ to $95.7\%$ and the map
lookup from $95.0\%$ to $98.3\%$, a few points either way against the thirty-eight
which the high-force corner gains. The cost is in measurement. An $8.75$ ms window
holds fewer than four nominal periods, so every regime call rests on
correspondingly less evidence.

Raising the permitted increment separates them. The map lookup recomputes its
target from the map at every step, and it climbs from $47.7\%$ to $88.6\%$ as the
limit is relaxed eightfold. The network gains $4.6$ points and then saturates at
$45.5\%$. The labels on which it was trained are themselves clipped to the
nominal increment, so it never learned to ask for a larger step, and relaxing the
limiter offers it room which it does not use. Inside the band, neither change
costs the lookup anything, since it holds $95.0\%$ at every increment, and
$95.0\%$, $95.0\%$, and $98.3\%$ across the three intervals. It does cost the gated recurrent unit of Section~\ref{sec:architecture}, however, which falls from
$94.4\%$ to $80.7\%$ to $74.3\%$ on A, and from $94.5\%$ to $65.7\%$ to $63.6\%$ on
E, as the interval shortens. Therefore, shortening the interval trades in-band
regulation for corner recovery, and it does so only for a controller which
carries state.

Repeating the increment ladder with the gated recurrent unit makes the saturation
stronger, rather than weaker. At twice, four times, and eight times the nominal
cap, it scores identically to every digit on all four strings, since the
increments which it asks for already lie inside twice the cap, so relaxing it
further is literally a no-op. Over the same range, the lookup gains $25.0$ points
on A and $27.3$ on E. The feedforward baseline of Section~\ref{sec:training}, a single sklearn fit, at least moved by $4.5$ points at the first rung. This one moves by $-3.0$, $+3.9$, $-3.8$, and $+0.4$ on the four strings, as means over three seeds, and none of the four can be distinguished from zero, at $p = 0.24$ to $0.92$ on standard errors of $2.3$ to $6.2$ points.

Two features of the ladder do not carry across strings. Shortening the interval
helps monotonically only on E, while G rises and then collapses, and D and A both
rise at the middle rung and fall at the shortest. On D, that fall may be partly
due to the diagnostic, as G's collapse is, since the shortest interval gives the
classifier $1.5$ periods there, and the lookup falls with it. Furthermore, the
periods column of Table~\ref{tab:ratelimit} is specific to A. Namely, the same
intervals give $6.9$, $3.4$, and $1.7$ periods per control interval on G, against
$15.4$, $7.7$, and $3.9$ on A, which is the column as printed. At $8.75$ ms on G,
both controllers read $0.0\%$, the lookup included, which is the regime
diagnostic failing at one nominal period per analysis window, rather than either
controller failing. Therefore, the shortest interval in the ladder does not exist
as a measurement on the lowest string.

Therefore, the number of decisions bounds what any controller can do from this
corner, while the size of each one bounds only a controller which recomputes its
target. For the network, the binding constraint is the rule which generated its
labels, which is the ceiling that we reach from a different direction in
Section~\ref{sec:capacity}.

\begin{table}[htbp]
\caption{Percentage of control steps in Helmholtz motion from the high-force corner, as the
permitted increment and the control interval are varied, with the nominal setting
first. Bold marks the best value in each column. The periods column is the number
of nominal periods in each control interval. The classifier reads the last $60$
percent of that interval, so the evidence behind one regime call is $0.6$ of the
figure shown, and the shortest interval gives it $2.3$ periods, rather than
$3.9$.}
\label{tab:ratelimit}
\centering
\begin{tabular}{|l|c|c|c|c|}
\hline
& \textbf{Steps} & \textbf{Periods} & \textbf{Network} & \textbf{Map} \\
\hline
Nominal, $\Delta t = 35$ ms & $44$ & $15.4$ & $40.9\%$ & $47.7\%$ \\
\hline
\multicolumn{5}{|l|}{{\it Permitted increment}} \\
\hline
$\times 2$ & $44$ & $15.4$ & $45.5\%$ & $63.6\%$ \\
\hline
$\times 4$ & $44$ & $15.4$ & $45.5\%$ & $79.5\%$ \\
\hline
$\times 8$ & $44$ & $15.4$ & $45.5\%$ & $\mathbf{88.6\%}$ \\
\hline
\multicolumn{5}{|l|}{{\it Control interval}} \\
\hline
$17.5$ ms & $87$ & $7.7$ & $59.8\%$ & $74.7\%$ \\
\hline
$8.75$ ms & $175$ & $3.9$ & $78.9\%$ & $82.2\%$ \\
\hline
\end{tabular}
\end{table}

\subsection{Summary}

Table~\ref{tab:results} collects the four episodes described above in one place,
for the A-string feedforward baseline of Section~\ref{sec:training}. It reports the
number of control steps which each stroke took, the fraction of those steps in
Helmholtz motion, and the mean stick fraction over the stroke. They are single episodes of one fit, and the twenty-seed comparison is in Section~\ref{sec:architecture}.

\begin{table}[htbp]
\caption{Closed-loop performance of the A-string feedforward baseline of
Section~\ref{sec:training}, on the four episodes plotted in
Figures~\ref{fig:closed} to~\ref{fig:ic}. The twenty-seed comparison is in Table~\ref{tab:board}.}
\label{tab:results}
\centering
\begin{tabular}{|l|c|c|c|}
\hline
\textbf{Condition} & \textbf{Steps} & \textbf{Helmholtz} & \textbf{Mean $\eta$} \\
\hline
Start inside playable region & $58$  & $94.8\%$ & $88.7\%$ \\
\hline
Cold start $(0.05\,$N$,\,0.02)$ & $238$ & $89.5\%$ & $88.8\%$ \\
\hline
Hot start $(4.00\,$N$,\,0.42)$  & $44$  & $40.9\%$ & $88.5\%$ \\
\hline
With mid-stroke disturbance & $91$ & $75.8\%$ & $89.7\%$ \\
\hline
\end{tabular}
\end{table}

\subsection{Why No Direct Comparison with Prior Work Is Reported}
\label{sec:nocompare}

The closest prior work is {\it Vivi}, the virtual violinist of {\it Percival,
Bailey, and
Tzanetakis}~\cite{r-m:percival-2011,r-m:percival-2013-1,r-m:percival-2013-2},
which also closes a learned loop around a bowed-string physical model. We report
no head-to-head number against it, for three reasons.

A shared ingredient is not a shared measurement, and the bowing-imitation work
of {\it Jin, et al.}~\cite{r-m:jin-2024} shows the distinction plainly. They adopt a
Stribeck friction law of the same family as the one which we use here, and they
apply it to shape the damping of a single-degree-of-freedom oscillator, whereas
here it drives a distributed string. Their objective is the distance between a
produced trajectory and a recorded human one, measured by Euclidean, Fr\'echet,
and related metrics. Nothing in that arrangement produces or scores Helmholtz
motion, so a comparison with this paper is undefined, rather than unperformed,
since the two report no result in common.

First, Vivi and the controller here do not share a plant. Vivi drives a modal formulation after {\it Demoucron}~\cite{r-m:demoucron-2008}, whereas the model here is a finite-difference scheme with implicitly resolved friction and rigid terminations. A bow force of a given magnitude does not mean the same thing in the two, since the bow-force limits depend on the wave impedance, and it is not matched between the two.

Second, they do not share a command set. Vivi generates the finger position,
bow-bridge distance, bow velocity, and bow force, all specified at the contact
point. The controller here commands the normal force at the frog together with
the bow velocity, lets the contact force follow from the transmission factor of
Equation~\ref{eq:alpha}, and holds the contact point fixed.

Third, they do not share an objective. Vivi is trained and assessed on the timbre
judged from the audio which it produces, and its reported outcome is musical,
namely, that it performs a beginner repertoire at a level compared with that of a
first-year student. The objective here is the oscillation regime, measured
mechanically as the stick fraction of Equation~\ref{eq:stick}. A percentage of
control steps in Helmholtz motion and a judgment of tone quality are not
commensurable, and placing them in adjacent columns would imply a rigor which
the comparison does not have.

A comparison remains possible in principle, by the following route. Vivi is
publicly available, and scoring its output with the diagnostics of
Section~\ref{sec:diagnostics} would put both systems on one scale. Reviving the
published implementation is a project in itself, since it rests on a toolchain
from the era of Qt4 and Python~2, with a separate audio back end. The harder half
is the plant. A fair test needs the two string models to be matched closely
enough that the result measures the controllers, rather than the models which
they drive, and that matching is the real work.

The other learning-based work cited, that of {\it Luan and
Scavone}~\cite{r-m:luan-2025}, models the friction nonlinearity, rather than
controlling it, so it offers no controller to compare against.

\subsection{Comparison with Alternative Controllers}
\label{sec:baselines}

What may be compared is what the learned policy buys over simpler controllers
which drive the same plant from the same initial conditions.
Table~\ref{tab:baselines} reports three such controllers. The first holds the
initial commands for the whole stroke. The second is the labeling rule itself,
which steps toward the nearest playable pair using only $s$, $F_{\mathrm{frog}}$,
and $v_b$. It is open loop with respect to the oscillation, and it reads an exhaustive tabulation of the plant at $1089$ operating points. Both halves are important. In steady-state regulation, wherever that tabulation is correct, the entry which the lookup reads already gives Helmholtz motion, and the learned controller is trained to reproduce those entries, so the lookup is a strong baseline, rather than a handicapped one. The ceiling belongs to the tabulation, and
not to the plant, and in Section~\ref{sec:reach}, we measure where the two part
company. The third is the trained network with its string-state inputs, $r$,
$\eta$, $\Delta f$, and $\Delta A$, frozen at nominal values. On the other three strings, the in-band start is ($0.64$ N, $0.08$ m/s) on G, ($0.64$ N, $0.10$ m/s) on D, and ($1.43$ N, $0.34$ m/s) on E. The disturbance start is each string's own in-band start with the same mid-stroke override. The hot start is the same on every string, and so is the cold start, except on G, whose force floor puts it at $0.40$ N.

\begin{table}[htbp]
\caption{Percentage of control steps in Helmholtz motion on each of the four open strings, one episode per cell. A dagger marks a run which failed to reach the tip within the $400$-step limit, and
a regime score on such a run is not meaningful. A double dagger marks inputs frozen
at inference, rather than removed, which the ablation of Table~\ref{tab:ablation}
shows to be a different question. The two learned rows are one fit of the gated recurrent unit, from the grid family of Section~\ref{sec:architecture}, and the other architectures appear in Table~\ref{tab:board}.}
\label{tab:baselines}
\centering
\footnotesize
\setlength{\tabcolsep}{4pt}
\begin{tabular}{|l|l|r|r|r|r|}
\hline
\textbf{Controller} & \textbf{Start} & \textbf{G} & \textbf{D} & \textbf{A} & \textbf{E} \\
\hline
No control,      & band & $94.3$ & $91.2$ & $80.3$ & $87.0$ \\
commands         & cold & $0.2^\dagger$ & $32.8^\dagger$ & $15.0^\dagger$ & $86.2^\dagger$ \\
held             & dist & $27.3^\dagger$ & $19.5^\dagger$ & $5.5^\dagger$ & $5.2^\dagger$ \\
\hline
Map lookup,      & band & $\mathbf{98.2}$ & $96.1$ & $95.0$ & $\mathbf{98.1}$ \\
open loop        & cold & $\mathbf{94.8}$ & $99.5^\dagger$ & $99.8^\dagger$ & $99.5^\dagger$ \\
                 & dist & $\mathbf{94.3}$ & $\mathbf{92.4}$ & $\mathbf{77.5}$ & $79.3$ \\
\hline
Network,         & band & $90.6$ & $95.2$ & $81.0$ & $66.1$ \\
inputs           & cold & $93.2$ & $94.8^\dagger$ & $55.2$ & $52.0$ \\
frozen$^\ddagger$ & dist & $88.0$ & $79.0$ & $29.1$ & $23.7$ \\
\hline
Network,         & band & $81.9$ & $\mathbf{97.2}$ & $\mathbf{96.7}$ & $96.4$ \\
full             & cold & $86.7$ & $86.9$ & $42.8^\dagger$ & $\mathbf{95.1}$ \\
inputs           & dist & $71.2$ & $78.2$ & $70.2$ & $\mathbf{79.7}$ \\
\hline
\end{tabular}
\end{table}

Holding the commands is not sufficient even from a good start, since $\alpha(s)$
falls as the bow advances, and the contact force falls with it. It fails outright
elsewhere, stalling from the cold start and under disturbance on every string,
which accounts for eight of its twelve runs.

Stroke completion has to be read alongside the regime score, since the two may
point in opposite directions. The map lookup reaches $99.8\%$ from the cold start
on A while traveling only to $s = 0.70$ before the step limit, against the $0.98$
which ends a stroke. In other words, holding a low bow speed keeps the string in
Helmholtz motion while the bow barely moves. Scoring regime maintenance without
also requiring the stroke to finish rewards exactly that. Counted over the twelve
combinations of string and start, the lookup completes nine, the learned
controller eleven, and its frozen variant eleven, against four for the held
commands.

Where the rule is right, the learned controller does not outperform it on
steady-state regulation, and this holds well beyond the four open strings. The
comparison was extended to the forty positions of Section~\ref{sec:fingering},
four of them open, at twenty seeds, with the starts matched so that both sides
average over the same ones and no cell is excluded. There, the lookup leads in all $160$ cells. By a one-sample $t$-test on the seeds, its lead is significant at $p < 0.05$ in $158$, and in $156$ after Holm correction~\cite{r-m:holm-1979} over the $160$. The four which do not survive are the gated recurrent unit at four positions on the A string, where the lead is $0.9$ to $3.8$ points. That comparison is not the
one which it looks like. The lookup reads the playability map for the actual
fingered pair, so it is handed the answer for a plant to which it was never asked
to transfer, while the network is transferring. Read as an oracle bound, rather
than as a baseline, it says that the map dominates wherever the controller has
not seen the position. Referring the amplitude target to the training pitch narrows the gap without closing it. On the gated unit, the lookup's lead stays significant at $p < 0.05$ in $38$ of the $40$ cells, the same count as without the correction, and in $36$ after Holm correction over the $160$.

On the open strings, against the gated recurrent unit, the lookup leads in Helmholtz fraction from the high-force corner and under disturbance on all four strings, by $4.5$ to $11.1$ and $4.2$ to $12.9$ points, and each lead survives Holm correction over the starts on its string. Inside the band, it leads significantly only on E, by $2.2 \pm 0.4$ points, and on G, D, and A the difference cannot be distinguished from zero. From the cold start, the lookup reaches the tip only on G, where it leads by $5.8 \pm 1.7$ points, at $p = 0.0033$, its only cold-start win earned by regulating, rather than by never arriving.

The lookup's advantage on the cold start is of a different kind, and it should
not be read from the percentage. It reaches $99.5$, $99.8$, and $99.5$ on D, A,
and E, and on all three, the bow never reaches the tip, which is what the dagger
records. It fails to finish the cold start on three of the four open strings and
on $29$ of the $37$ notes of the earlier single-episode design. On that design,
the learned controller fails on $14$ once its amplitude target is referred to the
sounding pitch, and on $15$ without that correction. On A, the bow reaches only
$s = 0.70$, and on D and E, it stops just short, at $0.97$. Holding a low speed
keeps the string in Helmholtz motion while the bow fails to arrive, which is the
same degenerate behavior which Section~\ref{sec:rl} finds a reward search
exploiting. The advantages of the learned controller are of a different kind. It
needs no map stored or searched at run time, and it produces rate-limited
commands, instead of nearest-neighbor jumps. It completes two more of the twelve
runs, and it extends to observations by which no map is indexed.

Freezing the four string-state inputs is not a measure of what those inputs
contribute, and the four-string result shows why, more clearly than one string
could. Freezing improves the disturbance response on G and D, by $16.8$ and $0.8$
points, and it destroys it on A and E, by $41.1$ and $55.9$. That is the same
split by string which the retrained ablation of Section~\ref{sec:ablation} finds,
from a different experiment. Namely, removing the string state helps on G and
hurts on A and E. Therefore, on this protocol, the disturbance start of
Table~\ref{tab:baselines}, two methods which disagree about the size of the effect
agree about which strings it helps.

That agreement does not survive a change of protocol, which is worth stating,
rather than leaving it for a reader to find. Under the contact disturbance of
Section~\ref{sec:contactdist}, measured at twenty seeds rather than on one
episode, freezing the same inputs {\it hurts} on G, by $13.5$ points at
$p = 0.0001$, which is the largest effect of the four strings and the opposite sign
to the $16.8$ above. Neither measurement is wrong. They use different
disturbances, different protocols, and different families of fit, and they have
not been pooled. The choice of measurement here decides which way an effect
points, rather than whether it exists, which is the third instance in this paper
of the caution in Section~\ref{sec:contactdist}. What freezing measures on its own
is the sensitivity to out-of-distribution input, since a trained network, given
constant values for inputs which varied during training, is being shown
combinations which it never saw.

Taken with the attack result of Section~\ref{sec:acoustic}, the two places in
which the learned controller does better than the rule form a pattern, rather
than a pair of exceptions. It reaches the acoustic threshold from the high-force corner on five of five seeds on G, D, and A, where the lookup never reaches it. In addition, when
the friction characteristic is cut hard enough to move the playable band, it holds
the stroke together where the lookup does not. On the single episode of
Table~\ref{tab:contactdist}, which uses the feedforward baseline of
Section~\ref{sec:training}, it holds $80.6\%$ of the steps after the onset, against
the lookup's $25.7\%$. Retrained under twenty seeds, the gated recurrent unit leads by $20.1 \pm 20.9$ points on the same window, positive on seventeen of
twenty seeds at $p = 0.0004$, and the feedforward network leads by
$44.9 \pm 10.3$ on twenty of twenty. The two figures are not the same quantity,
and the comparison which makes sense is the like-for-like one, since the episode
is itself a feedforward sklearn fit. The seeded feedforward mean is
$44.9 \pm 10.3$, against the episode's $54.9$, so the two agree to within about a
fifth. The gated recurrent unit's $20.1$ is lower than either, which is a fact
about the architecture, rather than about the measurement.

The advantage is not monotone in severity, which is the part that a reader would
not guess, and it is not monotone only for the architectures which carry state.
At the two mildest reductions, after the onset, the recurrent architectures cannot be distinguished from the lookup, while the feedforward one trails it. At $0.55$, the lookup is ahead on all three, by $7.0$ points for the gated recurrent unit and $41.0$ for the feedforward network.
Only at $0.40$ does the network lead, and there, it leads on every architecture
and on both windows.

The mechanism is visible in what the lookup does, rather than in what it scores.
Its commands do not read the string, so its trajectory is identical to six digits
at every friction level, $60$ steps ending at $s = 0.977$ in all four cases. Only
the regime classification moves underneath it. Both conditions are ones which the
map cannot describe, namely, a transient, for which a steady-state tabulation has
no entry, and the plant moving away from the tabulation. The learned controller does better than the rule where the plant leaves the model, and in one more case. In
Section~\ref{sec:reach}, we measure a plant which is stationary and settled, and a
tabulation which is wrong about it.

In Section~\ref{sec:contactdist}, we take the second of those conditions, the
plant moving away from the tabulation, at twenty seeds on four strings. What this
section has shown is narrower, namely, the single episode of
Table~\ref{tab:contactdist}, on one string, which carries the same caution as the
rest of that table.

\subsection{Architecture Comparison}
\label{sec:architecture}

\begin{figure}[htbp]
\centerline{\includegraphics[width=\figwidth]{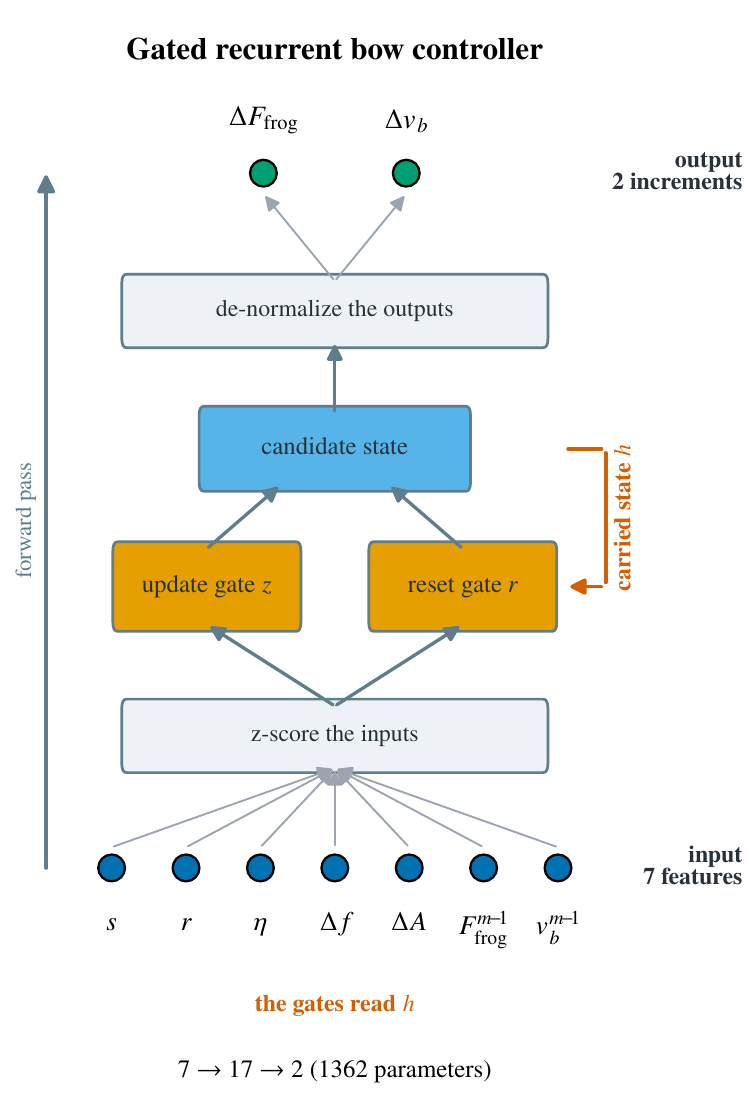}}
\caption{Architecture of the gated recurrent bow controller, drawn to the same scale as
Figure~\ref{fig:nn}. Both gates are computed from the input and from the carried
state, so the state may influence how much of itself survives the next step. The
input features are those of Equation~\ref{eq:features}.}
\label{fig:archgru}
\end{figure}

\begin{figure}[htbp]
\centerline{\includegraphics[width=\figwidth]{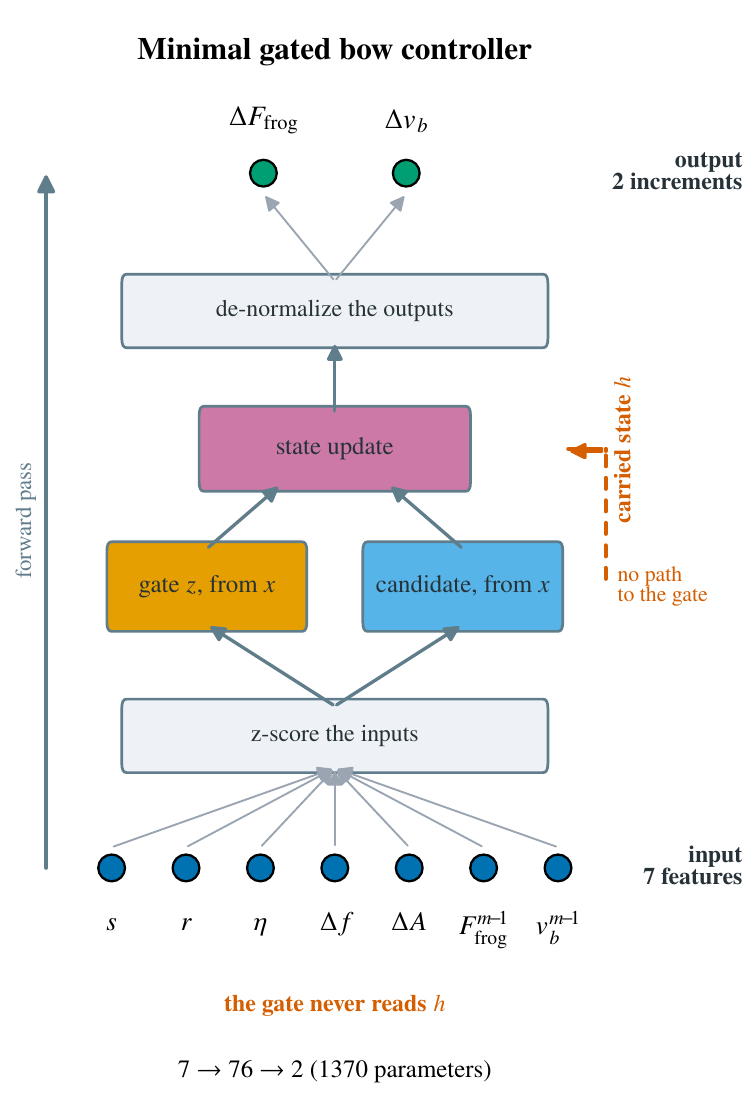}}
\caption{Architecture of the minimal gated variant. The gate is computed from the input
alone, so no path carries the state into it, and the state cannot influence how
much of itself survives. That is the omission behind the failure above. Removing
the state from the gate also makes each unit less expensive, so at a matched
parameter budget, this cell is $76$ units wide, against the gated recurrent unit's
$17$.}
\label{fig:archmingru}
\end{figure}

\begin{figure}[htbp]
\centerline{\includegraphics[width=\figwidth]{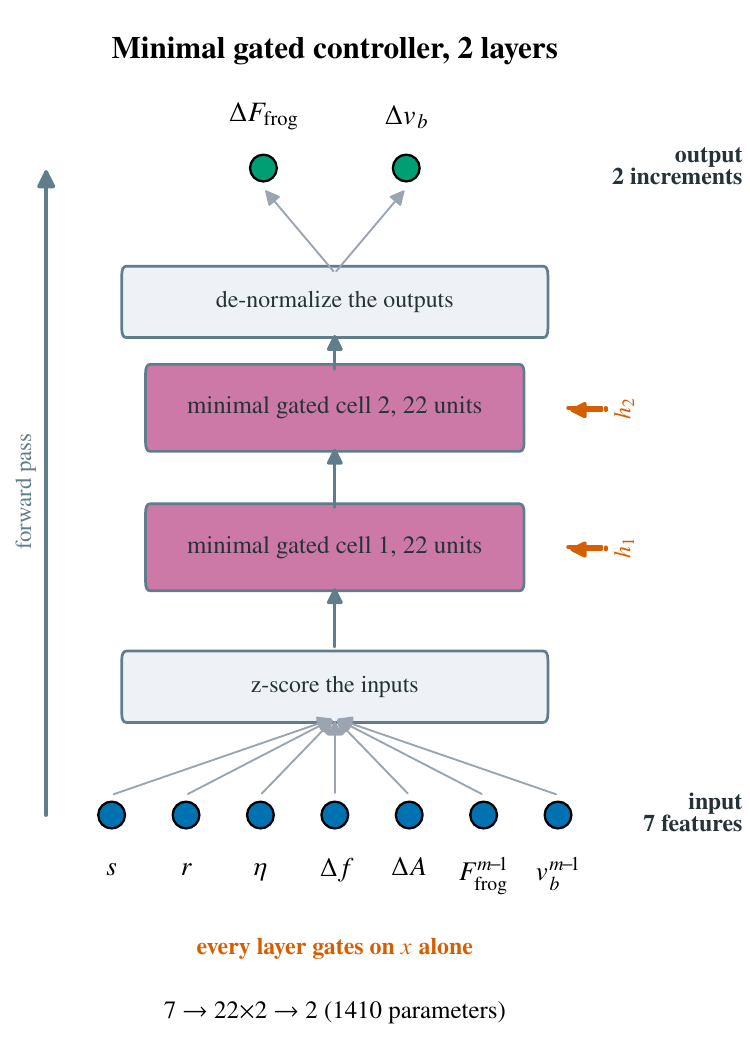}}
\caption{The minimal gated cell stacked to two layers, each of which carries its own state.
The parameter budget is spent on depth, rather than on width, at $22$ units per
layer.}
\label{fig:archdeep2}
\end{figure}

\begin{figure}[htbp]
\centerline{\includegraphics[width=\figwidth]{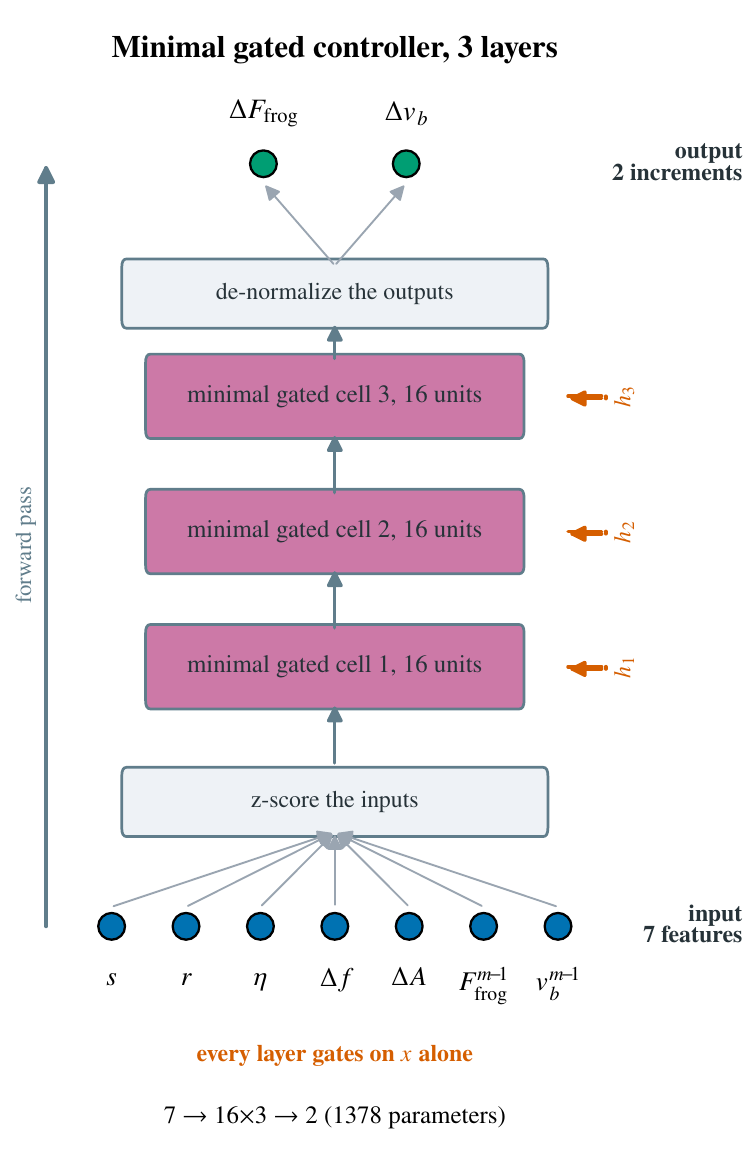}}
\caption{The same cell at three layers of $16$ units each. Every layer still gates on its
input alone, so stacking adds carried states without giving any of them influence
over its own gate.}
\label{fig:archdeep3}
\end{figure}

\begin{figure}[htbp]
\centerline{\includegraphics[width=\figwidth]{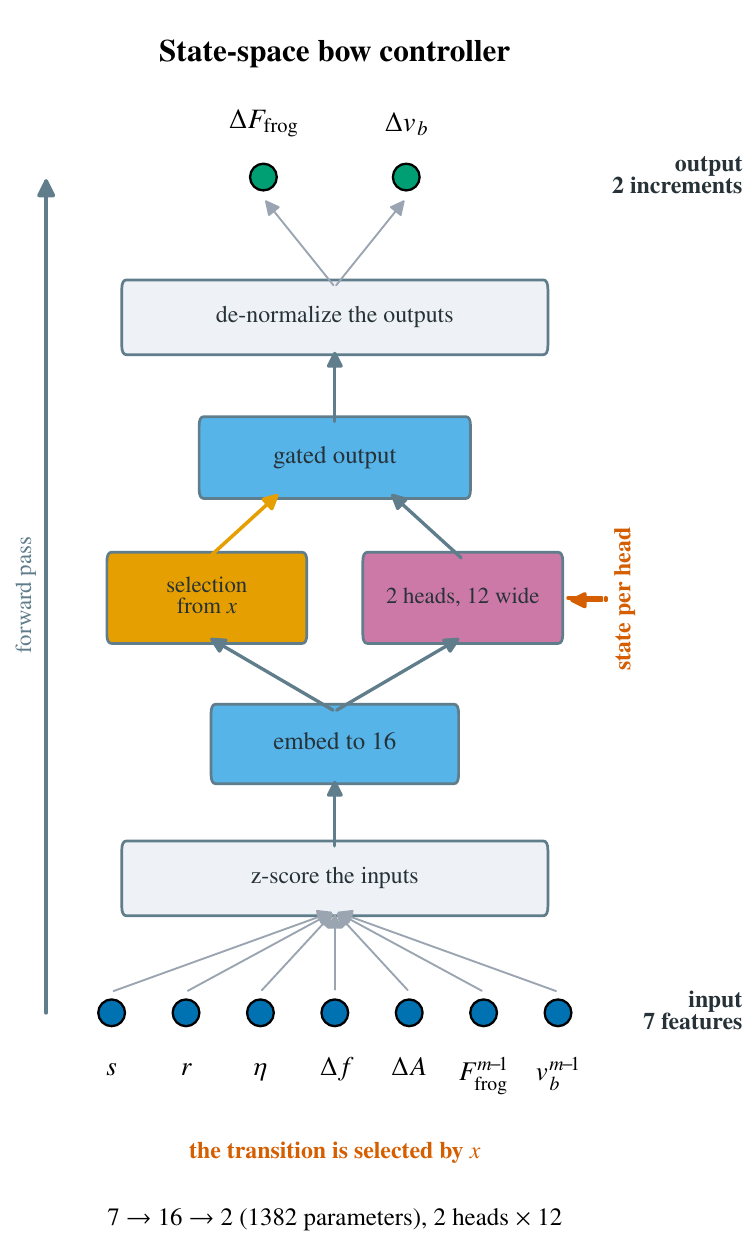}}
\caption{Architecture of the state-space controller. The transition applied to each state
head is selected by the input, rather than fixed, which is the mechanism that it
shares with a gate, and the state is carried in two heads of width $12$.}
\label{fig:archmamba}
\end{figure}

Since the labels depend on neither the oscillation nor its history, this comparison measures the inductive biases of networks fitted to the same open-loop labels, and not whether the task needs a memory of the oscillation. Six architectures were trained at a
matched parameter count on all four open strings, with twenty seeds each, and with
every fit on one device. They are the feedforward network of
Section~\ref{sec:training}, a {\it gated recurrent unit}~\cite{r-m:cho-2014} (Figure~\ref{fig:archgru}), its minimal variant~\cite{r-m:feng-2024}, whose gates do not depend on the hidden state (Figure~\ref{fig:archmingru}), two deeper minimal variants at two and three layers (Figures~\ref{fig:archdeep2} and~\ref{fig:archdeep3}), and a {\it state-space model} with input-selective transitions~\cite{r-m:gu-2024,r-m:dao-2024} (Figure~\ref{fig:archmamba}). The parameter counts span $1362$ to $1410$, which is a $3.5\%$ spread. A matched parameter count is a weak proxy for matched capacity across families, since it makes the gated recurrent unit $17$ units wide and the minimal variant $76$. Every fit was trained in PyTorch with Adam, at a learning rate of $3\times10^{-3}$ annealed on a cosine schedule~\cite{r-m:loshchilov-2017} over $4000$ epochs, on mini-batches of $64$ sequences for the recurrent and state-space models and of $256$ samples for the feedforward network, with no early stopping. This protocol differs from the training of the feedforward baseline of Section~\ref{sec:training}, which is a single fit with early stopping. The twenty-seed set of this comparison, trained on a GPU, is the {\it board family}. The {\it grid family} is a set of the same architectures, trained by the same protocol on a CPU, with twenty seeds for the feedforward network, the gated recurrent unit, and the minimal variant. A CPU and a GPU give different networks from the same seed, so the two families are never pooled.

\begin{figure*}[htbp]
\centerline{\includegraphics[width=\figwidewidth]{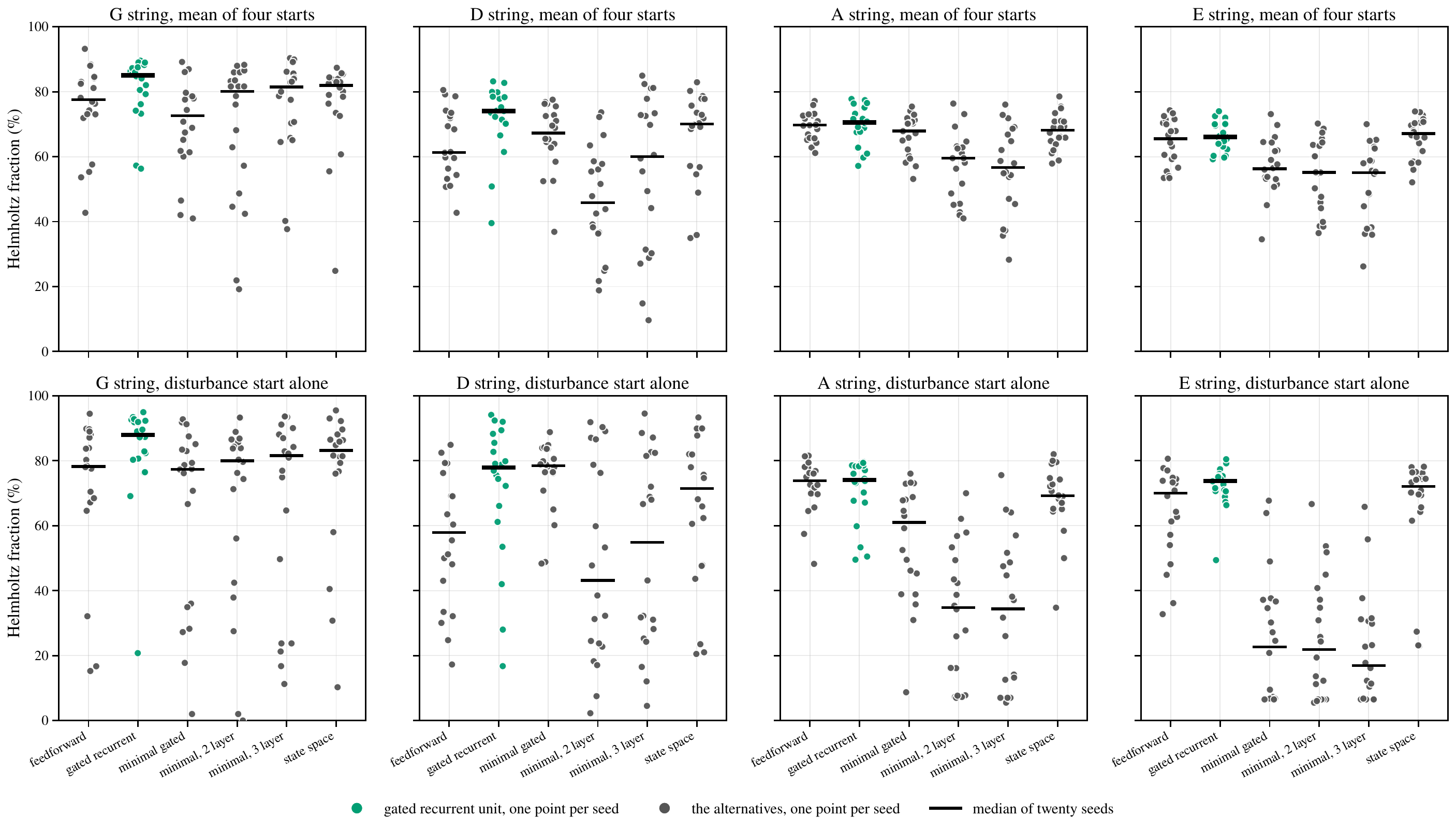}}
\caption{Every seed of the six-architecture board, twenty per cell, with the median marked.
Top row: the mean over four starts, which is what Table~\ref{tab:board} reports.
Bottom row: the disturbance start alone. Points are shown instead of a box plot,
since the finding is a cluster, and not a shifted mean. Namely, on E, the minimal
gated variant places eight of twenty seeds below nine percent on the disturbance
start, which a box plot would render as outliers. The gated recurrent unit and the
feedforward network overlap almost completely on A, which is the tenth-of-a-point
margin reported above.}
\label{fig:seedspread}
\end{figure*}

Figure~\ref{fig:seedspread} shows every seed behind Table~\ref{tab:board}. Two runs of the same controller, which differ only in a numerically negligible perturbation of the input, may diverge on the Helmholtz fraction. Over ninety-six cells spanning four strings, six architectures, and four starts, the divergence of a cell is taken as the larger of its two perturbation sizes. It is exactly zero in ninety-two of them, and it reaches $4.79$ points in the worst, with a bootstrap interval on the worst case of $[0.33, 4.79]$. Three of the four cells which move at all are on the disturbance start.

A perturbation a thousand times larger produces about three times the divergence, where a smooth sensitivity would produce a thousand times as much. We call this amplification by discontinuous switching, rather than chaotic amplification, since most cells do not move at all and the effect saturates. Differences below about four points between single episodes are read as ties. That bar is a heuristic tie margin, and not a test of equivalence. It sits near the top of a mostly-zero distribution, rather than at a typical divergence. It is not its maximum, since one cell, the feedforward network on E from the disturbance start, exceeds it.

For means over twenty seeds, whose standard errors run from $0.98$ to $5.43$ points across the twenty-four architecture and string cells, the comparisons rest on those standard errors and on the tests reported with them.

\begin{table}[htbp]
\caption{Mean Helmholtz fraction over four starts, at twenty seeds per cell, with the
standard deviation across seeds and the median in parentheses. Every fit was made on one device. Bold marks the highest mean, and Table~\ref{tab:completion} gives completion beside it.}
\label{tab:board}
\centering
\footnotesize
\setlength{\tabcolsep}{2.5pt}
\begin{tabular}{|l|c|c|c|c|}
\hline
\textbf{Arch.} & \textbf{G} & \textbf{D} & \textbf{A} & \textbf{E} \\
\hline
Feedforward & $75.1 \pm 13.2$ & $63.6 \pm 11.1$ & $69.5 \pm 4.4$ & $64.3 \pm 7.2$ \\
 & $(78)$ & $(61)$ & $(70)$ & $(65)$ \\
\hline
Gated recurrent & $\mathbf{81.3} \pm 9.8$ & $\mathbf{71.7} \pm 10.6$ & $\mathbf{69.6} \pm 6.2$ & $\mathbf{66.2} \pm 4.5$ \\
 & $(85)$ & $(74)$ & $(71)$ & $(66)$ \\
\hline
Minimal gated & $69.6 \pm 14.2$ & $66.3 \pm 10.2$ & $66.5 \pm 6.3$ & $56.9 \pm 8.7$ \\
 & $(73)$ & $(67)$ & $(68)$ & $(56)$ \\
\hline
Minimal, 2 layer & $68.3 \pm 22.0$ & $46.6 \pm 16.5$ & $57.3 \pm 10.4$ & $54.5 \pm 11.1$ \\
 & $(80)$ & $(46)$ & $(60)$ & $(55)$ \\
\hline
Minimal, 3 layer & $75.6 \pm 15.0$ & $55.3 \pm 24.3$ & $55.9 \pm 13.7$ & $51.0 \pm 12.3$ \\
 & $(81)$ & $(60)$ & $(57)$ & $(55)$ \\
\hline
State space & $76.2 \pm 14.6$ & $66.3 \pm 13.9$ & $67.8 \pm 5.5$ & $65.8 \pm 6.4$ \\
 & $(82)$ & $(70)$ & $(68)$ & $(67)$ \\
\hline
\end{tabular}
\end{table}

The gated recurrent unit leads on the mean of all four strings. Averaged over the
four starts, its margin over the feedforward network survives Holm correction on
none of the four strings. From the in-band start, it leads on all four and is significant on their mean, ahead on seventeen of the twenty seeds at $p = 0.002$, before any correction for the choice of start. Across the forty positions of Section~\ref{sec:fingering}, four of them open, it is ahead at thirty-seven from the same start, at $p = 0.0015$ averaged over them. The other three lie within one standard error of zero. Positions on one string are one plant relabeled (Section~\ref{sec:fingering}), so that count is not independent evidence. Its margin over the minimal variant survives on G, at
$p = 0.0074$, and on E, at $p = 0.0010$, which are the two strings on which its
margin over the minimal variant is largest, at $11.67$ and $9.22$ points. The
minimal variant's latching is worst on E, then A, then G. Paired over the same twenty seeds by a two-sided t-test, its margin over the feedforward network is $6.16$ points on G at
$p = 0.122$, $8.12$ on D at $p = 0.037$, $0.12$ on A at $p = 0.947$, and $1.86$ on
E at $p = 0.296$.

The margin on A is a tenth of a point, and it should be read as a tie. Against the
minimal variant, it gains $11.67$ points on G at $p = 0.0025$, $5.44$ on D at
$p = 0.148$, $3.06$ on A at $p = 0.140$, and $9.22$ on E at $p = 0.00025$. Against
the feedforward baseline, the minimal variant on A reads $-2.94$ at
$p = 0.094$.

That last figure reverses an earlier reading of this comparison, and the reason
why it reverses is the finding itself. Measured on A over the three starts which
exclude the disturbance, the minimal variant beats the feedforward network inside
the band by $3.58$ points at $p = 0.0038$, and at the high-force corner by $3.70$
at $p = 0.011$. It loses the disturbance start by $17.36$ points at $p = 0.0002$,
and on E, it scores $24.6\%$ there.

\begin{table}[htbp]
\caption{Mean Helmholtz fraction on the disturbance start alone, at twenty seeds per
cell.}
\label{tab:disturbstart}
\centering
\begin{tabular}{|l|c|c|c|c|}
\hline
\textbf{Architecture} & \textbf{G} & \textbf{D} & \textbf{A} & \textbf{E} \\
\hline
Feedforward & $71.1$ & $55.9$ & $\mathbf{72.2}$ & $64.0$ \\
\hline
Gated recurrent & $\mathbf{83.0}$ & $71.4$ & $70.9$ & $\mathbf{72.2}$ \\
\hline
Minimal gated & $64.4$ & $\mathbf{75.3}$ & $54.8$ & $24.6$ \\
\hline
Minimal, two layers & $66.7$ & $48.9$ & $33.3$ & $25.2$ \\
\hline
Minimal, three layers & $66.5$ & $52.1$ & $33.0$ & $21.8$ \\
\hline
State space & $75.2$ & $65.2$ & $68.2$ & $65.1$ \\
\hline
\end{tabular}
\end{table}

Table~\ref{tab:disturbstart} isolates that start. The defining simplification of
the minimal variant is that its gates are computed from the input alone, and not
from the hidden state, which Figures~\ref{fig:archgru} and~\ref{fig:archmingru}
contrast directly. That is why it can latch after the abrupt override of the bow velocity. On E, eight of its twenty seeds hold
between $3\%$ and $9\%$ Helmholtz motion for the remainder of the stroke once the
override lifts. Forcing a state reset at every control step recovers every one of them, to between $54.8\%$ and $75.9\%$, with every stroke completed. The reset raises eleven of the other twelve seeds as well. The architectures which survive the
disturbance are the two which keep a state-aware or an input-selective gate.

Depth does not repair it. The two-layer and three-layer minimal variants never
significantly improve on the single layer, and they significantly worsen it on D,
by $19.70$ points, and on A, by $9.24$, both surviving Holm correction. They are
the two weakest architectures on the disturbance start on D and on A. On G, the
minimal single-layer variant is weaker than either of them, and on E, it sits
between them, below the two-layer variant and above the three-layer one. Depth compounds the latching clearly on D and A, at $-26.3$ and $-21.5$ points for the two-layer variant, both at $p = 0.0012$. On G and E, the differences, $+2.3 \pm 8.3$ and $+0.6 \pm 6.5$ points, cannot be distinguished from zero, at $p = 0.78$ and $0.92$. Nothing about depth
helps on any string.

The state-space model is never the winner, and it never collapses in the mean.
It places second or third on every string, and it holds $65.1\%$ on the E
disturbance, where the minimal variant manages $24.6\%$. Individual seeds do go
low. In fact, its seed spread on G is $14.6$ points, the widest in its own row of
Table~\ref{tab:board}, so the claim is about the distribution, and not about every
draw from it.

\begin{table}[htbp]
\caption{Cost per control step. The multiply-accumulate count is the hardware-independent
column, while the microsecond column measures this implementation, rather than the
architecture.}
\label{tab:archcost}
\centering
\begin{tabular}{|l|c|c|c|c|c|}
\hline
\textbf{Architecture} & \textbf{Par.} & \textbf{MAC} & \textbf{Tr.} & \textbf{$\mu$s} & \textbf{$\times$} \\
\hline
Feedforward & $1378$ & $1312$ & $64$ & $25.4$ & $1.00$ \\
\hline
Gated recurrent & $1362$ & $1258$ & $68$ & $53.3$ & $2.10$ \\
\hline
Minimal gated & $1370$ & $1216$ & $152$ & $50.9$ & $2.01$ \\
\hline
Minimal, two layers & $1410$ & $1320$ & $88$ & $80.6$ & $3.18$ \\
\hline
Minimal, three layers & $1378$ & $1280$ & $96$ & $111.8$ & $4.41$ \\
\hline
State space & $1382$ & $1920$ & $23$ & $132.3$ & $5.22$ \\
\hline
\end{tabular}
\end{table}

Three readings of Table~\ref{tab:archcost} bear on the choice of architecture. The gated
recurrent unit is cheaper than the feedforward network on both counts, rather than
dearer, with fewer parameters than the feedforward network, at $1362$ against
$1378$, and fewer multiply-accumulates, at $1258$ against $1312$, while leading on
all four strings. The microsecond column measures a NumPy implementation in which
the per-step latency is dominated by interpreter overhead, which is why the
minimal variant has the fewest multiply-accumulates of anything in the table and
still costs twice as much as the feedforward network. The state-space model's
multiply-accumulate count exceeds its parameter count, $1920$ against $1382$,
since its recurrence carries $592$ operations which are not weights.

None of this decides the question. Against the cost of one finite-difference
control step, the cheapest controller occupies between $0.042\%$ and $0.142\%$, and
the dearest between $0.221\%$ and $0.743\%$, so every architecture in
Table~\ref{tab:archcost} is under $1\%$ on every string.

Completion has to be read beside the means of Table~\ref{tab:board}, and Table~\ref{tab:completion} gives it. The feedforward network completes more strokes on three of the four starts, and it ties on the hot start, at $80$ of $80$. One of the three carries most of the difference, namely, sixteen
more strokes of eighty from cold, against five on the disturbance start and one
inside the band. On the other three starts together, the two are at $237$ and
$231$ of $240$. That start commands $v_b = 0.02$ m/s, which is a $32.5$ s
frog-to-tip stroke, and in Section~\ref{sec:coldplayable}, we show that the target
regime is out of reach only on G, where it lies below the commanded force floor.
On D and A, it lies between the force rows of the map, one control step above the
starting force, and on E, the start is on it. Therefore, on every string, the cold
start tests reaching Helmholtz motion from a start which no player would use,
rather than regulation inside the playable band. On the disturbance start, where
the commanded bow speed is overridden at mid-stroke, the gated recurrent unit leads the feedforward network by $11.9$, $15.5$, and $8.2$ points on G, D, and E, and on A it trails by $1.3 \pm 2.3$ points, at $p = 0.59$. None of these margins survives Holm correction over the five comparisons on its string. The asymmetry is not confined to disturbances which the
commands can see. In Section~\ref{sec:contactdist}, we reduce the limiting
friction instead, which is what rosin depletion or a heating contact would do, and
which leaves $s$, $F_{\mathrm{frog}}$, and $v_b$ untouched. At twenty seeds, we
find that both recurrent architectures show the effect over the whole stroke below
$p = 0.001$, and the feedforward network reaches significance on neither window.
That contrast is about whether an architecture uses its oscillation inputs at
all, so it does not by itself say how the feedforward network performs under a
contact disturbance. What says that is Table~\ref{tab:lookupcontact}. At the
intermediate reduction, scored after the onset, the feedforward network trails
the tabulation by $41.0$ points, with no seed of twenty going the other way,
against the gated unit's $7.0$, with three. Over the whole stroke, the gated unit's figure falls to $3.2$, with a standard error of $1.7$, at $p = 0.076$, so that comparison is one which the window decides, and the feedforward network's is not.

We prefer the gated recurrent unit, and the preference carries no significance. It has the highest mean on every string, and it avoids the latching described above, but at twenty seeds it cannot be distinguished from the feedforward network or the state-space model, and the feedforward network completes more strokes.

\subsection{The Training Context Window Bounds the Comparison}
\label{sec:context}

Every fit in Table~\ref{tab:board} was trained on sequences capped at thirty
control steps. Since the recurrent architectures differ in what they can retain,
that cap is a condition on the comparison, rather than an incidental setting. A sweep over six windows on two strings, with three seeds at each of the four shorter windows and six at the two longest, measures its effect. The minimal variant stops at the second longest, which is
where the comparison is made.

The advantage of the gated recurrent unit over the minimal variant is largest at
the window which the comparison used, and it is absent at a longer one. On A, the
difference is $-13.4$ points at thirty steps and $-0.2$ at four hundred. On E, it is
$-14.3$ and $+0.2$. At three seeds this is a direction, and on A the twenty-seed comparison of Table~\ref{tab:board}, at the same cap, puts the margin at $3.06$.

This does not show the minimal variant winning at the longer window. It states a condition on the comparison. Namely, the accuracy argument for the gated recurrent unit over the minimal variant holds at the training window which was used, and it weakens as the window grows.

The practical argument, however, moves the other way. One minimal-variant fit at
the longer window costs $3216$ seconds, against $51$ for the gated recurrent unit,
a factor of $63$, so a board of the size reported here is not affordable in that
family at that window. That same factor is why this sweep covers two strings,
whereas the capacity ladder of Section~\ref{sec:capacity} covers four. Extending it
to the other two would cost between six and eleven times the ladder, and almost
all of that cost is the minimal variant at the long window, which is the cell of
which the comparison is made. Therefore, the scope is a costed choice, rather than
an oversight. The extension is carried in Section~\ref{sec:futurearch}.

The training loss falls with the window in almost every cell of the sweep, by
$76\%$ on the A-string feedforward network, and it rises only slightly in three of
the six cells run at the longest window, while the closed-loop score moves in
every direction. Therefore, the divergence between the supervised loss and the
control quality reported in Section~\ref{sec:capacity} holds along the context
axis as well, and the supervised loss cannot be used to select a training window
either.

\subsection{The Cold Start Against the Map's Helmholtz Threshold}
\label{sec:coldplayable}

The cold start commands $v_b = 0.02$ m/s, the lowest speed which the action bounds
permit, at each string's own force floor, which is $0.40$ N on G and $0.05$ N
elsewhere. On $0.65$ m of playing hair, that is a $32.5$ s frog-to-tip stroke, far
outside ordinary playing. Read at that speed, the properly settled map holds no
Helmholtz cell in that column on three of the four strings. On E, it resolves a
single cell, at $0.05$ N.

An empty column is not proof that Helmholtz motion is unavailable there.
Both edges of a map built from rest are linear in the bow speed (Section~\ref{sec:schelleng}), so as the speed falls, the window slides toward zero force and narrows, rather than
vanishing, and a window narrower than one cell of the force grid will not be
resolved. Fitting both bounds through the origin against the resolved columns of
each string's properly settled map, and evaluating them at $0.02$ m/s, gives
$[0.072, 0.159]$ N on G, $[0.058, 0.124]$ on D, $[0.048, 0.099]$ on A, and
$[0.041, 0.088]$ on E.

A direct scan settles what the fit only bounds. In this scan, the force is stepped
at the cold-start speed in $5$~mN increments, at the map's own settling window, on
D, A, and E. It finds Helmholtz motion between the rows of the map, namely, at
$0.060$ to $0.075$~N on A, at $0.055$~N and from $0.075$ to $0.095$~N on D, and on
E, in a run from $0.050$ to $0.065$~N which covers the start itself. The column is
ragged at that resolution, with Helmholtz and raucous motion alternating from one
step to the next, which is the low-speed raggedness of
Section~\ref{sec:schelleng}. On D, Helmholtz motion also appears below the fitted
window, so the fitted windows bound the band, rather than trace it. A control
step moves the force by at most $0.05$~N, so on D and A, the target is one step
away from the start.

On G, that window lies entirely below the commanded force floor of $0.40$ N. The
floor on G is $0.40$ N, where the other three strings carry $0.05$, and it is a set
constant, rather than a quantity derived from either bow-force law. It is also
higher than the string requires. Probing below it, at the map's own five lowest
speeds, returns Helmholtz motion in thirty-five of seventy cells, including at the
cold-start speed. Therefore, the cold start is unplayable on G since the command
set forbids the force, while the string itself can sustain the motion there. The
upper bound of the window does not rise to meet the floor until the bow speed
reaches about $0.050$ m/s. On G, the controller is forbidden by its own action
bounds from reaching Helmholtz motion at the cold-start speed, and this is a
statement about the command set, rather than about the grid resolution.

What occupies the low-speed column instead is constant sticking, between $23$ and
$26$ of the $33$ force cells on the four strings. At the cold-start speed, the bow
drags the string, rather than releasing it, over most of the force range. The
remaining cells form a single run at the low end of the range, where the string
is released into multiple slipping, raucous motion, or sub-harmonic motion, and on
D, A, and E, into Helmholtz motion in the narrow band which the scan above
resolves.

That a bow speed may be too low for a note is not a new observation.
For a string bowed at its midpoint, with a small dissipative force at the contact, {\it Friedlander}~\cite[pp.~524--525]{r-m:friedlander-1953} derived a stability condition for the periodic motion, and he found that it fails below a bow speed which his analysis locates only qualitatively. He reported that the string then produces noise, rather than a note, adding that the effect did not appear to have been noted and could easily be verified in practice. What is new here is the threshold as a measured quantity, per string and against the model's own command set, as well as the regime which the string occupies below it, which is sticking, rather than the non-periodic motion which his argument suggests.

\begin{table}[htbp]
\caption{Strokes reaching the tip, out of twenty seeds on each of four strings. The final
column is restricted to the three starts other than the cold start.}
\label{tab:completion}
\centering
\small
\begin{tabular}{|l|c|c|c|c|c|}
\hline
\textbf{Architecture} & \textbf{Band} & \textbf{Cold} & \textbf{Hot} & \textbf{Dist.} & \textbf{Three} \\
\hline
Feedforward & $80/80$ & $70/80$ & $80/80$ & $77/80$ & $\mathbf{237/240}$ \\
\hline
Gated recurrent & $79/80$ & $54/80$ & $80/80$ & $72/80$ & $231/240$ \\
\hline
Minimal gated & $79/80$ & $69/80$ & $80/80$ & $68/80$ & $227/240$ \\
\hline
State space & $78/80$ & $59/80$ & $80/80$ & $73/80$ & $231/240$ \\
\hline
\end{tabular}
\end{table}

Therefore, Table~\ref{tab:completion} reports completion in two ways. Pooled over
all four starts, the feedforward network completes every stroke on $68$ of $80$
seeds, against the gated recurrent unit's $51$. Restricted to the three playable
starts, the gap is $237/240$ against $231/240$, and on those same three starts,
the gated recurrent unit leads the feedforward network on the mean Helmholtz
fraction on all four strings, at $71.5\%$ against $66.4\%$. The advantage of the
feedforward network in reliability rests largely on the cold start, a start which
no player would use.

The cold start is retained, rather than dropped. It measures reaching Helmholtz
motion from a start which no player would use, and that is out of reach only on
G, where the force floor forbids it. This is a different question from regulation
inside the playable band, and the paper should not let one number stand for
both.

\subsection{Feature Ablation}
\label{sec:ablation}

The policy was retrained from scratch with each feature or feature group removed, rather than masked at inference. Every condition was evaluated on the same ten initial conditions, comprising the four named cases of Table~\ref{tab:results} and six drawn at random from the command box, so the comparison is paired. Table~\ref{tab:ablation} reports the paired difference against the full model. Since the labels do not depend on the string-state inputs, this measures input representation. It asks how removing an input changes what a fit to the same labels does in closed loop, and not what that input contributes as trained feedback.

\begin{table}[htbp]
\caption{Feature ablation, given as the full model's mean Helmholtz fraction minus that of
the ablated model, over ten paired starts, so that a positive entry means that
removing the group hurts. The policy is retrained from scratch in every one of the cells, across four strings and four architectures. Each cell is one fit, from seed $0$, and its $p$-value is from a paired t-test over the ten starts. MLP, GRU, and minGRU denote the feedforward network, the gated recurrent unit, and its minimal variant, and SSM the state-space model.}
\label{tab:ablation}
\centering
\footnotesize
\setlength{\tabcolsep}{3.5pt}
\begin{tabular}{|l|c|r|r|r|r|}
\hline
\textbf{Removed} & \textbf{Str.} & \textbf{MLP} & \textbf{GRU} & \textbf{minGRU} & \textbf{SSM} \\
\hline
             & G & $-1.1$  & $+1.0$  & $-5.2$  & $-22.3$ \\
all four     & D & $-3.7$  & $+11.5$ & $+1.6$  & $+16.7$ \\
string-state & A & $+12.0$ & $+8.0$  & $+15.6$ & $+14.0$ \\
             & E & $+1.2$  & $+15.1$ & $+15.1$ & $+11.7$ \\
\hline
             & G & $+6.9$  & $+0.8$  & $+17.1$ & $-9.9$ \\
previous     & D & $-1.8$  & $+6.9$  & $+3.6$  & $+12.9$ \\
commands     & A & $+2.3$  & $+3.2$  & $+6.0$  & $+5.8$ \\
             & E & $+3.6$  & $+4.4$  & $+14.1$ & $+4.0$ \\
\hline
             & G & $+3.4$  & $-2.1$  & $+7.5$  & $-7.6$ \\
stroke       & D & $-7.3$  & $+7.6$  & $-3.8$  & $+5.7$ \\
position $s$ & A & $+5.1$  & $+2.9$  & $+8.9$  & $-2.8$ \\
             & E & $+2.5$  & $-3.8$  & $+9.4$  & $+3.8$ \\
\hline
\end{tabular}
\end{table}

Removing the previously applied commands hurts in fourteen of the sixteen cells,
and it is the one group whose sign is close to consistent. Removing the stroke
position hurts in ten. Removing the four string-state features together hurts in
twelve cells. The four in which it helps are concentrated, rather than scattered,
since three of the four are on the G string. The two in which the help reaches $p < 0.05$ are both on G, at $5.2$ points for the minimal gated variant at $p = 0.030$, and at $22.3$ for the state-space model at $p = 0.026$, uncorrected and over the ten starts of a single fit. The other two
are inside the noise, at $p = 0.803$ on G and $p = 0.687$ on D.

That structure is the result. Removing the string state hurts on A and on E, by $12.4$ and $10.8$ points pooled over their four cells each, and it helps on G. The individual cells on A and E run from $1.2$ to
$15.6$, and four of the eight clear $p < 0.05$ on their own. The spread of the effect across strings is $19.3$ points, against $6.8$ across architectures. The individual features within the group behave in the same way. Their signs turn over between strings more than between architectures, and no single one of them carries the group.

Two limits bound that contrast. Each cell is one fit, from seed $0$, and its
$p$-value is over the ten starts, so it measures variation across starts, and not
across training seeds, which Section~\ref{sec:variance} finds interacting strongly
with the start. The closed-loop window is also fixed in seconds, so the four strings
differ in how many periods each regime call sees, as well as in their physics. Under
this protocol, how much the string-state inputs are worth to a fit varies more
across strings than across architectures. Whether that belongs to the strings, or to
the fixed window and command box, is not separated here.

One conjecture fits the ordering, and it is offered as a conjecture, rather than
as a result. G is the string whose Helmholtz band sits highest in force, and
whose commanded floor was raised accordingly. A schedule over the stroke position
and the current commands may therefore be closer to sufficient there, with the
extra inputs acting as a distraction. On A and E, the band is lower in the same
coordinates, and the string state may carry information which the commands do
not. Regressing the per-string effect on the establishment threshold gives a correlation of $-0.944$ at $p = 0.056$ over four points, which is suggestive but not
established, and the rank correlation is weaker still, since A and E invert the
predicted order. In Section~\ref{sec:schelleng}, a cause for the impedance exponent is named only because it follows from the equation of motion, rather than from a correlation over four strings, and the same standard is applied here.

The scope of the result is limited in one more way. The disturbances tested are a step change in the commanded velocity and adverse initial conditions, both of which are recoverable without the string-state inputs. A
disturbance acting on the string, rather than on the command, is not visible in
$s$, $F_{\mathrm{frog}}$, or $v_b$. In Section~\ref{sec:contactdist}, we measure
that case at twenty seeds and find an effect on both recurrent architectures over
the whole stroke, although whether it reaches the steps after the onset depends
on the architecture.

\subsection{Capacity of the Function Approximator}
\label{sec:capacity}

A natural question is whether a larger or more expressive network would do better.
It may be answered directly, over a ladder of parameter budgets which spans the
architectures that this paper actually uses, holding the dataset, the
normalization, and the evaluation starts fixed.

Two questions sit behind it, namely, whether the approximator family matters, and
whether the budget within a family matters. Table~\ref{tab:capacity} takes the
first, over five families spanning a linear map to exact recall, on the A string, fitted to the same labels and evaluated from the same four starts. Exact recall is a nearest-neighbor lookup which returns the label of the closest training sample.

\begin{table}[htbp]
\caption{Approximator family against fit quality and closed-loop performance, on the A
string. The two $R^2$ columns are in-sample, against the commanded force increment
and the commanded speed increment. The four right-hand columns give the percentage
of control steps in Helmholtz motion from each start. The two families which carry
a training seed, the wide network and the forest, were checked across twenty
seeds, and every value here falls inside the observed range.}
\label{tab:capacity}
\centering
\footnotesize
\setlength{\tabcolsep}{3.5pt}
\begin{tabular}{|l|r|r|r|r|r|r|}
\hline
 & \multicolumn{2}{c|}{\textbf{$R^2$}} & \multicolumn{4}{c|}{\textbf{Helmholtz \%}} \\
\hline
\textbf{Family} & $\Delta F$ & $\Delta v_b$ & band & cold & hot & dist \\
\hline
Linear & $0.585$ & $0.451$ & $\mathbf{96.6}$ & $57.1$ & $11.4$ & $67.4$ \\
MLP $32$--$32$ & $0.812$ & $0.703$ & $94.8$ & $89.5$ & $40.9$ & $75.8$ \\
MLP $256^3$ & $0.815$ & $0.692$ & $93.7$ & $77.9$ & $42.2$ & $78.1$ \\
Random forest & $0.971$ & $0.951$ & $91.8$ & $81.7$ & $38.6$ & $\mathbf{80.6}$ \\
Exact recall & $1.000$ & $1.000$ & $\mathbf{96.6}$ & $\mathbf{97.3}$ & $\mathbf{47.7}$ & $73.9$ \\
\hline
\end{tabular}
\end{table}

Some capacity is required, and very little of it suffices. The linear map holds
the band but loses the cold start, at $57.1\%$ against $89.5\%$ for the smallest
network, a gap of $32.4$ points. Above that, the single draws printed for the three learned families span $3.0$ points on the band. At twenty seeds, however, the wide network leads the forest there by $4.5 \pm 1.1$ points, at $p = 0.0002$ by Welch's $t$-test. The linear map and exact recall both sit at $96.6$
there, so the two ends of the capacity range meet on the start which is the
easiest to hold.

Fit quality does not order them. The in-sample $R^2$ rises from $0.585$ to exactly
$1.000$ across the table, while the band score runs $96.6$, $94.8$, $93.7$,
$91.8$, and $96.6$, ending where it started. A family which reproduces its labels
perfectly scores the same on the band as the linear map which reproduces them
worst. That is the same divergence between the supervised loss and the
closed-loop quality which the ladder finds against budget, reached here against
family instead.

Exact recall is the useful limit. It is at least as good as every other family tested from three of the four starts, which suggests that the ceiling is set by the labels, rather than by the fit. No learned
family reaches it from cold, and the rule which generated those labels is the
subject of Section~\ref{sec:baselines}.

The ladder in Table~\ref{tab:ladder} covers four parameter budgets, from about
$1400$ to $12000$, on all four strings, for four architectures, at three seeds
each, which amounts to one hundred and ninety-two fits.

\begin{table}[htbp]
\caption{In-sample training mean-squared error and closed-loop performance against the parameter budget, as the mean of three seeds. Each cell gives the training error and the percentage of control steps in Helmholtz motion, averaged over the four starts. The budget of Table~\ref{tab:board} is the leftmost column. The abbreviations are those of Table~\ref{tab:ablation}.}
\label{tab:ladder}
\centering
\footnotesize
\setlength{\tabcolsep}{3pt}
\begin{tabular}{|l|l|c|c|c|c|}
\hline
 & & \multicolumn{4}{c|}{\textbf{parameter budget}} \\
\hline
\textbf{Str.} & \textbf{Arch.} & $1.4$k & $3$k & $6$k & $12$k \\
\hline
G & GRU    & $0.185/85.4$ & $0.155/70.8$ & $0.134/77.7$ & $0.101/36.0$ \\
  & minGRU & $0.206/87.5$ & $0.177/73.6$ & $0.158/69.9$ & $0.148/84.4$ \\
  & SSM    & $0.209/70.9$ & $0.177/74.1$ & $0.158/62.8$ & $0.131/47.8$ \\
  & MLP    & $0.123/90.5$ & $0.102/68.2$ & $0.094/69.4$ & $0.089/57.0$ \\
\hline
D & GRU    & $0.240/78.1$ & $0.211/66.9$ & $0.166/72.6$ & $0.120/57.7$ \\
  & minGRU & $0.268/69.4$ & $0.233/68.6$ & $0.211/65.0$ & $0.194/58.4$ \\
  & SSM    & $0.270/74.6$ & $0.232/59.4$ & $0.204/70.1$ & $0.173/63.0$ \\
  & MLP    & $0.167/57.0$ & $0.139/68.0$ & $0.124/61.3$ & $0.117/70.4$ \\
\hline
A & GRU    & $0.194/73.2$ & $0.170/69.8$ & $0.130/63.1$ & $0.080/59.7$ \\
  & minGRU & $0.207/59.8$ & $0.190/62.1$ & $0.180/68.4$ & $0.174/64.8$ \\
  & SSM    & $0.220/73.3$ & $0.204/64.8$ & $0.190/65.1$ & $0.171/69.6$ \\
  & MLP    & $0.165/72.2$ & $0.148/63.1$ & $0.133/68.8$ & $0.120/70.8$ \\
\hline
E & GRU    & $0.187/70.4$ & $0.166/61.9$ & $0.137/70.6$ & $0.087/66.5$ \\
  & minGRU & $0.213/56.1$ & $0.197/56.3$ & $0.188/56.5$ & $0.183/55.5$ \\
  & SSM    & $0.219/66.4$ & $0.205/68.7$ & $0.198/54.0$ & $0.176/52.3$ \\
  & MLP    & $0.176/67.1$ & $0.158/64.4$ & $0.145/59.5$ & $0.132/60.8$ \\
\hline
\end{tabular}
\end{table}

The in-sample training mean-squared error falls with the budget in every one of the sixteen cells, by between $14.3$ and $59.0$ percent. The closed-loop quality does not follow it. The mean falls in fourteen of the sixteen cells, by more than four points in nine, and it rises in two. At three seeds, however, the standard error of a change runs from $2.2$ to $19.3$ points, and only one change, a fall of $11.7$ points for the state-space model on D, is significant by Welch's $t$-test, at $p = 0.0062$. None survives Holm correction over the sixteen. The mean change is $-11.1$ points, and two of the nine falls are catastrophic, at $-49.4$ and $-33.5$.

The two exceptions should be named, rather than left for a reader to find. On A,
the minimal gated variant gains $5.0$ points, from the lowest starting score on
that string, which is the easy case to explain. On D, the feedforward network
gains $13.4$, and it is not the easy case. It starts at $57.0$, which is low for
that string, and it finishes at $70.4$, the best D-string cell at the largest
budget. That is the largest rise in the table and the seventh largest change of
the sixteen in either direction, and it exceeds three of the nine falls.
Neither rise is significant at three seeds, so the ladder shows no cell in which extra capacity measurably helps.

The effect is also not uniform across strings. Per string, the mean change is
$-27.2$ on G, $-7.4$ on D, $-3.4$ on A, and $-6.2$ on E, so G carries most of it
and A is nearly flat. Measured on G and A alone, the mean would have been
$-15.3$, and adding D and E made the effect smaller, rather than larger, since
those two strings happened to be the extremes of the range.

The seed spread moves in the same way, and it is not visible in the loss at all.
Comparing the smallest budget with the largest, the standard deviation across the
three seeds rises in thirteen of the sixteen cells. In four of them, it rises by a
factor of four or more, namely, from $2.4$ to $24.5$ on the G-string feedforward
network, $5.2$ to $24.9$ on the G-string gated unit, $4.5$ to $19.7$ on the
D-string gated unit, and $2.9$ to $12.8$ on the A-string gated unit. Extra capacity
lowers the mean, and it also makes the outcome less reproducible.

Three caveats belong with the ladder. Three seeds establish a direction, and not
an effect size, so the change in no single cell should be quoted as a precise
figure. The ladder holds the training window at thirty control steps, so it has
not been crossed with the sweep of Section~\ref{sec:context}. In addition, the
budgets span a range which the paper's own comparison does not use, since every
architecture in Table~\ref{tab:board} sits in the leftmost column.

Therefore, Table~\ref{tab:ladder} says that a search driven by the quality of fit
selects badly here. It would pick the largest budget in every cell, and that is
the better controller in two of sixteen. Selection has to run on the closed-loop
measure.

Therefore, what limits the controller is the supervision, rather than the model.
The targets are nearest-neighbor jumps on a discretized map, so they are
discontinuous in the state, and fitting them more closely buys a policy which
inherits the discontinuity. The ceiling on performance is set by the supervision, the map and the greedy rule built on it, rather than by the capacity of the model which fits them, and it is reached well below the largest budget tested. This paper does not separate the map from the rule.

\section{Robustness and Scope}

\subsection{Sensitivity to the Transmission Factor}
\label{sec:alpha}

The transmission factor $\alpha(s)$ of Equation~\ref{eq:alpha} is assumed, rather
than measured, so its influence must be quantified. The controller was trained
under the nominal $\alpha_{\mathrm{tip}} = 0.5$ and then run on plants whose true
value differs, which is what a mis-specified transmission would look like in
practice. Table~\ref{tab:alpha} reports the result.

\begin{table}[htbp]
\caption{Mean percentage of control steps in Helmholtz motion over the three starts, when
the true transmission factor differs from the nominal $0.5$ under which the
controller was trained, for each architecture on each open string. Bold marks the
best value in each row.}
\label{tab:alpha}
\centering
\footnotesize
\setlength{\tabcolsep}{4pt}
\begin{tabular}{|l|l|r|r|r|r|r|}
\hline
 & & \multicolumn{5}{c|}{$\alpha_{\mathrm{tip}}$} \\
\hline
\textbf{Str.} & \textbf{Arch.} & $1.00$ & $0.75$ & $\mathbf{0.50}$ & $0.35$ & $0.25$ \\
\hline
G & MLP    & $92.5$ & $\mathbf{94.5}$ & $\mathbf{94.5}$ & $90.1$ & $78.7$ \\
  & GRU    & $86.1$ & $\mathbf{86.3}$ & $79.9$ & $84.5$ & $\mathbf{86.3}$ \\
  & minGRU & $65.3$ & $63.3$ & $71.2$ & $\mathbf{88.0}$ & $85.6$ \\
  & SSM    & $43.3$ & $40.9$ & $37.2$ & $39.7$ & $\mathbf{64.1}$ \\
\hline
D & MLP    & $37.4$ & $\mathbf{48.1}$ & $41.1$ & $31.8$ & $24.1$ \\
  & GRU    & $83.6$ & $71.8$ & $87.4$ & $\mathbf{91.7}$ & $90.1$ \\
  & minGRU & $71.3$ & $66.7$ & $\mathbf{80.0}$ & $65.6$ & $62.8$ \\
  & SSM    & $57.6$ & $72.5$ & $84.8$ & $\mathbf{89.8}$ & $82.3$ \\
\hline
A & MLP    & $65.1$ & $69.2$ & $\mathbf{82.2}$ & $81.7$ & $79.2$ \\
  & GRU    & $36.6$ & $62.5$ & $69.9$ & $70.7$ & $\mathbf{71.6}$ \\
  & minGRU & $37.7$ & $52.4$ & $\mathbf{81.6}$ & $72.5$ & $79.8$ \\
  & SSM    & $63.3$ & $67.6$ & $76.9$ & $77.1$ & $\mathbf{78.9}$ \\
\hline
E & MLP    & $47.4$ & $44.0$ & $78.9$ & $74.6$ & $\mathbf{84.1}$ \\
  & GRU    & $47.6$ & $77.5$ & $\mathbf{90.4}$ & $84.0$ & $87.5$ \\
  & minGRU & $39.9$ & $72.6$ & $\mathbf{82.9}$ & $69.2$ & $80.5$ \\
  & SSM    & $52.8$ & $75.2$ & $76.7$ & $\mathbf{78.7}$ & $70.6$ \\
\hline
\end{tabular}
\end{table}

The response is asymmetric, and the asymmetry is the robust part of this table.
Over-estimating the loss is expensive. A bow with no loss at all,
$\alpha_{\mathrm{tip}} = 1.00$, is the worst or the second worst setting in eleven
of the sixteen combinations of string and architecture, and in those eleven, it
costs between $8$ and $44$ points against that cell's best. Under-estimating it,
on the other hand, is inexpensive, and in several cells, it is an improvement.

The reason for the expensive direction is visible in Figure~\ref{fig:closed}. Since the schedule reaches the controller through its labels (Section~\ref{sec:bowforces}), the controller raises the frog force as the stroke proceeds, compensating for a contact force which its labels expect to fall. When that fall does not occur, the
compensation becomes an overshoot into raucous motion.

What does not survive at four strings is the claim that the nominal value is
best. It is the best setting in only five of the sixteen cells, and on G, the
feedforward network is as good at $0.75$, while the gated unit is worst at the
nominal value and indistinguishable across the other four. The differences among
$0.50$, $0.35$, and $0.25$ are mostly a few points, and many of them are inside the
interpretability bar of Section~\ref{sec:architecture}. The assumption is not
free. The risk which it carries runs in one direction, and erring toward too little
assumed loss is the safer choice. Therefore, the defensible statement is the
one-sided one.

\subsection{A Disturbance the Commands Cannot See}
\label{sec:contactdist}

The disturbances used in the ablation of Section~\ref{sec:ablation} were a step in the commanded velocity and adverse initial conditions, both of which are visible in the commands themselves. A change in the contact condition is not. Here, the limiting friction is reduced at mid-stroke, as it would be when the rosin depletes or the contact heats, which moves the playable band while leaving $s$, $F_{\mathrm{frog}}$, and $v_b$ untouched. The labels never respond to such a change, so a network which responds does so through inputs which its labels did not use.
Table~\ref{tab:contactdist} reports the result.

\begin{table}[htbp]
\caption{Percentage of control steps in Helmholtz motion on the A string
after a mid-stroke reduction of the friction characteristic, for the full controller and for
the same controller with its four string-state inputs frozen, scored over
the steps following the onset and over the whole stroke.}
\label{tab:contactdist}
\centering
\footnotesize
\setlength{\tabcolsep}{4pt}
\begin{tabular}{|c|c|c|c|c|}
\hline
 & \multicolumn{2}{c|}{\textbf{after onset}} & \multicolumn{2}{c|}{\textbf{whole stroke}} \\
\hline
\textbf{Friction} & \textbf{Full} & \textbf{Frozen} & \textbf{Full} & \textbf{Frozen} \\
\hline
$100\%$ & $100.0$ & $90.9$ & $94.8$ & $70.7$ \\
\hline
$70\%$ & $100.0$ & $97.0$ & $94.8$ & $74.1$ \\
\hline
$55\%$ & $100.0$ & $60.6$ & $94.8$ & $53.4$ \\
\hline
$40\%$ & $80.6$ & $87.9$ & $83.6$ & $69.0$ \\
\hline
\end{tabular}
\end{table}

The answer depends on which window is scored, and at twenty seeds, that
dependence is itself the finding.

Over the whole stroke, the full controller leads at every reduction, by between
$14.7$ and $41.4$ points. Over the steps following the onset, it leads clearly at
two reductions, ties at a third by the four-point bar, and trails at the fourth by
$7.2$ points. Therefore, the same episodes support opposite conclusions according
to the interval chosen, and neither is the obvious choice. The disturbance is a
persistent change of condition, rather than a transient from which to recover,
which argues for the whole stroke. On the other hand, the post-onset window is
what isolates the response to it, which argues for that window.

Retraining under twenty seeds settles which of those readings is statistically
supported, and the two do not agree. Scored over the whole stroke, the gap is
$14.7 \pm 15.0$ points for the minimal gated variant, at $p = 0.0003$, and
$18.5 \pm 17.8$ for the gated recurrent unit, at $p = 0.0002$. Scored over the
steps after the onset, the same twenty fits give $18.4 \pm 24.6$ for the minimal
variant, which is still significant at $p = 0.0034$, and $3.8 \pm 35.4$ for the
gated unit, which is indistinguishable from zero at $p = 0.64$. The feedforward
network reaches significance on neither window, at $p = 0.16$ and $p = 0.27$.

For the gated recurrent unit, the choice of window decides whether the effect exists at all. That is the same disagreement which the
single-episode table above shows, now measured with twenty retrainings behind it,
rather than one.

\begin{table}[htbp]
\caption{Network minus lookup on the A string at twenty seeds, in points of Helmholtz
fraction, for each friction level and each scoring window. A positive entry favors
the learned controller. The lookup is seed-independent by construction, and it is
run once per onset. Bold marks entries which differ from zero at $p < 0.05$ by a one-sample $t$-test over the twenty seeds. Their standard errors run from $0.25$ to $6.68$ points, apart from two entries which are zero on every seed. The architectures are those of the grid family, with the abbreviations of Table~\ref{tab:ablation}, and not the single sklearn fit of Table~\ref{tab:contactdist}, which is the early-stopped feedforward baseline of Section~\ref{sec:training}.}
\label{tab:lookupcontact}
\centering
\footnotesize
\setlength{\tabcolsep}{4pt}
\begin{tabular}{|c|c|c|c|c|c|c|}
\hline
 & \multicolumn{3}{c|}{\textbf{after onset}} & \multicolumn{3}{c|}{\textbf{whole stroke}} \\
\hline
\textbf{Fric.} & \textbf{MLP} & \textbf{GRU} & \textbf{minGRU} & \textbf{MLP} & \textbf{GRU} & \textbf{minGRU} \\
\hline
$1.00$ & $\mathbf{-4.6}$ & $-1.1$ & $-2.0$ & $\mathbf{-6.6}$ & $+0.2$ & $-1.1$ \\
\hline
$0.70$ & $-4.1$ & $+0.0$ & $+0.0$ & $\mathbf{-6.0}$ & $\mathbf{+0.9}$ & $+0.1$ \\
\hline
$0.55$ & $\mathbf{-41.0}$ & $\mathbf{-7.0}$ & $\mathbf{-17.5}$ & $\mathbf{-27.4}$ & $-3.2$ & $\mathbf{-10.1}$ \\
\hline
$0.40$ & $\mathbf{+44.9}$ & $\mathbf{+20.1}$ & $\mathbf{+38.0}$ & $\mathbf{+22.5}$ & $\mathbf{+12.6}$ & $\mathbf{+22.5}$ \\
\hline
\end{tabular}
\end{table}

Table~\ref{tab:lookupcontact} sets the same twenty fits against the map lookup
instead, which is a different question from the one above. That question asks what
freezing the inputs costs, whereas this one asks whether reading the string beats
consulting an exhaustive table of steady states. Both are answered at twenty
seeds, and neither is answered by the other.

If we read down the friction column, the two families of architecture behave
differently. The feedforward network simply crosses over once. The lookup leads it at $1.00$, $0.70$, and $0.55$, significantly except after the onset at $0.70$, at $p = 0.059$, and only at $0.40$ does the network lead the lookup. The two recurrent architectures are not monotone. After the onset, they cannot be distinguished from the lookup at $1.00$, at $p = 0.10$ and $0.11$, and they match it on every seed at $0.70$. They lose to it at $0.55$, by $7.0$ points for the gated recurrent unit at $p = 0.023$ and $17.5$ for the minimal variant at $p = 0.0059$, and they beat it at $0.40$, by $20.1$ and $38.0$. Over the whole stroke at $1.00$ and $0.70$, only the gated unit's $+0.9$ at $0.70$ differs from zero, at $p = 0.0021$. Therefore, the
crossover is a single one for a memoryless controller, and a reversal followed by
a crossover for one which carries state.

That reversal is itself window-dependent for the gated recurrent unit, in the same way as the full-against-frozen contrast above. The minimal variant shows it on both windows. The gated recurrent unit shows it only after the onset, since over the whole stroke, its figure at $0.55$ is $-3.2$, with a standard error of $1.7$, at $p = 0.076$. A reader who chooses the whole-stroke window sees no significant reversal on the gated recurrent unit. One reading,
offered as a reading and not as a measurement, is that an intermediate change
moves the plant enough to disturb a controller which is watching the string, but
not enough to invalidate a table which is not. The lookup's inability to react is
then briefly an advantage. At $0.40$, it stops being one.

Repeating the same comparison on all four strings, using the board family rather than the grid family on which Table~\ref{tab:lookupcontact} is built, shows where the learned controller overtakes the tabulation. It does not show that the oscillation inputs are the reason. With its four string-state inputs frozen, the network overtakes the failing tabulation by the same margin, at $74.6$ against $41.9$, and freezing costs about ten points only where the tabulation holds. The advantage lies in the learned map from stroke position and commands, which is consistent with the supervision property of Section~\ref{sec:training}. In the four-string comparison, every stroke starts from A's in-band pair, $1.6$~N and $0.3$~m/s, rather than from the in-band pair of its own string. The friction falls at one of six stroke positions, from $s = 0.25$ to $0.50$ in steps of $0.05$, and each seed's score is the mean over the six. At $0.40$, the network leads on D, A, and E, by $14.9$ to $59.6$
points after the onset, depending on the architecture. On G, it loses on all three
architectures and on both windows, by $8.6$ to $14.1$, with between two and eight
seeds of twenty positive, and with every cell significant at $p < 0.01$. That is a
clean reversal on one string, and not a noisy cell.

\begin{table}[htbp]
\caption{The lookup's own Helmholtz fraction after the onset, per string and friction
level. The last two columns extend the grid below the range used elsewhere in this
paper, and they were fixed before they were run. The tabulation's score is not
monotone in the friction retained, and $0.40$ is a local minimum on D, A, and E.}
\label{tab:lookuparm}
\centering
\footnotesize
\setlength{\tabcolsep}{4pt}
\begin{tabular}{|c|c|c|c|c|c|c|}
\hline
\textbf{String} & $1.00$ & $0.70$ & $0.55$ & $0.40$ & $0.25$ & $0.10$ \\
\hline
G & $100.0$ & $100.0$ & $59.6$ & $96.0$ & $94.0$ & $60.7$ \\
\hline
D & $58.8$  & $96.0$  & $90.6$ & $26.7$ & $77.4$ & $30.0$ \\
\hline
A & $100.0$ & $100.0$ & $98.9$ & $41.4$ & $94.0$ & $58.6$ \\
\hline
E & $100.0$ & $100.0$ & $98.4$ & $22.8$ & $96.6$ & $26.1$ \\
\hline
\end{tabular}
\end{table}

Table~\ref{tab:lookuparm} explains the exception, rather than excusing it. At the
four reported factors, the lookup collapses on D, A, and E at $0.40$, to between
$22.8$ and $41.4$, and on G, it does not. It dips at $0.55$, and it recovers to
$96.0$. Therefore, the learned controller is not beating a collapsing table on G,
since the table does not collapse there. G also inverts the milder reduction,
giving the network a $12.0$-point lead at $0.55$, where A gives the lookup $7.7$.
Both are read off the four-string board, rather than the grid family of
Table~\ref{tab:lookupcontact}, which puts the same A cell at $7.0$.

Stated that way, the relation could be a coincidence across four strings at one
factor, so it was put to a test which the relation could fail. The test is a
regression of the controller's own score on that of the tabulation, over the four
friction factors reported here. If the controller simply inherits the competence
of the tabulation, the slope is one. If it degrades more slowly than the
tabulation does, the slope is below one, and the two curves must cross. The fitted slope is $0.319$. The three architectures share each tabulation score, so the standard error is $0.082$ with one point per string and friction level, rather than the $0.065$ of the forty-eight cells. Either way, the slope lies more than eight standard errors below one, and the fitted lines cross at a tabulation score of $88$. That crossing is the
overtaking threshold, estimated rather than asserted.

The Helmholtz fraction cannot exceed $100$, and in eighteen of these cells, the
tabulation sits exactly there, where the controller cannot overtake it, by
arithmetic. Those cells could bias the slope below one on their own, so the fit
was repeated without them. The slope then moves to $0.224 \pm 0.093$, further from
one rather than closer, so the saturation was attenuating the effect and not
producing it. The D string carries no saturated cell at all, and its slope of
$0.180$ is four standard errors below one, with no exclusion needed.

The levels behind it need no model. Over the fifteen cells in which the
tabulation scores below $60$, it averages $41.9$ and the controller $74.4$. Over
the thirty-three in which it scores $90$ or above, it averages $98.2$ and the
controller $90.9$. Both halves of the claim are in those four numbers, and the
same slope is below one on every string taken separately, from $0.180$ on D to $0.627$ on G, by between $3.9$ and $10.5$ standard errors.

Two cautions bear on the slope. A learner which behaves as the teacher plus a
perturbation would also give a slope below one, since the perturbation costs where
the teacher is perfect and may help where it fails, and no null controller of that
kind is scored here. The tabulation's score also enters from a single run per onset,
and error in a regressor attenuates a least-squares slope toward zero. What the
slope locates is where the controller overtakes the tabulation, and not why.

\begin{table}[htbp]
\caption{The G string, ordered by the tabulation's own score, rather than by how much
friction was removed. The gated recurrent unit overtakes in exactly the two conditions in which the tabulation fails, and those two are not adjacent in
friction.}
\label{tab:gcells}
\centering
\footnotesize
\begin{tabular}{|c|c|c|c|}
\hline
\textbf{Friction} & \textbf{Tabulation} & \textbf{Controller} & \textbf{Margin} \\
\hline
$0.55$ & $59.6$  & $71.6$ & $\mathbf{+12.0}$ \\
\hline
$0.10$ & $60.7$  & $67.2$ & $\mathbf{+6.5}$ \\
\hline
$0.25$ & $94.0$  & $91.6$ & $-2.4$ \\
\hline
$0.40$ & $96.0$  & $82.2$ & $-13.8$ \\
\hline
$1.00$ & $100.0$ & $95.3$ & $-4.7$ \\
\hline
$0.70$ & $100.0$ & $98.5$ & $-1.5$ \\
\hline
\end{tabular}
\end{table}

Table~\ref{tab:gcells} is the clearest form of the result, and it is on the string
which appeared to be the counterexample. Two further friction factors, $0.25$ and
$0.10$, were fixed before they were run, and they were added below the grid used
elsewhere. If G's six conditions are ordered by the tabulation's score, rather than
by friction, the controller overtakes in exactly the two where the tabulation
fails. Those two are $0.55$ and $0.10$, which are not adjacent, since $0.40$ and
$0.25$ lie between them and the tabulation is healthy in both. No ordering by how
much friction was removed produces that column, since the score of the tabulation
is not monotone in friction on any string. What the controller tracks is whether
the tabulation still describes the plant, and not how far the plant has moved.

No mechanism is claimed for the non-monotonicity itself, since six sampled factors
do not determine the shape between them.

One string is not accounted for. D's tabulation is the weakest of the four even
when undisturbed, scoring $58.8$ at full friction, where the other three score
$100.0$, and the network's lead on D is the largest of those reported above, at
$59.6$ points. Part of that lead may therefore be a statement about D's map,
rather than about the disturbance, and this paper does not separate the two. The
$0.55$ and severity readings above are readings, whereas this one is an
acknowledged gap.

The following is offered as a reading, and not as a measurement. The four friction
factors were chosen on A, and they are not a shared scale across strings which sit
at different places on the Schelleng diagram, so the G column may be sampling a
different part of the same curve, rather than describing a different curve.

No cell of the twenty-four separates the two fit families at $p < 0.05$ by Welch's $t$-test over the seeds. The three largest differences, all at $0.40$, lean toward the board family as the more conservative one, by between $4.7$ and $5.3$ points, with standard errors of $3.3$ to $6.2$.

Two things limit what follows from this. The whole-stroke window is the less
specific of the two, since it includes the steps before the onset. There, the
frozen variant already differs from the full one, because its inputs are held from
the beginning of the stroke, rather than from the disturbance. So part of that gap
is not a response to the disturbance at all, and part of its extra stability is
only the effect of averaging over more steps. Furthermore, the per-seed sign still
turns over, in sixteen of the sixty whole-stroke cells, so what is established is a
mean, and not a property of every fit.

What the experiment does establish is narrower than the table suggests, and wider
than nothing. Freezing the string-state inputs measurably lowers a recurrent controller's Helmholtz fraction across a stroke in which the contact has changed, on two architectures at twenty seeds. Whether it lowers the specific response to the change, measured only after the onset, is established for the minimal variant, and not for the gated one. Since the labels do not depend on those inputs, this measures how a network uses them, and not a feedback law which it was trained to learn.

\subsection{Variance Across Seeds and Initial Conditions}
\label{sec:variance}

The closed-loop figures of Tables~\ref{tab:results} and~\ref{tab:baselines} are single episodes. To separate the training
noise from the effect of the initial condition, the policy was retrained under
five seeds, and each was evaluated on the same ten randomized starts. The grand
mean is $77.5\%$ of the control steps in Helmholtz motion, with a $95\%$ confidence
interval of $\pm 1.6$ points on four degrees of freedom, and all fifty strokes reach
the tip. The spread of the ten per-start means is $19.4$ points, against $1.3$
points for the five per-seed means, a ratio of $14.9$.

That ratio is a statement about an average of ten starts, and it does not transfer
to any single figure. A grand mean over ten starts cannot suppress seed variance,
since the seed effect of a fit is shared across all ten of its starts. It does not
need to here. Decomposed, that effect is under a point, and what the mean
suppresses is the much larger start-to-start term. Measured at twenty seeds, the
same architecture gives per-cell standard deviations with a median of $10.3$
points on a three-start mean, and of $17.8$ on the contact-disturbance gap of
Section~\ref{sec:contactdist}. Averaging over starts conceals a source of
variation which a single-episode number carries in full, so the per-condition
figures in this paper should be read as single draws, and not as properties of
those conditions.

\subsection{Reinforcement Fine-Tuning}
\label{sec:rl}

In Section~\ref{sec:capacity}, we located the ceiling in the supervision, which
suggests optimizing against the simulator instead of against the teacher. The
output layer was fine-tuned by a {\it cross-entropy method} on the reward,
\begin{equation}
J = 100\,\frac{\#\{m : r_m = r_{\mathrm{H}}\}}{M} - 60\,\bigl[\text{stroke unfinished}\bigr]
\label{eq:reward}
\end{equation}
where $r_{\mathrm{H}}$ is whichever label the classifier assigns to Helmholtz
motion, the regime defined in Section~\ref{sec:diagnostics}, so that the reward
counts the steps in that regime and carries no dependence on the numbering. The
reward is evaluated on two starts, the in-band case and the high-force corner. The
completion penalty is necessary. Without it, the optimum is to creep at a low bow
speed, holding Helmholtz motion while never reaching the tip, which is the same
degenerate behavior into which the map lookup falls in
Section~\ref{sec:baselines}. Only the $66$ output-layer parameters were searched,
which keeps a derivative-free method tractable and leaves the learned features
intact.

Repeated for four architectures on each of the four strings, the fine-tuning does
not improve on the policy from which it starts. The mean change over the sixteen
cells is $-3.4$ points, with seven cells improving, at $p = 0.12$.

The mechanism is clearer than the net effect, and it is significant where the net
effect is not. The search optimizes on two starts, so eight of the ten are held
out. Across the same sixteen cells, the trained starts gain $4.0$ points on average
and improve in fifteen of sixteen, while the held-out starts lose $5.2$ and improve
in five. The paired difference between the two groups is $9.2$ points at
$p = 0.0008$. In other words, fine-tuning systematically improves the conditions
which it is shown and degrades the ones which it is not.

Individual cells are not quotable here. The search reproduces exactly for its
first five iterations and then diverges through amplification by discontinuous switching, so a change
of host moves a per-cell result by about $5$ points on average, and it reverses its
sign in a third of the cells. The group-level contrast above survives that change
of host, whereas the per-cell figures do not.

The same pattern appears again from a different architecture and a different
search. A {\it residual controller}, in which a small learned correction is added to
the lookup's command instead of replacing it, was optimized against the simulator
on six random starts, and it was evaluated on the four named ones which it never
saw. The string on which training improved most, by $41.2$ points, is the string on
which the held-out performance fell furthest. The two strings between them do not
order in the same way, so the trade is visible at the extremes, rather than across
the set. That a second method, with a different inductive bias, shows the same trade is consistent with a property of the training distribution, rather than of either search. Lifting the imitation ceiling may require optimization over a distribution of initial conditions which is wide enough to represent the task, at a simulation cost proportionally greater than was spent here. With four strings, however, the string which gains most would coincide with the one which loses most by chance one time in four. The finding of
Section~\ref{sec:capacity} is unaffected.

\subsection{Acoustic Assessment}
\label{sec:acoustic}

Every measure so far is mechanical. Since the model gives the transverse force
which the string exerts on the bridge, $T\,\partial u/\partial x$ at $x=0$, which is
what drives the body, each condition may also be rendered as audio and assessed by
descriptors from the speech and voice literature~\cite{r-m:beigi-2011}. The body itself is not modeled, so these describe the drive of the string, rather than radiated sound, and they are acoustic, rather than perceptual, measures.

\begin{figure*}[htbp]
\centerline{\includegraphics[width=\figwidewidth]{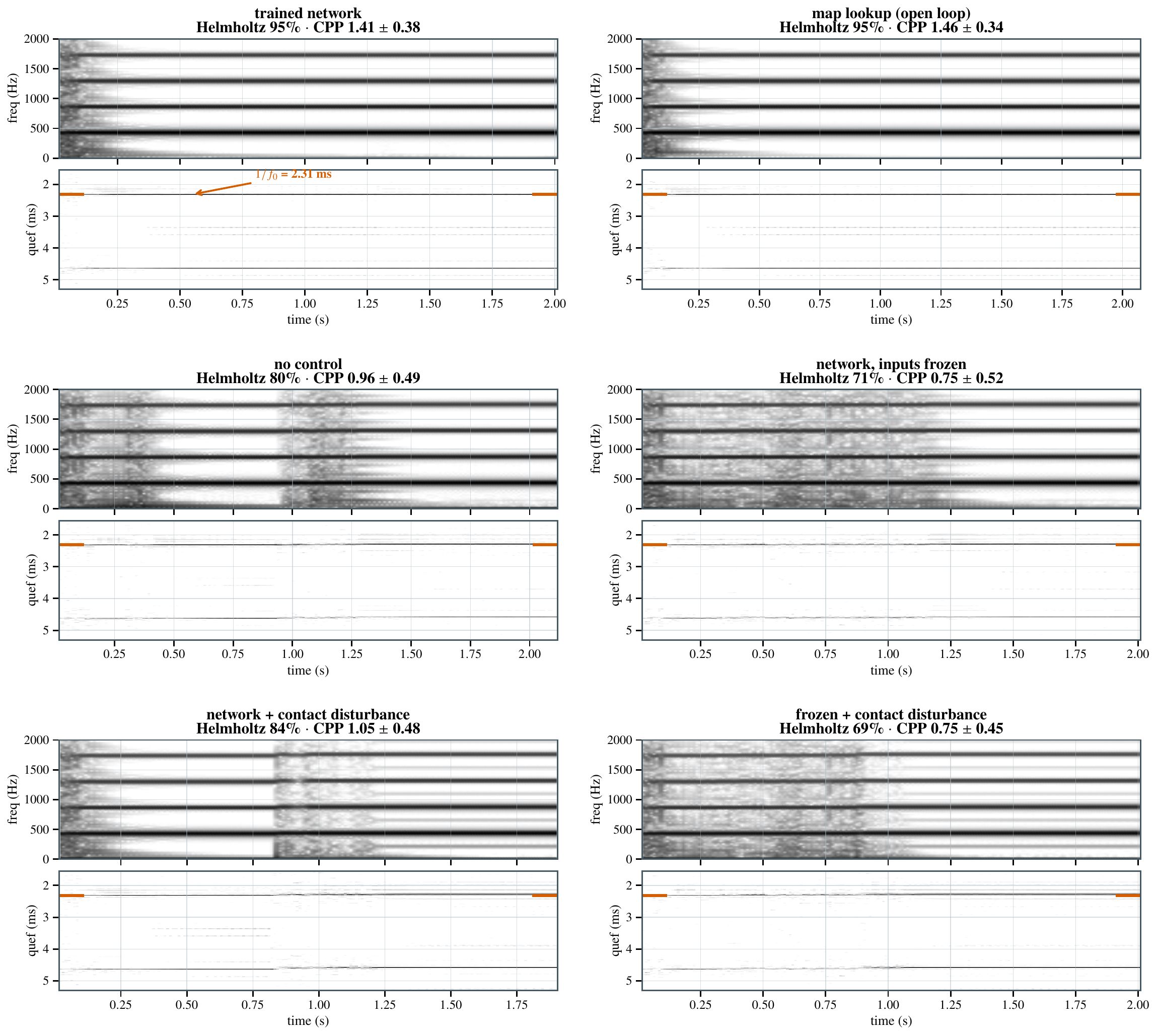}}
\caption{Short-time spectrum (upper panel of each pair) and cepstrum (lower panel) of the
bridge-force recordings. The string is an open violin A, of nominal
$A_4 = 440$~Hz, sounding at $432.4$~Hz. The recordings are generated at a sampling rate
of $48$~kHz, but the spectrogram is displayed only up to $2$~kHz, which gives a
better visual contrast, since the first four harmonics are then resolved
individually, whereas over the full band, they blur together. In the cepstrogram,
the dark band at a quefrency of $2.31$~ms, the slip period, marks the periodicity at the fundamental. It is continuous under the network and the map lookup, it dims and
wanders without control, and after the contact disturbance at about $1.0$~s, it
survives under the full controller, while breaking up when the string-state inputs
are frozen. The spectrogram shows the same event as a broadening of the harmonics
and a spread of energy between them.}
\label{fig:cepstrogram}
\end{figure*}

Figure~\ref{fig:cepstrogram} shows the short-time spectrum and the real cepstrum of
each recording, the latter computed as the inverse transform of the log magnitude spectrum~\cite{r-m:bogert-1963,r-m:noll-1967},~\cite[Ch.~5]{r-m:beigi-2011}. Writing $C(q)$ for
that cepstrum at the quefrency $q$, and $\widehat{C}(q)$ for the least-squares line
fitted to it over the search band, the {\it cepstral peak prominence}~\cite{r-m:hillenbrand-1994} is the height
of the cepstral peak above that line,
\begin{equation}
\mathrm{CPP} = C(q^\star) - \widehat{C}(q^\star),
\qquad
q^\star = \operatorname*{arg\,max}_{q_1 \le q \le q_2} C(q)
\label{eq:cpp}
\end{equation}
where the search band runs from $q_1 = 0.8$ ms to $q_2 = 6$ ms, which brackets the
periods of interest, and the frames are $50$ ms long, at a $10$ ms hop. Those are
the frames on which the tabulated values are computed, while the cepstrogram above
is displayed on a slightly shorter frame, chosen for the figure. A sharply
periodic signal puts a tall narrow peak at the period, so that the prominence is
large. Motion which is periodic, but not sharply so, leaves the peak close to the
line. Table~\ref{tab:cpp} collects the result against the mechanical measure.

\begin{table}[htbp]
\caption{Cepstral peak prominence and harmonicity of the bridge-force recordings on the A
string, given as the mean and the standard deviation across $50$~ms frames, against
the mechanical measure. These six conditions are one cell of the sixteen over
which the correlations quoted in the text are pooled.}
\label{tab:cpp}
\centering
\footnotesize
\begin{tabular}{|l|c|c|c|}
\hline
\textbf{Condition} & \textbf{CPP} & \textbf{Harm.} & \textbf{Helmholtz} \\
\hline
Map lookup                 & $1.460 \pm 0.337$ & $0.988$ & $95.0\%$ \\
\hline
Trained network            & $1.414 \pm 0.378$ & $0.981$ & $94.8\%$ \\
\hline
Network, contact disturbed & $1.050 \pm 0.484$ & $0.962$ & $83.6\%$ \\
\hline
No control                 & $0.959 \pm 0.486$ & $0.946$ & $80.3\%$ \\
\hline
Network, inputs frozen     & $0.751 \pm 0.524$ & $0.940$ & $70.7\%$ \\
\hline
Frozen, contact disturbed  & $0.750 \pm 0.454$ & $0.936$ & $69.0\%$ \\
\hline
\end{tabular}
\end{table}

The three findings do not point in the same direction.

The first is a check on the diagnostic, and it is measured over seventy-two distinct renderings. Six conditions are rendered on each of four strings, and four of the
six depend on the controller, while the map lookup and the uncontrolled stroke do
not, giving $4 \times 4 \times 4$ plus $4 \times 2$. The cepstral peak prominence
tracks the mechanical Helmholtz percentage at a Spearman correlation of $+0.803$
over seventy-two renderings, at $p = 2 \times 10^{-17}$ if they are treated as independent. Over the twenty-four means of string and condition, which do not count the four controllers as replicates, it is $+0.912$, at $p = 5.5 \times 10^{-10}$. A simpler harmonicity
measure, the fraction of spectral energy which falls on harmonics of the detected
fundamental, tracks it at $+0.324$ over the same seventy-two points. Therefore, a
regime measure defined on the stick fraction of the relative velocity predicts an
acoustic property of the resulting sound, which supports the diagnostic of Section~\ref{sec:diagnostics}, and the sharpness criterion carries
that support where the periodicity criterion largely does not. Multiple slipping
is still periodic, so a measure of whether the signal is periodic separates the
regimes less sharply than one of how sharply periodic it is. In other words, a
stick test is what an oscillation classifier needs, and a sharpness criterion is
its acoustic counterpart.

Both measures rank the six conditions of Table~\ref{tab:cpp} perfectly, at a
correlation of $+1.000$ over six points, which is not evidence of much on its own.
That agreement is a property of six points in a single cell, rather than of the
measures themselves, and pooling over all seventy-two separates them. Six
conditions are too few to distinguish a criterion which works from one which
happens to order them.

The second finding concerns the contact disturbance. Under it, the full controller
holds a prominence of $1.050$, against $0.750$ with the string-state inputs frozen,
and Figure~\ref{fig:cepstrogram} shows the band at $2.31$~ms surviving in the first
case and breaking up in the second. That ordering agrees with the whole-stroke
column of Table~\ref{tab:contactdist}, rather than with the post-onset column which
reverses, which is consistent, since both are measured over the whole recording.
It is one episode, and in Section~\ref{sec:contactdist}, we measure the same
contrast at twenty seeds, with a per-seed spread wide enough that a single episode
carries no interval, so it is offered as agreement between two whole-stroke
measures, and not as evidence for a gap.

The third finding does not favor the controller. In the steady state, the network
and the map lookup are indistinguishable, at $1.414$ and $1.460$, a difference of
about an eighth of one standard deviation, and the lookup is nominally ahead. The
acoustic evidence agrees with Section~\ref{sec:baselines}, namely, that where the
rule is right, the learned policy matches it, rather than improving on it.

The six recordings above are one operating point driven in six ways. Rendering the
playability space instead asks whether the regime labels of
Section~\ref{sec:diagnostics} correspond to a difference which an acoustic measure detects.
Table~\ref{tab:regimes} scores one fixed-command stroke from the interior of each
region of Figure~\ref{fig:playmap}, together with a point below the mapped force
range at which the bow never grips the string.

\begin{table}[htbp]
\caption{One fixed-command stroke from the interior of each region of
Figure~\ref{fig:playmap}, and a point below the mapped force range, at which the
bow never grips. The cepstral peak prominence and the harmonicity are means across
$50$~ms frames of the bridge force. Bold marks the best value in each of those two
columns.}
\label{tab:regimes}
\centering
\begin{tabular}{|l|c|c|c|c|}
\hline
\textbf{Regime} & $F_N$ (N) & $\eta$ & \textbf{CPP} & \textbf{Harm.} \\
\hline
Helmholtz           & $0.91$ & $0.891$ & $\mathbf{1.49}$ & $0.996$ \\
\hline
Multiple slipping   & $1.41$ & $0.882$ & $0.07$ & $0.245$ \\
\hline
Raucous             & $3.75$ & $0.931$ & $0.09$ & $0.175$ \\
\hline
Sub-harmonic, ALF   & $1.78$ & $0.970$ & $0.08$ & $0.047$ \\
\hline
Constant sticking   & $2.77$ & $0.989$ & $0.06$ & $0.009$ \\
\hline
Bow never grips     & $0.01$ & $0.000$ & $0.18$ & $\mathbf{1.000}$ \\
\hline
\end{tabular}
\end{table}

The cepstral peak prominence separates Helmholtz motion from every other regime by
a factor of between eight and twenty-four. Across the six controller conditions of
Table~\ref{tab:cpp}, it spans a factor of $1.9$, since those conditions all sit
near the Helmholtz band. Therefore, sampling the regimes, rather than the
controllers, widens the separation by an order of magnitude.

Harmonicity supplies a counterexample, rather than a weak correlation. The point at
which the bow never grips scores $1.000$, above the $0.996$ of Helmholtz motion,
while its stick fraction is $0.000$ and its cepstral peak prominence is $0.18$
against $1.49$. A free string rings at its natural frequency and is perfectly
periodic, so a measure of whether the signal is periodic rates it as ideal. The two
columns pick different rows.

The stick fraction alone does not order the table either. The non-Helmholtz
regimes carry stick fractions from $0.882$ to $0.989$, against $0.891$ for
Helmholtz motion, which is why the classifier of Section~\ref{sec:diagnostics} tests
the slips per period and the jitter as well. Three attacks from the
acceleration-force plane of Section~\ref{sec:guettler} are rendered alongside these,
at nearly one bow force, so that the acceleration is what differs, namely, one
clean attack, one whose first slips are delayed by fifty-one, and one which never
settles. Neither acoustic measure separates them. The cepstral peak prominence
spans $0.09$ across all three, where it spans $0.06$ to $1.49$ across the steady
regimes, and the harmonicity rates the attack which never settles at $0.988$,
against the clean attack's $0.991$, while singling out the delayed one at
$0.738$. These are single renderings with no seed, so the figures bound nothing
tightly, but the ordering is wrong, rather than noisy. What the acoustic measures
validate is steady motion, and the attack is measured by settling time below,
rather than by a mean over the transient.

\begin{figure}[htbp]
\centerline{\includegraphics[width=\figwidth]{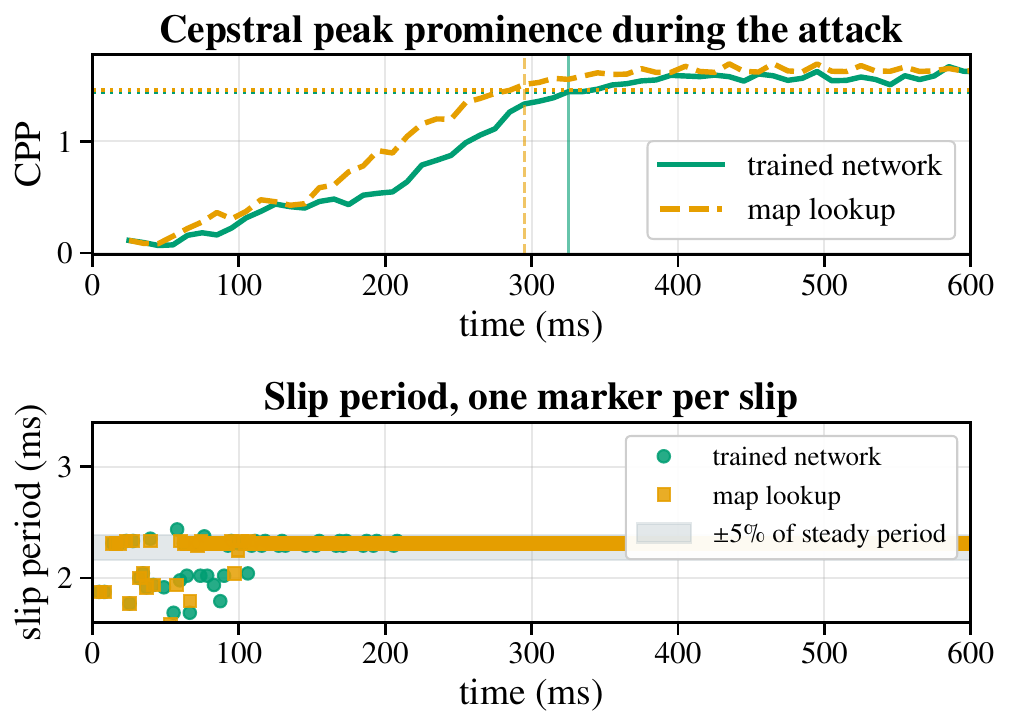}}
\caption{The attack of the feedforward baseline and the map lookup from the in-band start on the A string, measured in two ways. Top: the cepstral peak prominence of the bridge force, with dotted lines at $90\%$ of each steady value and vertical lines where each first reaches it. Bottom: the interval between successive slips, one marker per slip, against a band of $\pm 5\%$ around the steady period. The lookup reaches each threshold first, and the two measures settle on timescales differing by a factor of $2.9$ for the
network and $4.4$ for the lookup.}
\label{fig:attack}
\end{figure}

Both families of measure favor the lookup in the attack. Figure~\ref{fig:attack}
follows one stroke from the in-band start. The cepstral peak prominence rises ahead
of the network on $97.4\%$ of the frames over the first $400$~ms, and it reaches
$90\%$ of its steady value at $295$~ms, against $325$~ms. On the same episode, the
slip period settles to within $5\%$ of its final value at $67$~ms for the lookup,
against $113$~ms for the network. Therefore, both panels of
Figure~\ref{fig:attack} favor the lookup.

Repeating that comparison for the gated recurrent unit of Section~\ref{sec:architecture}, across four strings and five seeds, splits the answer by start. On the twelve cells covering the three starts which begin in or near the playable band, the lookup still reaches the acoustic threshold ahead of the seed mean in eleven. By a one-sample $t$-test on the seeds which reach it, five of those leads are significant at $p < 0.05$, and one survives Holm correction over the twelve. The mechanical measure no longer separates the two consistently. The lookup settles significantly first on D and E from the in-band start, whose attack the disturbance start shares, and the network on the G and E cold starts. After Holm correction, only the network's lead on the G cold start remains. The total variation of the commands gives nine ties of twelve, in the sense that the five seeds straddle the lookup's value.

At the high-force corner, the ordering reverses, and it reverses on every string.
The gated recurrent unit reaches the acoustic threshold on all five seeds on G, D,
and A, at $1282$, $1272$, and $1191$~ms, where the lookup never reaches it at all,
and on E, it settles at $1305$~ms, against the lookup's $1485$. The corner is one
start at which a learned controller does something which the rule it imitates
does not, and the cold start is another, where in Section~\ref{sec:reach}, we find
it reaching motion which the map cannot represent.

Two cautions belong with those figures. The acoustic transient is barely
determined for this controller. Namely, across the cells where two or more seeds
settle, the seed standard deviation of the threshold time has a median of $0.37$
of the mean and a worst case of $0.87$. Fourteen of eighty seed-runs never reach
the threshold. Furthermore, the gated recurrent unit is not a uniform improvement on
the feedforward baseline at the onset. On the A string, its mechanical transient
is faster at every start, $77$ against $113$~ms inside the band and $93$ against
$144$ from cold, while its acoustic one is slower, at $945$ against $325$~ms and
$10493$ against $7085$.

The acoustic measure reverses the ordering on every string only from the start which the teacher cannot handle. Everywhere in or near the playable band, the lookup reaches the acoustic threshold ahead of the seed mean in eleven of the twelve cells, and the mechanical measure does not separate the two consistently.

What the two measures retain is a difference of timescale, rather than of
verdict. For the feedforward network and the lookup, from the in-band and cold
starts, the mechanical measure settles within $45$ to $144$~ms, and on three of
those four, the acoustic measure settles within $185$ to $325$~ms, between $2.9$
and $4.4$ times longer. The fourth is the feedforward network's cold start, where
the mechanical measure reads $144$~ms and the acoustic one reads $7085$~ms. Where
both settle promptly, the acoustic measure is the slower by a similar factor, and
where they do not, it is the acoustic measure which runs away. The gated recurrent unit above and the disturbance start both fall on the second side of that
line. Neither range covers the high-force corner, where the mechanical measure
reaches $847$~ms and the acoustic one $1015$~ms, so a choice between them is a
claim about what a listener attends to, rather than a detail of method.
The {\it Guettler diagram}~\cite{r-m:guettler-2002-2} of this plant, in Section~\ref{sec:guettler}, maps attacks from rest at a constant acceleration and a constant force. At $\beta = 1/10$, its band of clean attacks is narrow and diagonal. It does not bound the attacks of a controller with time-varying commands, and the controller here commands increments from a bow already in motion, outside that protocol.

\subsection{Fingered Notes: Chromatic Positions from G3 to G6}
\label{sec:fingering}

Everything above is measured on open strings. In this section, we evaluate forty
positions instead, four of them open and thirty-six fingered. Each string runs from
its own open pitch up to the open pitch of the next string, seven semitones in each
case, except for E, which has no string above it and runs fifteen semitones to G6.
Therefore, the three boundary pitches, D4, A4, and E5, are each played twice, once
open and once stopped on the string below. Together, they cover every chromatic
semitone from G3 to G6, three octaves, which include every note a violinist can reach
without leaving the first four positions. Stopping the string with a finger
shortens the sounding length at an unchanged tension and mass density, so the wave
speed is unchanged, and the fundamental rises as the reciprocal of the length.

Two conventions are possible for the bow-bridge distance under fingering. The bow
may stay at a fixed absolute distance from the bridge, in which case $\beta$ rises
as the string shortens, or it may track the shortening string at a constant
$\beta$. We use the second convention here. Schelleng's limits give $F_{\max} \propto Z v_b/\beta$ and $F_{\min} \propto Z^2 v_b/(R\beta^2)$, where on rigid supports the termination resistance $R$ scales as $Z Q$ (Section~\ref{sec:schelleng}). At an unchanged impedance and quality factor, holding $\beta$ constant leaves each string's force bounds valid at every position, and it requires no change to the command set. Every result in this
section depends on that choice.

Under a rigid stop and a constant $\beta$, fingering changes nothing about the
dynamics except for the time scale. The sample rate is set proportional to the
fundamental, and the grid spacing proportional to the length. Therefore, the Courant number, the bow-point index, the wave impedance, and both damping coefficients in discrete form are all invariant. The damping is invariant only because each stopped length is given the same modal quality factors as its open string, which is an assumption of the model. The friction law is a function of
the absolute relative velocity, which is itself invariant. In other words, two
positions on one string are the same simulation, with the time axis relabeled.
When F5 is compared against G6, fourteen semitones apart on the E string, whose own
run spans fifteen from E5, the relative velocity histories agree to
$6.8 \times 10^{-13}$~m/s over sixteen operating points.

The regime classifier thresholds only quantities derived from the relative
velocity, so the classification is position-invariant within a string whenever
the analysis window is set in string periods. That condition is not incidental,
and it is the same one at which Section~\ref{sec:params} arrives from the open
strings. A window fixed in seconds does not rescale with the time axis, so a
stopped position is measured over a different number of periods than the open
string, and the classifier sees a different steady state. Maps built in that way
disagree with their own open string by as much as $27$ percent of their cells, and
the disagreement grows with the pitch ratio. Built at $140$ settling periods, as
the four-string maps of Figure~\ref{fig:fourplaymap} are, the open-string map is the
map for every stopped position on that string, and each of the thirty-six stopped
positions then agrees with it in every one of the $1089$ cells. This holds for the
rigid stop assumed throughout, in which the finger is a hard support which
reflects the wave completely. A real finger, however, is flesh, of finite
impedance, and it absorbs at the stopping point. That loss depends on how much
string lies behind the finger, and so it differs between positions, which breaks
the invariance exactly, rather than approximately. Every result in this section is
a statement about a rigid-stop instrument, and the flesh-finger model is left to
future work.

Since the plant is exactly invariant, every point of transfer loss when an
open-string controller is deployed at a stopped position is attributable to the
controller, rather than to the string. That is an unusually clean attribution, and
it is why the transfer arm is reported. Each of the forty positions was evaluated
under four architectures and four starts, at twenty trained-network seeds, which
is $12\,800$ episodes on each of the two seeded arms. With the $160$ episodes of
the lookup arm, the study comprises $25\,760$ episodes in all. The unit of
replication is the seed. The loss is taken within each seed, against that
string's own open pitch, rather than against the first note of its run, and every
figure in this section is over the episodes which reached the tip. Both
conventions are important. Scored against the first note of each run, the same
quantity reads $4.42$, $3.55$, $4.62$, and $7.40$, and those figures reproduce only
if the stalled episodes are pooled in.

So measured, the mean Helmholtz fraction falls by $4.21 \pm 0.46$ points for the
gated recurrent unit, $1.07 \pm 1.06$ for the minimal variant, $3.96 \pm 0.86$ for
the feedforward network, and $5.05 \pm 0.64$ for the state-space model. Those pool
the completed episodes equally. Pooling the positions equally instead moves the
gated unit to $4.47$, and the minimal variant on the amplitude arm from $-0.27$ to
$+0.15$, on a standard error near $1.15$, and it changes no conclusion here. The
gated unit sits outside its own standard error under both weightings, at
$0.65 \pm 0.45$ by episode and $0.55 \pm 0.47$ by position, and the other three are
inside under both. The minimal variant's figure is not distinguishable from zero.
Referring the amplitude target to the training pitch removes the loss. On that
arm, the same four figures are $0.65 \pm 0.45$, $-0.27 \pm 1.16$, $0.38 \pm 0.74$,
and $-0.05 \pm 0.50$, all within $0.7$ points of zero, and three of the four inside
their own standard error. The gated unit is the exception, at $1.4$ standard
errors, which is the largest residual that the correction leaves. Therefore, what
the transfer arm measures is not a loss of control competence across fingered
positions. It is what a displacement target trained at the open pitch does when it
is carried to a stopped one without being referred back to that pitch.

\begin{figure*}[htbp]
\centerline{\includegraphics[width=\figwidewidth]{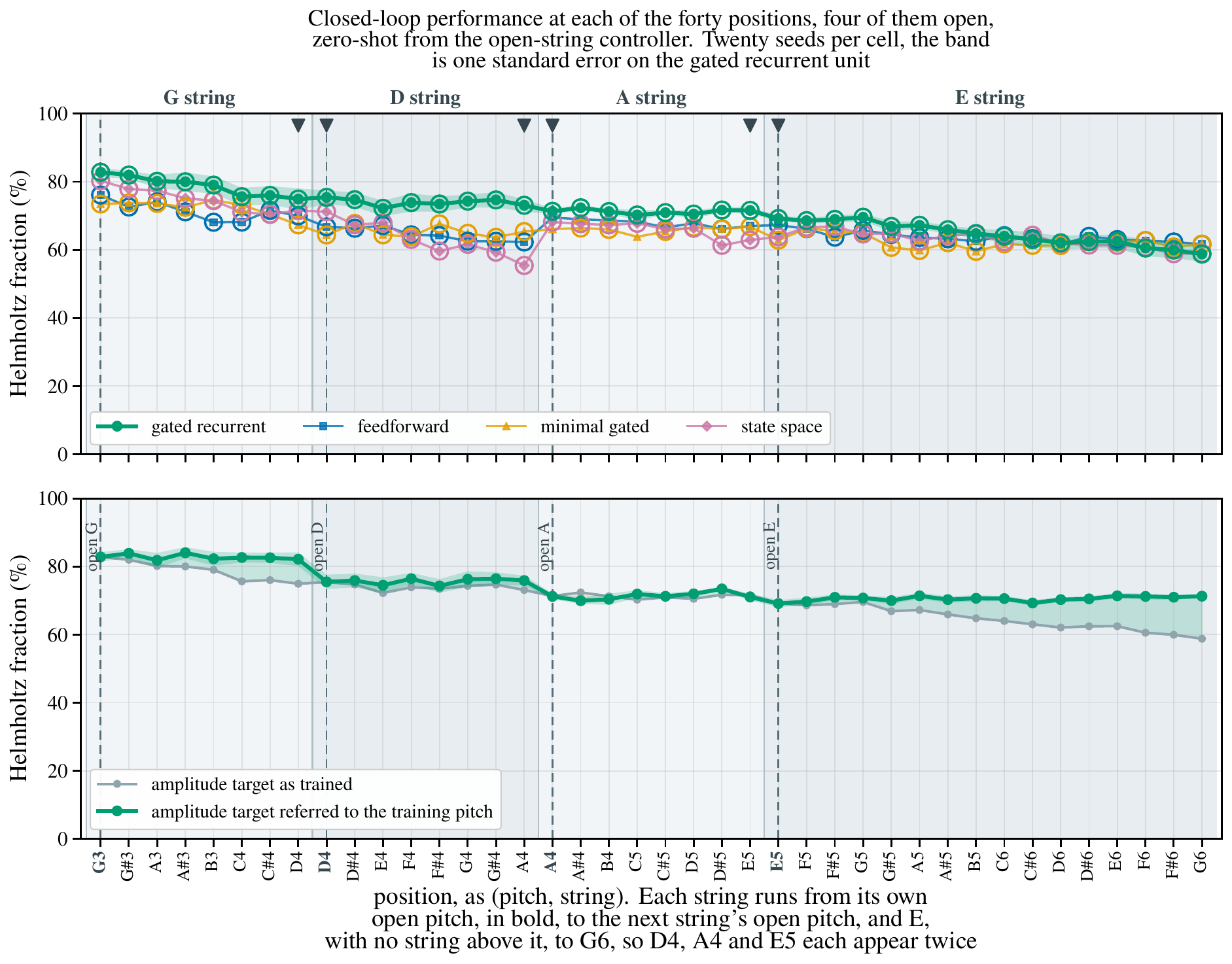}}
\caption{Closed-loop performance at each of the forty positions, four of them open,
deployed zero-shot from the open-string controller at twenty seeds per cell. Upper
panel: all four architectures, with the gated recurrent unit drawn heavier and a band of one standard error around it. Lower panel: the gated recurrent unit with the amplitude target as trained, against the same target referred to the training pitch, with the shading marking what the correction recovers. Each string runs from its own open pitch, in bold on the axis, up to the open pitch of the next string, except for E, which runs fifteen semitones to G6, so D4, A4, and E5 each appear twice, once open and once stopped on the string below, where the arrowheads mark them. A point drawn as an open circle had at least one start-and-seed cell excluded, since the stroke never reached the tip, and those cells are dropped from the within-seed mean, rather than averaged in.}
\label{fig:fingerbynote}
\end{figure*}

One number per architecture is not a summary of the sixteen string and
architecture cells behind it, and it is not their average either. Each figure
differences the four open pairs collectively against the thirty-six fingered
ones, so the fourteen-point spread between open strings enters it, and E
contributes fifteen of the thirty-six. No weighting of the per-string rows below
reproduces it. On the zero-shot arm, the state-space model runs from
$9.16 \pm 1.82$ on D to $0.84 \pm 1.32$ on E. The A string is within a point of zero for two of the four architectures. The minimal variant is negative on D, at
$-0.99 \pm 2.21$, its fingered notes scoring above its own open string. Eight of the sixteen cells differ from zero at $p < 0.05$ by a one-sample $t$-test over the twenty seeds, all with the fingered notes below the open string, and five survive Holm correction over the sixteen.
On the amplitude arm, the same table is flat everywhere, with fifteen of sixteen
within $2.4$ points of zero and thirteen of them negative.

\begin{table}[htbp]
\caption{Transfer loss per string on the zero-shot arm, given as the score at that string's
own open pitch minus the mean over its fingered notes, taken within each seed and
averaged, with the standard error across twenty seeds in parentheses. A positive
entry means that the fingered notes score below the open string. Every entry is
over the episodes which reached the tip. Excluded start-and-seed cells are counted
per position, rather than summed, since a note which stalls under three starts and
one which stalls under none are different results with the same mean.}
\label{tab:fingering}
\centering
\scriptsize
\setlength{\tabcolsep}{3pt}
\begin{tabular}{|l|c|c|c|c|c|}
\hline
\textbf{String} & \textbf{Notes} & \textbf{GRU} & \textbf{MLP} &
\textbf{minGRU} & \textbf{SSM} \\
\hline
G & $7$  & $4.57$ $(1.34)$ & $5.30$ $(2.38)$ & $1.25$ $(2.15)$ & $6.30$ $(1.36)$ \\
\hline
D & $7$  & $1.65$ $(1.32)$ & $2.55$ $(1.68)$ & $-0.99$ $(2.21)$ & $9.16$ $(1.82)$ \\
\hline
A & $7$  & $0.08$ $(0.85)$ & $1.87$ $(0.52)$ & $0.19$ $(1.19)$ & $2.54$ $(0.99)$ \\
\hline
E & $15$ & $4.86$ $(1.23)$ & $3.78$ $(1.27)$ & $0.63$ $(1.34)$ & $0.84$ $(1.32)$ \\
\hline
\end{tabular}
\end{table}

Table~\ref{tab:fingering} gives it per string.

Three pitches are played twice, once as an open string and once as the fourth
finger on the string below, which is the only matched-pitch comparison in this
work. The pitch is the same, and so is the fit, but the plant is different. When
the difference is taken within seed and architecture on the amplitude arm, eleven
of the twelve cells are positive, and the shorter string wins. D4 stopped on G, at
a sounding length of $0.2202$~m, beats D4 open on D, at $0.3304$~m, the full string
length to within the rounding described below, by between $6.64 \pm 2.76$ and
$9.85 \pm 3.04$ points across all four architectures. The comparison is read on the
amplitude arm, since the zero-shot arm carries the amplitude bias, which is what
puts A4 at $-7.20$ and $-12.76$ there. Nothing about D4 on the zero-shot arm is
resolved at twenty seeds, where standard errors of $3.7$ to $4.1$ sit on
differences of $0.3$ to $3.2$, and it is not quoted.

The three duplicated pitches are also where the assignment of a note to a string
stops being arbitrary. The chromatic chain is anchored at $196.0$~Hz, while the
open pitches are rounded to $294$ and $660$, so D4 and E5 fall $1.9$ cents below
their own string's nominal open value, and A4 falls just above it. Therefore, a
rule which places each pitch on the highest open string at or below it pushes D4
and E5 down to the string below and leaves A4 where it is, which is an artifact of
rounding, rather than a statement about fingering.

Figure~\ref{fig:fingerbynote} shows the structure which those pooled figures hide.
Performance does not decay smoothly with pitch. It is highest on the lowest string,
and it settles into a band between roughly $60$ and $75$ percent from the middle of
the range upward, with the spread between architectures much wider at the bottom
than at the top. At one episode per cell, the feedforward network appeared to
collapse on the D string, below $50$ percent across five consecutive semitones, and
that excursion does not survive twenty seeds. On the same eight D-string
positions, it runs between $62.3$ and $67.1$. Therefore, the largest
per-architecture excursion in the single-episode figure was a property of the
draw. Failures to reach the tip are concentrated in the upper octave, where the
four starts increasingly disagree. Of $12\,800$ zero-shot episodes, $1\,402$ did not
complete, and the cold start accounts for $936$ of them, while the E string
accounts for $794$. The in-band and high-force starts together account for $78$.
The lookup baseline states it most cleanly, since $32$ of its $40$ pairs stall from
cold, and nothing else stalls at all. That concentration on the cold start appears
in a population which was not built to test it. For the lookup and the network
alike, it is the slow bow speed against the step limit which
Section~\ref{sec:baselines} describes, since their stalled strokes average about
$0.03$~m/s. They differ in what the string does at that speed. The lookup holds
Helmholtz motion there, on the three strings where it stalls, while the network
holds it on about half of its steps and spends almost all of the rest in multiple
slipping or raucous motion. On G, where the lookup never stalls, the network's
stalls hold the force at its $0.40$~N floor, mostly at the cold-start speed, where
we find the Helmholtz target out of reach in Section~\ref{sec:coldplayable}, and
they spend most of their steps in multiple slipping.

The G-string loss has a single identified cause. The amplitude feature supplied to
the controller is the logarithm of a target displacement over the measured
displacement, and the target is a fixed displacement. Under Helmholtz motion, the bow point sticks to the bow for a fraction $1-\beta$ of each period, so it moves $(1-\beta)v_b/f_0$ peak to peak. The product $A f_0$ of the amplitude and the fundamental is then the velocity $(1-\beta)v_b$, independent of the pitch, so the amplitude itself falls as the reciprocal of the fundamental. A target fixed as a displacement, $A^\star$, therefore biases $\Delta A$ by the logarithm of the frequency ratio, whatever the amplitude scale. The bias reaches $+0.90$ over the three octaves which the design spans. Within a
single string, which is where the controller transfers, it is $+0.18$ on G, D, and
A, each of which spans seven semitones, and $+0.38$ on E, which spans fifteen. In
other words, the controller is told that the note is too quiet when the motion is
ideal.

Three couplings to pitch are candidates, namely, the amplitude target, a
pitch-error clip which saturates at high positions, and the number of string
periods per control step. Over the forty positions on four strings, paired within
seed, referring the amplitude target back to the training pitch is worth
$3.56 \pm 0.54$ points on the gated unit, $3.58 \pm 0.73$ on the feedforward
network, and $5.09 \pm 0.58$ on the state-space model. On the minimal variant, it is
$1.34 \pm 0.89$, which is not resolved, so this is a result about three of the four
architectures, and not all four.

One number per architecture hides a spread across the strings as wide as the
spread across architectures. The gated unit gains $0.28 \pm 0.48$ points on A and
$6.36 \pm 1.32$ on E. The state-space model gains between $4.63$ and $6.84$ on every
string. The minimal variant's pooled $1.34$ is $3.32 \pm 1.70$ on G and near zero on
the other three. Therefore, how much the amplitude reference is worth depends on
the string at least as strongly as on the network which reads it, which is the
same shape as the ablation of Section~\ref{sec:ablation}.

A five-arm ablation on the G string separates the other two. Over eight positions,
four architectures, and twenty seeds, paired within each architecture and seed,
since every arm reuses the same fit, correcting the amplitude target recovers
$4.76 \pm 0.78$ points of transfer loss on that string. Correcting the pitch-error clip returns $-0.24 \pm 0.48$, at $p = 0.62$, and holding the periods per control step constant returns $0.21 \pm 0.62$, at $p = 0.74$. Their $95\%$ intervals, $[-1.19, +0.71]$ and $[-1.01, +1.42]$, exclude an effect of $1.5$ points in either direction, against the $4.76$ of the amplitude target. Correcting all
three together recovers $4.21 \pm 0.92$ against $4.73$ for the sum of the three
singly, a difference of $-0.52$ which is inside that interval, so the three do not
interact, and the amplitude term is the whole effect. An earlier single-seed
ablation put the clip at $-1.45$ points and the periods term at $-2.02$. Both fall
outside the intervals measured here, both came from one fit, and both overstated
the harm.

The strongest form of the evidence is not an ordering, since the ordering is not
resolved. Namely, the three largest losses and the three largest recoveries differ
by less than one combined standard error, $0.2$ to $0.6$ between neighbors. The
evidence is rather that the correction removes each loss by its own amount. After
it, the three architectures which lost ground are left at $-0.01 \pm 1.01$,
$+0.40 \pm 1.16$, and $+0.05 \pm 0.84$ points, each within $0.3$ standard errors of
zero, and the minimal variant, which did not clearly degrade, is overcorrected to
$-2.07 \pm 1.66$. The baseline losses on G are $6.30$ points for the state-space
model, $5.30$ for the feedforward network, $4.57$ for the gated recurrent unit, and
$1.25$ for the minimal variant. Applying the correction returns $6.32$, $4.91$,
$4.52$, and $3.32$ points. Three of the four are resolved, at $p = 0.0003$,
$0.015$, and $0.0014$, and the minimal variant is not, at $p = 0.066$, which is the
architecture whose baseline loss was never distinguishable from zero.

Applied across the thirty-seven notes of the earlier single-episode design, one
episode per cell, the correction moves the mean from $67.41$ to $70.44$, a gain of
$3.02 \pm 0.76$ at $p = 7.1 \times 10^{-5}$. That population is not the
forty-position one used above, and its unit of replication is the episode, rather
than the seed. It was tuned on G, so the other three strings are out of sample, and
it replicates on D, at $5.26$ with $p = 0.017$, but not on A or E, at $1.14$ and
$1.46$ with $p$ above $0.2$. That is what the mechanism predicts, since A and E
barely degraded, and there is little to recover. The correction improves $289$ of
$528$ cells, which is $54.7\%$, namely, a small consistent shift on top of large
per-cell variation, rather than a uniform improvement.

The correction is specific to pitch changes within one string. Across strings,
the impedance, the force floor, and the playable region genuinely differ, and
pooled over every ordered pair of distinct strings, at four architectures and
twenty seeds, $960$ cells in all, the same rescaling returns $+0.66 \pm 0.50$
points, which is not distinguishable from zero.

That null is a cancellation, and not an absence, which is the more useful
statement. By ordered pair, the effect runs from $-7.11$ points, moving a
controller from G to E, to $+10.72$, moving one from E to G. Its standard deviation
across the twelve pairs is $5.24$, and several of them sit four standard errors from
zero in both directions. The correction does a great deal across strings. It helps
substantially in some directions, it hurts substantially in others, and it averages
away.

Therefore, the rescaling is not inert on a different plant, and the reason to keep
it within a string is that its sign is not predictable across strings. The effect
correlates with the pitch distance moved, counting a move to a lower string as
positive, at $+2.68$ points per octave with a correlation of $+0.57$. That fits the two extremes
and leaves two pairs of the same octave shift carrying opposite signs, so it is a
direction worth naming, and not a mechanism.

The gated recurrent unit of Section~\ref{sec:architecture} transfers to stopped positions without modification. It is not the least affected of those which do lose ground. The three
lose $3.96$, $4.21$, and $5.05$ points, the feedforward network the least of them,
and the two smallest differ by $0.25$ points against a combined standard error of
$0.97$. The amplitude bias is present for every architecture, and no ordering among
the three is resolved.

\section{Discussion}

\subsection{What a Provably Invariant Plant Buys}

The fingered results of Section~\ref{sec:fingering} are worth separating from the
rest, since they rest on a different kind of argument. Everywhere else in this
paper, a controller is compared against another controller, or against itself
under a changed condition, and the comparison carries the variance of whatever the
plant did on that episode. Under a rigid stop at a constant bow-bridge fraction,
however, the plant does not vary at all. Stopping the string rescales time and
changes nothing else, to $6.8 \times 10^{-13}$~m/s over sixteen operating points. In
other words, two positions on one string are the same simulation relabeled.

This converts a transfer experiment into a clean attribution. When an open-string
controller loses ground at a stopped position, the loss cannot be due to the
environment, since the environment is provably identical, so all of it belongs to
the controller. Forty pairs across three octaves then act as probes of one policy
against a fixed plant, and the loss turns out to have a single cause, namely, an
amplitude target specified as a displacement, in a quantity which scales with the
reciprocal of the fundamental. Referring that target back to the training pitch
leaves every architecture within $0.7$ points of zero over twenty seeds, three of
the four inside their own standard error, and the gated unit at $1.4$. On the
minimal variant, there was no resolved loss to remove, so the correction is
established on the three architectures which lost ground, and it is consistent
with the fourth. A cause which accounts for the whole of an effect, wherever the
effect was resolved, is the strongest form which this evidence could take.

The same invariance is what makes the assumption behind it consequential. A rigid
stop is a hard support. A real finger is flesh, and a termination of finite
impedance absorbs by an amount which depends on how much string lies behind it, so
it breaks the invariance exactly, rather than approximately. Therefore, every
conclusion above is a statement about a rigid-stop instrument, including the useful
one that a single playability map serves a whole string. The flesh-finger model is
the obvious next test, and it is set out in Section~\ref{sec:futurearch}, where it
makes a falsifiable prediction, rather than an open-ended extension.

\subsection{Limitations}

The supervision sets what the results mean. The labels depend only on the
stroke position and the previous commands, and no teacher or interactive relabeling
in this paper responds to the oscillation. The two searches of
Section~\ref{sec:rl} are the exception, since they optimize against the simulator,
and the fine-tuning's reward counts the steps in Helmholtz motion. Every other
result here on the string-state inputs or on recurrence describes how networks
fitted to open-loop labels use those inputs, and none of them tests a feedback law.

The map is built from rest, and Section~\ref{sec:schelleng} shows that Helmholtz
motion, once established, persists far below the force needed to establish it. The
map's lower edge is not where established motion is lost, and the cells which this
paper calls mislabeled partly reflect that difference. The map is also settled at a
window set in seconds. The controller's plant runs at $N_x = 51$, coarser than the
band-edge sweeps of Section~\ref{sec:schelleng} at $N_x = 101$, and none of the
effects near the band edges has been checked under refinement.

The closed-loop features are fixed in seconds and in absolute units. The regime is
read over the last $60$ percent of each $35$ ms step, which spans about $4$ periods
on G and $14$ on E, the frequency error is in hertz, and the amplitude target is a
displacement. A comparison across strings may reflect those constants as well as
the strings.

The entries of Tables~\ref{tab:results}, \ref{tab:baselines}, and~\ref{tab:contactdist} are single episodes, so their individual figures should be read as single draws, and not as properties of their condition.
Section~\ref{sec:variance} bounds what that costs across initial conditions, but not
across fits. A grand mean over ten starts is stable to within $1.6$ points between
seeds, while the same architecture, retrained at twenty seeds, gives per-cell
standard deviations of $4.4$ to $23.8$ points on its three-start mean, with a
median of $10.3$. When this is decomposed on the gated recurrent unit, almost none of
it is due to fits differing in overall competence. The standard deviation across
starts is $20.96$ points, across seeds $0.87$, and the seed-by-start interaction
$12.31$. In other words, fits differ in which start they handle, rather than in how
well they play, and the per-start figures say which start, since the spread across
seeds is $2.2$ points in the band and $22.3$ from cold. That is the same fact which
Section~\ref{sec:variance} reports from the other side, and both are about the cold
start. Averaging over starts hides the seed variance which a single-episode figure
carries in full, so the risk lies in both the choice of start and the fit. No null
controller of imperfect imitation, smoothed lookup, horizon-aware planner, or
classical feedback rule is scored, so the ceiling's domain is located against one
teacher, and the map and the greedy rule built on it are not separated. No
comparison against prior work is reported, for the reasons set out in
Section~\ref{sec:nocompare}. The baselines of Section~\ref{sec:baselines} are single
episodes for the same reason, though the ablation of Section~\ref{sec:ablation} is
averaged over ten paired starts. Since the training labels are themselves derived
from the playability map, the network is best understood, where that map is
correct, as a smooth, rate-limited interpolation of a lookup policy. On the cases
where both complete the stroke, it does not improve on that lookup. Where the map is
wrong, however, it does more, and in Section~\ref{sec:reach}, we measure it reaching
motion which the map cannot represent.

The string model is a single polarization on rigid terminations, with no bending
stiffness, no body resonances, no torsional motion, no bow-hair compliance, and no
finite bow width. The rigid ends leave the string's own damping as the only loss,
so the maintenance limit measured here is Schelleng's minimum for a string with no
bridge resistance, and not for an instrument. The friction law is a static
characteristic with no internal state. Hysteretic and
thermal models give measurably different transient
behavior~\cite{r-m:galluzzo-2017}, and the dynamic branch has since been given a
formulation which is passive for any choice of parameters and has a unique
solution~\cite{r-m:matusiak-2025}, and it has been combined with the contact
temperature~\cite{r-m:vanwalstijn-2026}. Neither is implemented here. The thermal
law establishes Helmholtz motion more readily~\cite{r-m:woodhouse-2003}, so the
establishment threshold may move under a different law. The transmission factor is
assumed, and although Section~\ref{sec:alpha} bounds the cost of getting its shape
wrong, it has not been fitted to a measured stroke.
The acoustic assessment of Section~\ref{sec:acoustic} is computed on the bridge
force, so it describes the drive of the string, rather than radiated sound, and no
listening test was run. The contact point is fixed, so the third parameter which a
player controls is absent.

The quality factor is held fixed, the same on all four strings and at every stopped
position. That makes the four strings one non-dimensional problem, and it makes
their agreement a check on the code, rather than four findings. The establishment
threshold does not move with $Q$ at the measured losses, but the maintenance limit
goes as its inverse, so per-string values of $Q$ would shift the maintenance limit
from string to string.

Both limits are those of the discretized model. The constant $\kappa$ is not
converged, and it changes by up to about eleven percent under refinement. The
maintenance limit is measured on the A string alone, at the published grid and, on
six cells, at one refined grid. Its exponent on $v_b$, $+1.66$, is steeper than
Schelleng's and has no explanation here. The cells which carry that departure are
those with the largest $v_b/\beta^2$, and the two of them measured at the refined
grid are the two cells whose limit moved. At losses below the measured ones, the
form of the establishment threshold is not measured, since it changes with the
Courant number. Section~\ref{sec:schelleng} gives the dependence of both limits on
the grid and the time step. At a Courant number below one, the scheme is also
dispersive, so its upper partials fall flat of harmonic, with the opposite sign to
the inharmonicity which bending stiffness would add.

The bow does not reverse. The stroke position advances as
$s \leftarrow s + v_b \Delta t / L_{\mathrm{bow}}$, with $v_b$ confined to positive
values, so $s$ never decreases, and every episode in training, validation, and test
is a single down-bow from frog to tip. Therefore, a note in this work is exactly one
bow stroke, and its length is bounded by the hair divided by the bow speed. At the
fastest permitted speed, a full stroke takes $1.55$ s, which is a dotted quarter at
a quarter-note pulse of $60$. Anything shorter lies outside the present model,
which is every note at or below the quarter, and so does any note which outlasts
one bow. {\it D\'etach\'e} and {\it martel\'e} lie outside it for a second reason,
since they reverse the bow. This bounds what the controller results claim, rather
than qualifying them, since the regulation problem solved here is the interior of a
stroke, and not the joins between strokes.

The stopped-string results of Section~\ref{sec:fingering} assume a rigid
termination at the finger. A real finger, however, is flesh, of finite impedance,
so it absorbs at the stopping point, and the loss depends on how much string lies
behind it. That would break the exact scale invariance on which those results
rest, so the conclusion that one map serves a whole string is specific to the rigid
stop. It is the same class of omission as the rigid bridge, and the two are in fact
the same problem, at opposite ends of the sounding length.

Two measurement conventions in this paper are choices, rather than properties of
the model, and both affect numbers which are reported. The playability maps and the
band-edge sweeps are analyzed over a window set in string periods, rather than in
seconds, since a window fixed in seconds gives each string a different number of
periods to settle and makes the four strings incomparable. Under the fixed-duration
convention, the playable areas rank E, A, D, G, and at equal settling, they rank in
the opposite order. What is a property of G is that it settles slowest, not that
its playable region is smallest. The bow-bridge distance is held at a constant
fraction of the sounding length, so under fingering, the bow tracks the shortening
string, rather than staying at a fixed distance from the bridge. At the highest position reported, it sits $13.9$ mm from the bridge, outside the range of ordinary bowing.

\subsection{Future Work}
\label{sec:futurearch}

The extensions below are steps toward the playing simulation set out in
Section~\ref{sec:motivation}, and they divide into those of the plant and those of
the player. On the plant side, the rigid terminations have to go, namely, a bridge
carrying an admittance into a body and a sound post is needed at one end, and flesh
of finite impedance at the other. On the player side, the bow has to acquire an
angle and a tilt, it has to reverse, and it has to follow the dynamics of a phrase,
rather than hold one sustained tone. We take each of these in turn below, and the
controller work comes first, since it is the part which this paper can already
measure.

The clearest route to better control is to change the supervision, rather than the
architecture, and Sections~\ref{sec:context} and~\ref{sec:capacity} make the
argument. Namely, neither capacity, nor context length, nor training loss selects a
better controller, while the ceiling sits in the rule which generated the labels.
The teacher labels any state at no cost, so interactive aggregation in the manner of
{\it DAgger}~\cite{r-m:ross-2011} is directly applicable. A teacher whose labels
depend on the measured regime would give the oscillation inputs and memory a
supervised purpose. A classical feedback rule, which raises the force on multiple
slipping, lowers it on raucous motion, and holds it on Helmholtz motion, is the
baseline against which any learned feedback should be compared.
Section~\ref{sec:rl} takes a first step by optimizing against the simulator
instead, and it finds the search improving the two conditions which it is shown in
fifteen of sixteen cases and degrading those which it is not. The binding
constraint is the width of the training distribution, rather than the method, so
the next attempt should sample initial conditions across the command box at every
iteration, at a simulation cost roughly an order of magnitude above what was spent
here. A less expensive intermediate step is to improve the teacher, replacing the
greedy nearest-playable-pair rule with a horizon-aware planner over the map. The
greedy rule has no notion of finishing the stroke, which is precisely why it stalls
short of the tip from a cold start on three of the four strings. The hot-start
failure is likewise a horizon and rate-limit problem, which no change of
architecture addresses. Retraining on the properly settled map, and on a map built
from established motion, would test whether the ceiling moves with the supervision.

The case for a recurrent policy, whose margins in Section~\ref{sec:architecture}
over the feedforward network and the state-space model are not significant, would
be strengthened by a history-dependent contact model. Measured rosin friction
is hysteretic~\cite{r-m:smith-2000}, and the elasto-plastic family now offers a model
which is passive by construction and has a unique
solution~\cite{r-m:matusiak-2025}. The case would also be strengthened by a
controller which observes only the radiated sound, instead of the relative
velocity directly. A Kolmogorov-Arnold network remains worth revisiting, for its
readability rather than its accuracy. None of the bases tried improved on the feedforward network. The boundary being learned is the map's lower edge, which
is the establishment threshold of Section~\ref{sec:schelleng}, and a model whose
learned univariate functions may be plotted might expose that boundary in a form
comparable with that of {\it Schoonderwaldt, et
al.}~\cite{r-m:schoonderwaldt-2008}.

Features scaled by the Helmholtz kinematics, namely, an amplitude referred to
$(1-\beta)v_b/f_0$, a frequency error relative to $f_0^\star$, and a window set in
periods, should make transfer within a string exact under a rigid stop, and they
would separate the strings from the design constants in the ablation of
Section~\ref{sec:ablation}.

Three limits of the present study are concrete enough to be next steps, rather than
caveats. First, the bow does not reverse, so a note is exactly one stroke, and the
shortest which the controller can produce is $1.55$ s, a dotted quarter at a
quarter-note pulse of $60$. Reversal is the mechanism which makes shorter notes
possible, rather than an independent refinement. It changes the command box, the
friction regime near zero velocity, the episode definition, and the state-reset
logic at once, so the architecture comparison would have to be repeated, rather
than extended. Second, replacing the rigid finger with a flesh model, which we list
among the limitations as an assumption on which the fingered results rest, is the
natural first test of how far the invariance carries, and it yields a specific
prediction, rather than an open-ended extension. A flesh termination absorbs
differently at each position, since the length of string behind the finger
changes. The invariance should therefore fail progressively as the stopped length
shortens, and one map per string should become a family converging toward the
open-string map as the finger approaches the nut. This is falsifiable, and it is
the same modeling problem as the bridge admittance. Third, the training window was
capped at thirty control steps throughout, which Section~\ref{sec:context} shows is
the setting most favorable to the gated recurrent unit, so repeating the
comparison at a full-bow window would test that architecture where it is
weakest. Extending the context sweep itself to the remaining two strings belongs
beside that, and it is blocked by the same arithmetic, since both are dominated by
the cost of the minimal gated variant at long windows, which is the family that
the sweep exists to compare.

One architecture is worth naming, since it addresses the ceiling directly. A
residual controller adds a learned correction to the lookup's command, instead of
replacing it, so at initialization, it reproduces the tabulated rule exactly, and
the imitation ceiling does not apply to what it learns on top. A first attempt
behaves as that argument predicts where the conditions are right, gaining $9.0$
points on a held-out high-force start. It fails where they are not, since the lower
bound holds only at initialization, and it is destroyed by training on a start
distribution which does not cover the evaluation conditions. Therefore, the bound
has to be enforced, by regularizing the correction toward zero or by early
stopping on held-out starts, rather than merely initialized.

Section~\ref{sec:contactdist} suggests a different line of work. A change in the
contact condition is an unmeasured change in a plant parameter, and asking whether
a policy notices it implicitly is the harder of the two available questions. The
classical alternative is to estimate it. Namely, a {\it disturbance observer} driven
by the residual between the predicted and the measured oscillation would make the
change an observed signal, rather than a latent one. It would turn a question about
what a network can infer into one about what an estimator can track.

\begin{figure}[htbp]
\centerline{\includegraphics[width=\figwidth]{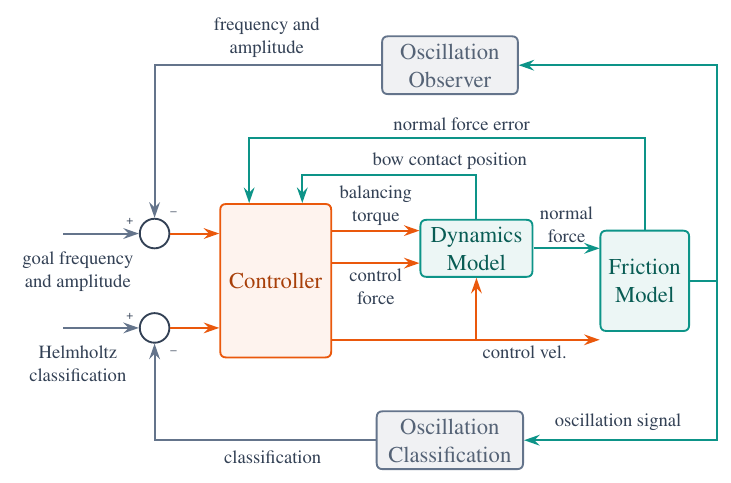}}
\caption{Prospective architecture for a controller which tracks frequency and amplitude
and commands a stabilizing torque.}
\label{fig:future}
\end{figure}

The natural extensions are three. The first makes the oscillation frequency and
amplitude true tracking targets, rather than diagnostic outputs. The second makes
the stabilizing torque an output of the controller, rather than an analytic
consequence of the commanded force. The third allows the contact point to move, so
that $\beta$ becomes a third control parameter. Figure~\ref{fig:future} sketches the
resulting architecture.

The longer aim is a complete parametric model of the instrument. The string here
terminates at rigid supports, which suppresses everything that gives a violin its
voice. Replacing those terminations with a modeled bridge, and coupling the bridge
in turn to the top plate, the sound post, and the back plate, would carry the
bowing force through the radiating structure, instead of stopping it at the
string. The bridge is the natural first step, since its admittance is what the
string actually sees, and it has been characterized as a small set of damped
resonances~\cite{r-m:woodhouse-1993-1}. The coupling itself is not new work, and
what already exists bounds what this step would add. The admittance is measured at
the bridge by hammer excitation, and the reliability of that measurement has been
tested~\cite{r-m:zhang-2014}. What a measured curve supports has been set
out~\cite{r-m:woodhouse-2012}. Such a curve reduces to a low-order passive modal
model, suited to time-domain synthesis~\cite{r-m:maestre-2013}. A bowed-string
model has already been run against a body response measured on the same
instrument~\cite{r-m:mansour-2016}. What this work would carry across the
termination is the parameter discipline of Section~\ref{sec:params}, rather than
the coupling, namely, every value from a published measurement and none fitted,
applied to the body as it is applied here to the string. The signal rendered in
Section~\ref{sec:acoustic} is precisely the input which such a body model would
take, so what is missing is the radiating structure itself, which is the larger
part of the problem. With those components parameterized, the controller developed
here would drive a synthesis model whose timbre depends on the geometry of the
instrument as well as on the bowing, so that the bow control and the instrument
model together form a parametric violin from which music may be generated
directly.

Three measurements would test the two limits further. The maintenance limit on the
other strings, and at further refinement, would confirm its scaling with $Z$ and
settle its dependence on $v_b$. A termination whose resistance does not scale with
$Z$, such as a bridge carrying a measured admittance~\cite{r-m:woodhouse-1993-1},
is the setting in which Schelleng's $Z^2$ is a prediction, and his form
predicts that the maintenance limit would then go as $Z^2$. A dynamic friction law
would show whether the establishment threshold belongs to the static
characteristic.

Extending the model across the violin family is a further step. Nothing in this
paper addresses any instrument but the violin, and the remarks below are a
direction, rather than a finding. The
smaller violins and the viola differ from the instrument modeled here mainly in
string length, tension, and wave impedance, so for those, a change of parameters
may suffice, though the measured parameters would have to be obtained. The cellos
in their several sizes, the double bass, and the contrabass are a stronger test,
and possibly not a parametric one, since bending stiffness is absent from the
scheme which we use here. It is exactly what grows on heavily wound low strings, so
a member which needs it needs a new term, rather than a new value. The same
extension would also test whether the scale invariance of
Section~\ref{sec:fingering} extends from stopped positions on one string to
instruments of different sizes.

Closing the loop on sound, rather than on mechanics, follows directly from
Section~\ref{sec:acoustic}. The descriptors are already computed there, and the
cepstral peak prominence tracks the mechanical measure closely enough to serve as
an objective in its own right, so the controller could be trained against what the
instrument sounds like, rather than against how the string moves. That would also
bear on the attack comparison left open in Section~\ref{sec:acoustic}, where the
measure which best predicts a listener's preference during the transient is not
established, and it would have to be determined by a listening test.
{\it Digital waveguide} models offer an efficient route to the audio rate which
real-time interaction requires~\cite{r-m:smith-2010}. A controller which listens and
adapts, rather than one which maintains a mechanical regime, is the arrangement
most analogous to a human player.

\section{Conclusion}

A finite-difference bowed-string model, with its friction, impedance and losses
taken from published measurement and its contact resolved implicitly, was mapped
over the bow force, the bow speed and the bow-bridge distance, with a classifier
which tests for a stick phase. The four open strings are one non-dimensional
problem, scaled in force by their impedance, so the information lies in the
dependence on $\beta$ and $v_b$, and in the constants. Driven at the midpoint of its
own band, each of the four open strings returns Helmholtz motion with a stick
fraction of $89.1\%$, against an ideal $90.0\%$. Sweeping the bow force and speed
recovers the Schelleng structure, with multiple slipping below the band and raucous
motion above it.

The upper edge of the playable band recovers Schelleng's maximum, and it is the
same whether it is approached from rest or by continuation. The lower edge splits
in two. From rest, Helmholtz motion appears only above $\kappa Z v_b/\beta$, with
$\kappa = 1.112 \pm 0.017$ at the discretization used, and raising the force slowly
lowers that little. The constant reproduces on sixty-four held-out operating
points, does not depend on the classifier's thresholds, does not move with the
quality factor at the measured losses, and changes by up to about eleven percent
under refinement. Once established, Helmholtz motion is kept far lower. The force
at which it is lost goes as $\beta^{-2}$ and as the inverse of the quality factor,
at $1.0$ to $2.7$ times Schelleng's minimum computed with the model's own losses,
and it holds under refinement of the grid. That limit is Schelleng's minimum. At the
measured losses, the establishment threshold lies $31$ to $108$ times above it, and
between the two, Helmholtz motion is kept if established and is not reached
otherwise. A map built from rest records the establishment threshold.

The model reproduces Guettler's three attack regions, and at the grid used, the
band of clean attacks closes as the bow moves from a tenth to a twentieth of the
string length from the bridge. That
comparison is against Guettler's description of the plane, and not against a
measured one, which is the weaker of the two forms which the test now takes. The
frog force which a robot commands is related to the contact force by a factor
which no free-body reading determines. Four readings of the bow disagree among
themselves in shape, magnitude, and sign by more than any of them disagrees with
the linear profile assumed here. That is why the assumption is tested by measuring
the cost of getting it wrong. Namely, a bow assumed to be lossless is the worst or
the second worst of five settings in eleven of sixteen combinations of string and
architecture.

Six architectures at a matched parameter count, on four strings at twenty seeds
each, give the gated recurrent unit the highest mean on every string, but it cannot
be distinguished from the feedforward network or the state-space model on any
string, and the feedforward network completes more strokes. Averaged over the four
starts, the gated unit's margins survive Holm correction on two strings against the
minimal variant and on none against the feedforward network or the state-space
model. From the in-band start, it leads the feedforward network on every string,
significantly on their mean, and at
thirty-seven of the forty positions, significantly when averaged over them. The
other three positions lie within one standard error of zero. Its margin over the
mean of the other five architectures is largest on the disturbance start, where
the commanded bow speed is overridden at mid-stroke, and the feedforward network's
advantage in completion is concentrated on the cold start, a start which no player
would use. The minimal gated variant, whose gates are computed from the input alone, can latch after an abrupt velocity override, since its gates cannot make the reset depend on the state itself.
Namely, eight of its twenty seeds on the E string hold below $9\%$ Helmholtz motion
for the rest of the stroke, and forcing a state reset at every step recovers all
eight. Depth compounds this on two strings and changes nothing on the other
two.

Three ways of choosing a controller were tried, and, as a direction at three seeds,
none of them works, namely, selecting by parameter count, by training-context
length, and by training loss.
Over a four-rung ladder of parameter budgets, measured on all four strings and four
architectures at three seeds each, the training loss falls in every one of the
sixteen cells. Meanwhile, the closed-loop quality does not rise. The mean falls in
fourteen of the sixteen cells, by more than four points in nine, and it rises in
two. At three seeds, however, the standard error of a change runs from $2.2$ to
$19.3$ points, and only one change, a fall of $11.7$ points for the state-space
model on D, is significant by Welch's $t$-test, at $p = 0.0062$. None survives Holm
correction over the sixteen. The mean change is $-11.1$ points, the seed spread
widens, instead of narrowing, in thirteen of the sixteen, and the effect is not
uniform across strings, running from $-27.2$ on G to $-3.4$ on A. The training loss
also falls with the length of the training context, while the closed-loop quality
moves in every direction. Therefore, a search driven by the quality of fit picks the largest model every time, although the largest model is not, in general, the better controller. That places the ceiling on the rule which generated the labels, rather than on the network which fits them.

The learned controller does not exceed that rule on steady-state regulation where
the rule is right. The rule is a strong baseline, rather than a handicapped one,
since it reads an exhaustive tabulation of the steady state which the plant reaches
from rest, at $1089$ operating points, and it is what generated the training labels
of the controller. Over forty positions on four strings, four of them open, at
twenty seeds, the lookup leads in all $160$ cells, significantly in $156$ after Holm
correction. That margin is an oracle bound, and not a baseline, since the lookup
reads the playability map for the very position at which it plays, while the
controller is transferring to it. What the controller does better is finish,
completing eleven of twelve combinations of string and start, against the
lookup's nine, and $23$ of $37$ fingered notes at the cold start, against the
lookup's $8$. It also does better in exactly one circumstance, and that
circumstance is the ceiling's own domain condition, rather than an exception to
it. Namely, the controller overtakes the tabulation wherever the tabulation fails,
and it is bounded by it wherever the tabulation holds. Regressing the score of the
controller on that of the tabulation gives a slope of $0.319 \pm 0.082$ over the friction sweep, more than eight standard errors below the unit slope which simply inheriting the teacher's competence would give, and the two cross at a tabulation
score of $88$. Where the tabulation scores below $60$, it averages $41.9$, against
the controller's $74.4$, and above $90$, the order reverses, at $98.2$ against
$90.9$. The clearest case is the G string, where the tabulation's score is not
monotone in the friction removed, and the controller overtakes in exactly the two
non-adjacent conditions in which it fails. On the undisturbed plant, the controller holds Helmholtz motion in cells which the map from rest calls unplayable, which is
consistent with the finding of Section~\ref{sec:schelleng} that established motion
persists far below the force needed to establish it. The rule's labels depend only
on the stroke position and the previous commands, so what the networks do with the
oscillation inputs is a matter of inductive bias, rather than of trained feedback.
Removing the four string-state inputs helps on G and hurts on A and E, and the
spread of that effect is $19.3$ points across strings, against $6.8$ across
architectures. Whether that belongs to the strings, or to the fixed window and
command box, is not separated here. A disturbance applied to the contact, rather
than to the command, isolates the one case which the commands cannot reveal, and
what it establishes depends on where it is scored. Over the whole stroke, freezing
the four string-state inputs separates both recurrent architectures from their full
form at twenty seeds, by $14.7$ and $18.5$ points, with $p$ below $0.001$. Over the
steps after the onset, which is the more specific measure, only the minimal variant
reaches significance, and the gated recurrent unit does not, at $3.8$ points and
$p = 0.64$. A single episode at the most severe reduction changes sign between the
two windows. For the gated recurrent unit, the choice of window decides whether the
effect exists at all. Optimizing against the simulator instead of the
teacher gives no net gain over sixteen architecture and string combinations, but it
gives a significant separation in mechanism, improving the two conditions which the
search is shown in fifteen of sixteen cases and degrading those which it is
not.

Stopping the string extends all of this across three octaves. Forty positions from
G3 to G6 were evaluated at twenty seeds each, four of them open. Under a rigid
termination at the finger, with the bow held at a constant fraction of the
sounding length, the dynamics are exactly invariant. In other words, every
coefficient of the discrete scheme is unchanged, and only the time axis is
rescaled. This is confirmed to $6.8 \times 10^{-13}$~m/s over sixteen operating
points. Therefore, one playability map serves every stopped position on a string.
Since the plant is provably identical, every point of loss when an open-string
controller is deployed at a stopped position belongs to the controller. Most of it
traces to one feature, namely, an amplitude target specified as a displacement,
where the displacement scales with the reciprocal of the fundamental and the
velocity does not. That invariance is specific to the rigid stop. A real finger is
flesh, rather than a hard support, and its finite impedance would add a
length-dependent boundary loss that breaks the invariance exactly. This is the same
omission as the rigid bridge, and it is left to future work.

Rendering the model as audio supports the diagnostic from a second direction.
Rendering one stroke from the interior of each regime separates Helmholtz motion
from every other regime by a factor of eight to twenty-four in the cepstral peak
prominence computed on the bridge force. It also supplies the counterexample which
a periodicity measure cannot, since a string which the bow never grips is
perfectly periodic, and it scores above Helmholtz motion on harmonicity, at a stick
fraction of zero. Over seventy-two renderings spanning four strings and four
controllers, the same measure tracks the mechanical Helmholtz percentage at a
Spearman correlation of $+0.803$, where a plain harmonicity measure reaches
$+0.324$. The acoustic evidence also agrees that, where the lookup is right, the
learned policy matches it, rather than beating it.

A regime classifier which does not test explicitly for a stick phase will label
small-amplitude periodic slipping as Helmholtz motion, since it has the right
period and the right spectral peak, while having none of the underlying physics.
The stick fraction is inexpensive to compute, and it should be checked before any
playability map is trusted.

Every extension named in Section~\ref{sec:motivation} changes the plant which a
controller sees, and two choices made here were made so that they would survive
it. The regime classifier tests for a stick phase, rather than for periodicity. A
stick phase is the property which a radiating body would leave intact, while
altering the spectrum on which a periodicity test keys. The commands are referred
to the frog, rather than to the contact point, which is where the controls of a
player remain fixed if the contact acquires an angle, a tilt, or a reversal.
Neither claim is tested here, since the terminations are rigid and the bow does
not reverse. Neither choice costs anything on one open string, and each is what allows this layer to be built on.

\section*{Code Availability}

The simulation, training and analysis code, with the stored results behind every figure and table, will be made available on publication.

\bibliographystyle{IEEEtran}
\bibliography{ms}

\end{document}